\documentclass[traditabstract]{aa} 
\usepackage{graphicx,natbib,longtable,txfonts,lscape,lineno}
\newcommand{\FRONE}{FRI{\sl{CAT}}}
\newcommand{\FRTWO}{FRII{\sl{CAT}}}
\newcommand{\WAT}{WAT{\sl{CAT}}}
\newcommand{\ergs}{\>{\rm erg}\,{\rm s}^{-1}}
\newcommand{\kms}{$\rm{\,km \,s}^{-1}$}

\usepackage{array}
\newcolumntype{P}[1]{>{\centering\arraybackslash}p{#1}}

\begin{document}

   \title{WAT{\sl{CAT}}: a tale of wide-angle tailed radio galaxies}

   \subtitle{}

   \author{V. Missaglia\inst{1,2,5}
          \and
          F. Massaro\inst{3,4,5,6}
          \and
          A. Capetti\inst{5}
          \and
          M. Paolillo\inst{1,7}
          \and
          R.P. Kraft\inst{2}
          \and
          R.D. Baldi\inst{8}
          \and
          A. Paggi\inst{3,4,5}
          }

   \institute{Dipartimento di Fisica ``Ettore Pancini'', Universit\`a di Napoli Federico II, via Cintia, 80126 Napoli, Italy
\and    
 Center for Astrophysics $\mid$ Harvard \& Smithsonian, 60 Garden Street, 02138 Cambridge, MA, USA
\and
Dipartimento di Fisica, Universit\`a degli
     Studi di Torino, via Pietro Giuria 1, 10125 Torino, Italy 
\and
Istituto Nazionale di Fisica Nucleare, Sezione di Torino, 10125 Torino, Italy
\and
INAF\ - Osservatorio Astrofisico di Torino, via Osservatorio 20, 10025 Pino Torinese, Italy
\and 
Consorzio Interuniversitario per la Fisica Spaziale (CIFS), via Pietro Giuria 1, 10125 Torino, Italy
\and
INFN, Sezione di Napoli, via Cintia, 80126 Napoli, Italy
\and
Department of Physics and Astronomy, University of Southampton, Highfield, SO17 1BJ, UK.}
   \date{\today}

   \abstract {
We present a catalog of 47 wide-angle tailed radio galaxies (WATs), the \WAT; these galaxies were\ selected by combining observations from the National Radio Astronomy Observatory/Very Large Array Sky Survey (NVSS), the Faint Images of the Radio Sky at Twenty-Centimeters (FIRST), and the Sloan Digital Sky Survey (SDSS), and mainly built including a radio morphological classification. We included in the catalog only radio sources showing two-sided jets with two clear ``warmspots'' (i.e., jet knots as bright as 20\% of the nucleus) lying on the opposite side of the radio core, and having classical extended emission resembling a plume beyond them. The catalog is limited to redshifts $z\leq 0.15$, and lists only sources with radio emission extended beyond 30 kpc from  the host galaxy. We found that host galaxies of \WAT\ sources are all luminous ($-20.5 \gtrsim M_r \gtrsim -23.7$), red early-type galaxies with black hole masses in the range $10^8\lesssim $ M$_{\rm BH} \lesssim 10^9$ M$_\odot$. The spectroscopic classification indicates that they are all low-excitation galaxies (LEGs). Comparing WAT multifrequency properties with those of FR\,I and FR\,II radio galaxies at the same redshifts, we conclude that WATs show multifrequency properties remarkably similar to FR\,I radio galaxies, having radio power of typical FR\,IIs.
   }  
      \maketitle

\section{Introduction}
We are now living in the golden age of wide and deep multifrequency surveys. This is extremely useful for investigating the properties  \citep[see, e.g.,][]{kauffmann03b}, nature, cosmological evolution \citep[see, e.g.,][]{2006AAS...209.8614B, 2014MNRAS.445..955B, 2007MNRAS.381.1548K}, and environments \citep[see, e.g.,][]{1988A&A...189...11M,2018arXiv181107949C,2018arXiv181107943H} of several classes of extragalactic sources, such as radio-loud active galaxies, having a low sky density (i.e., a small number of sources per square degree). Thus, we recently embarked on a project whose goal is to  create homogeneous and complete catalogs of different classes of radio galaxies, uniformly selected on the basis of their radio morphological properties \citep{fanaroff74}. The first three catalogs, restricted to the local Universe (at $z<0.15$) and covering the major classes of radio galaxies (namely FR\,I, FR\,II, and FR\,0)\ have been recently published \citep{2017A&A...598A..49C, 2017A&A...601A..81C, 2018A&A...609A...1B}, while the present work focuses again on a class of radio galaxies with extended radio emission beyond the host galaxy.

Both FR\,I and FR\,II radio galaxy catalogs were created on the basis of the radio morphological classification scheme of extragalactic radio sources with large-scale structures introduced by \citet{fanaroff74}. Thus, radio galaxies can be broadly divided in two main classes, on the basis of the ratio of distance between the regions of highest radio surface brightness on opposite sides of the central nucleus (associated with the host galaxy) to the total extent of the source, up to the lowest brightness contour in the radio images at 1.4 GHz. Edge-darkened radio sources are classified as FR\,Is, while edge-brightened sources as FR\,IIs. Then, in the FR0{\sl{CAT}}, we  presented a sample of 108 ``compact'' radio sources, called FR\,0s \citep{2015A&A...576A..38B}, with $z\le$ 0.05 and having a radio size $\lesssim$ 5 kpc, which mostly appear unresolved up to 0.2" and occasionally show resolved jets \citep{2019MNRAS.482.2294B}. These also show the typical optical spectrum of low-excitation radio galaxies.

Here we present the fourth catalog in the series, which focuses on an intermediate class of radio galaxies between FR\,Is and FR\,IIs, having extended radio emission well beyond the optical profiles of their host galaxies: the Wide-Angle Tailed radio galaxies catalog.
\begin{figure}
    \includegraphics[height=6.cm,width=8.4cm,angle=0]{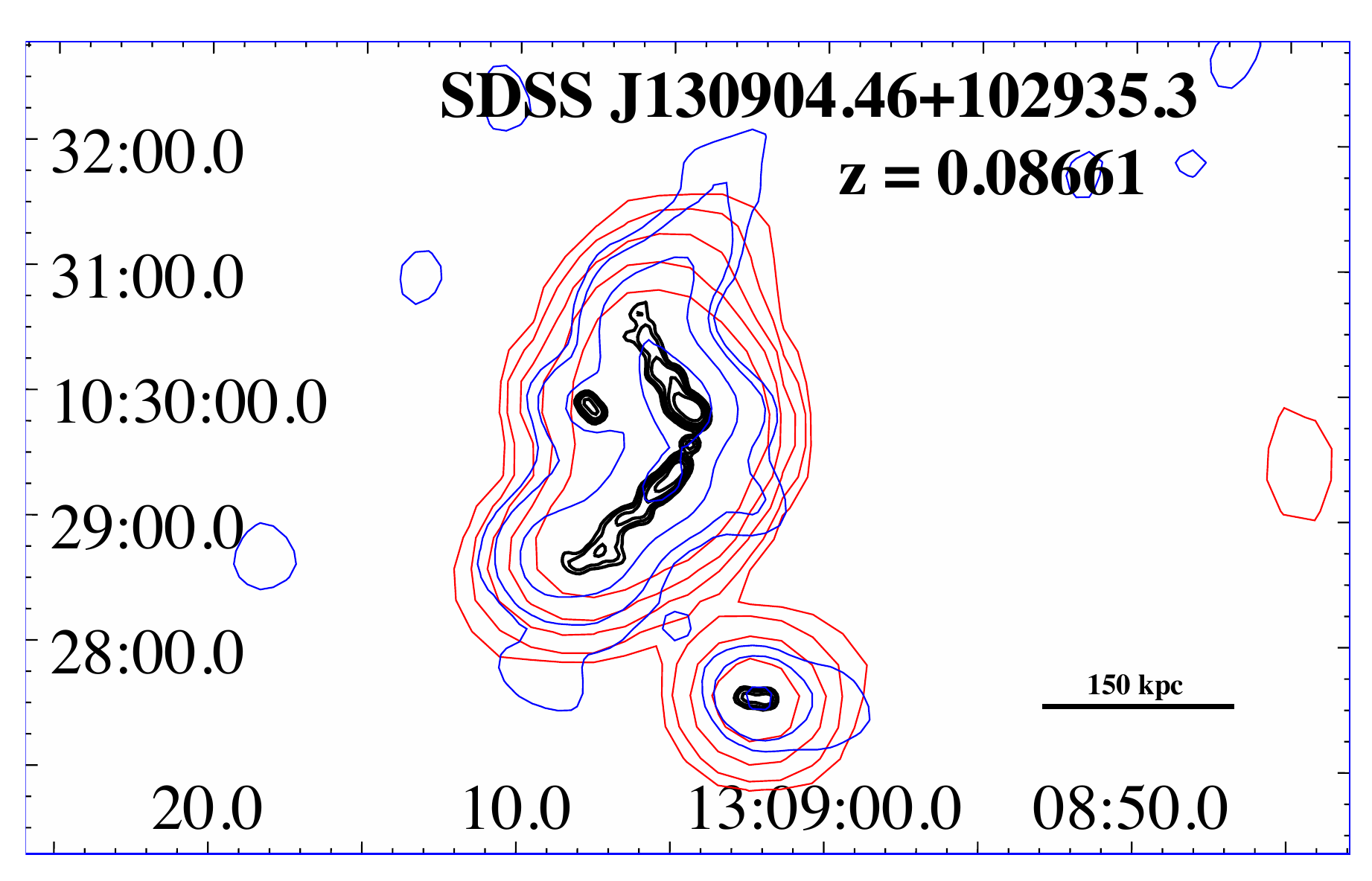}
    \caption{Radio contours of the \WAT\ source SDSS J130904.46+102935.31.4: 1.4 GHz FIRST (black), 1.4 GHz NVSS (red), and 150 MHz TGSS (blue). Contours levels and increase factors are summarized in Table \ref{tableA}. The field of view is 3'$\times$3'. The source name and redshift are shown in the upper right corner.}
    \label{images}
\end{figure}

Observed for the first time with the National Radio Astronomy Observatory (NRAO) interferometer at 2.7 GHz during a survey of sources in the Abell clusters \citep{owen76}, wide-angle tailed radio galaxies (hereafter WATs) were given this name because their radio morphology presents bright hotspots (called  ``warmspots'') closer to their radio core with respect to FR\,IIs and with extended radio plumes beyond them \citep{1990ApJS...72...75O}.
\citet{1993LNP...421....1L} defines WATs as a class of sources that initially have well-collimated jets on a kiloparsec scale and that suddenly flare into diffuse plumes, which may be significantly bent.

\citet{owen76}, noting that this source class has higher radio luminosity than the radio sources with narrower tails (also known as head-tailed radio galaxies), proposed that their curved radio structure could be due to the motion of the host galaxy through the intergalactic medium (IGM), moving more slowly than the head-tailed sources. This scenario is supported by the fact that WATs are generally associated with more dominant cluster galaxies \citep{1994AJ....107.1637B,1995PhDT........15P}.

\citet{2000MNRAS.311..649S} investigated the WAT radio structure using a sample selected from those lying in Abell galaxy clusters. These authors suggested that because the Universe evolves hierarchically with galaxy clusters forming through mergers of groups, WAT bent or curved tails are due to the ram pressure that originated in the merging processes.

In FR\,I showing twin-jet tails (i.e., double sided), their well-collimated jet flow undergoes significant bulk deceleration on scales of a few kpc \citep{1999MmSAI..70..129F,2013MNRAS.430.1976M,2013MNRAS.435..718M,2013MNRAS.434.3030B, 2014MNRAS.437.3405L,2016A&A...596A..12M,2018arXiv181200657M}, slowing down from relativistic speeds to transonic or subsonic regime with respect to the external medium \citep{2002MNRAS.336.1161L,1994AuJPh..47..669B}. This forms a jet that may in some cases make a smooth transition into a plume similar to those  observed in WATs \citep{1993LNP...421....1L}. The inner jets of FR\,Is are often also strong sources of X-ray emission \citep{2001MNRAS.323L..17H,2003ApJ...593..169H}. WAT jets should be similar, and so we could expect to detect synchrotron X-ray emission arising from knots and warmspots \citep[see also][]{2011ApJS..197...24M}. It is worth noting that for all WATs being observed in the X-rays, as for FR\,I jets, the regions where their jets decelerate are coupled with high-energy particle acceleration that yield to X-ray synchrotron emission (e.g., \citealt{2001MNRAS.323L..17H, 2001MNRAS.326L...7W,2002MNRAS.334..182H}).  On the other hand, WAT plumes should not be similar to regions at the edges of FR\,I radio structures since according to the model outlined by \citet{2004MNRAS.349..560H} in WATs the jet deceleration in the plumes takes place at a single discrete location, which relates the termination length (defined as the mean of the two linear distances between the core and what we have called warmspots located at the base of the plume) to cluster richness. The authors, assuming intrinsic symmetry of the jets and jet/counterjet luminosity ratios related to relativistic beaming effects, obtained a constraint on the beaming speed $\beta$ found to be weakly relativistic ($\beta$=0.3). Earlier, for a sample of 11 WATs, \citet{1993ApJ...408..428O} used their radio surface brightness and their spectral energy distributions to model the flow fields and the bending dynamics of the sources, and with the sidedness of the jets (side-to-side ratio of their luminosity) constrained the flow velocity. According to their results, the initial jet velocity in these galaxies is unlikely to be greater than 0.2$c$. The measured surface brightness ratios have been interpreted in the light of Monte Carlo simulations by \citet{1984BAAS...16..956O} that set an upper limit of 0.2$c$ to the bulk velocity of the beams in narrow-angle tailed (NAT) radio galaxies.
\citet{2005MNRAS.358.1394J} assume that the WATs analysed in their work lie on the plane of the sky, and find that the speed of fluid flow through the jets is again mildly relativistic, having approximately $\beta$ values of 0.3. (estimated from jet/counterjet ratios by \citealt{2004MNRAS.349..560H}). Jets in WATs do not appear significantly brighter than the counterjet, as one would expect from relativistic Doppler boosting.
In a following work, \citet{2006MNRAS.368..609J} measuring the distribution of the jet-sidedness ratios for the same WAT sample and assuming jets are beamed, obtained jet speeds in the range (0.3–0.7)$c$, in agreement with previous results.
\citet{2004Ap&SS.293....1G} showed that multi-epoch studies of radio-loud AGNs allow us to measure directly the apparent jet pattern velocity, which can be used to derive constraints on the intrinsic velocity of the pattern flow and the jet orientation with respect to the line of sight. Then assuming that jets are intrinsically symmetric, the jet/counterjet brightness ratio and the core dominance (comparison between the expected intrinsic core radio power, derived from the unboosted total radio power, and the observed core radio power) can also be used to constrain the jet bulk velocity and orientation with respect to the line of sight. 
As pointed out by \citet{2005MNRAS.359.1007H}, similarly to FR\,IIs, WAT jets remain weakly or mildly relativistic and well collimated until they undergo rapid deceleration at a distance of few tens of kpc from the nucleus, creating the so-called warmspots, the main feature of their radio morphology.

\citet{best05b}, \citet{baldi10b}, and \citet{best12}, to name recent examples, have analyzed properties of low redshift radio emitting AGNs using multifrequency information available thanks to the National Radio Astronomy Observatory/Very Large Array Sky Survey (NVSS), the Faint Images of the Radio Sky at Twenty-Centimeters (FIRST), and the Sloan Digital Sky Survey (SDSS). Here we just include radio morphological information to create the first catalog of WATs in the local Universe (hereafter the \WAT).
The catalog we have built is the first to have such a high level of completeness, and there are no other WATs catalogs with this population selected with such strict criteria. WATs present a number of interesting problems and open questions, and tackling them can help us  achieve a better understanding of the dynamics of extragalactic radio sources and their interaction with the intergalactic medium (IGM).

This paper is organized as follows. In Sect. 2 we describe the selection criteria of sources included in the \WAT; the radio and optical properties of the selected sources are then presented in Sect.\ 3. A comparison with FR\,I and FR\,II radio galaxies is discussed in Sect. 5. Section 6 is dedicated to our summary and conclusions.

A flat cosmology with $H_0=67.8 \, \rm km \, s^{-1} \,Mpc^{-1}$, $\Omega_{\rm M}=0.308$, and $\Omega_\Lambda=0.692$ \citep{ade16} has been adopted throughout the paper. For our numerical results, we use c.g.s. units unless stated otherwise. Spectral indices $\alpha$ are defined by the usual convention on the flux density, $S_{\nu}\propto\,\nu^{-\alpha}$. SDSS magnitudes are in the AB system and are corrected for  Galactic extinction. {\em WISE} magnitudes are instead in the Vega system and are not corrected for extinction since (as shown by, e.g., \citealt{dabrusco14}) such correction affects mainly the magnitude at 3.4 $\mu$m of sources lying at low Galactic latitudes (and by less than $\sim$3\%). Not all WATs are detected in all three {\em WISE} bands centered at 3.4, 4.6, and 12 $\mu$m. The detection rate is 78\% at 12$\mu$m, while it is 100\% at 3.4 and 4.6 $\mu$m.

\section{Sample selection}
\label{sample}
We searched for WATs in the sample of 18,286 radio sources built by \citealp{best12} (hereafter BH12) by limiting our search to sources classified as active nuclei (AGN) on the basis of their radio emission. The BH12 catalog was created on the basis of observations available in the archive of the SDSS (DR7; \citealt{abazajian09}),\footnote{{\tt http://www.mpa-garching.mpg.de/SDSS/}.} and combined with those of NVSS (\citealt{condon98}) and FIRST (\citealt{becker95}), having the same footprint of the SDSS. For the radio data sets a radio flux density threshold of 5 mJy in the NVSS was adopted and the BH12 sample was also cross-matched with optical spectroscopic catalogs  built by groups at the Max Planck Institute for Astrophysics and  Johns Hopkins University \citep{bri04,tre04}.  

As was done in previous catalogs of FR\,Is and FR\,IIs, we selected 3,357 sources with redshift $z\leq 0.15$. Then we visually inspected all FIRST images of each of these  sources and selected only those whose radio emission  extended beyond 30 kpc from the center of the optical host galaxy, at the sensitivity of the FIRST. This cut led to a selected sample of 741 sources. We adopted a threshold of 30 kpc, mainly due to the limited resolution of the FIRST survey, as an estimate of the maximum size of the host galaxies. At redshift $z=0.15$ this distance corresponds to 11$\farcs$4, thus ensuring that all 741 selected sources are well resolved with the 5$\arcsec$ resolution of the FIRST images. This allowed us to properly explore their morphology. 

We chose a reference surface brightness level of 0.45 mJy/beam, corresponding to about three times the typical rms of FIRST images, at $z=0.15$. As performed in previous radio galaxy catalogs, we also corrected the radio surface brightness for the cosmological dimming and we applied a $k$ correction by assuming a spectral index of 0.7, typical of lobes and plumes of radio galaxies. 

Then we selected WAT sources out of 741 objects on the basis of their radio morphology in FIRST images, considering strict criteria. Thus we included in the final sample only those radio sources showing two-sided jets with two clear warmspots (i.e., jet knots as bright as 20\% of the nucleus) lying on the opposite side of the radio core and having classical extended emission resembling a plume beyond. Four authors performed this analysis independently and only the sources for which there was agreement by at least three of us were included in the final sample. The final catalog of WATs lists 47 sources, the first of this kind, homogenous at low redshifts. 

In Fig. \ref{images} we present the radio image with surface brightness contours from FIRST, NVSS, and TIFR  Giant Metrewave Radio Telescope (GMRT) Sky Survey (TGSS) radio maps of a typical \WAT\ source. Images of all sources belonging to the \WAT\ are available in the Appendix, while their main properties (SDSS name, redshift, radio luminosity at 1.4 GHz and optical parameters) are listed in Table \ref{table2}. For each WAT, we provide both radio and [OIII] luminosities. 

Then given the measurements of the velocity dispersions, available in the SDSS database, we also computed their black hole masses M$_{BH}$ adopting the relation $\sigma_* - $M$_{\rm BH}$ of \citet{tremaine02}. The masses range between $7.9 \lesssim\log $M$_{\rm BH} \lesssim 9.1 $M$_\odot$, and their distribution peaks at $\sim10^{8.6} $M$_\odot$. The uncertainty on the M$_{\rm BH}$, on the order of a factor of 2, is mainly dominated by the dispersion of the correlation rather than by the uncertainty on $\sigma_*$ measurements.

According to \citet{montero09} and the redshift completeness of the SDSS, the \WAT, like the \FRONE\ and \FRTWO, is statistically complete at a level of $\sim$90\% in the optical band, whereas this extremely low level of incompleteness is only due to a random loss of $\sim$10\% of the potential spectroscopic targets \citep[see, e.g.,][]{zehavi02}. Therefore, there is the possibility that radio sources featuring extended radio emission below the 3$\sigma$
threshold in the FIRST images were missed.

\begin{figure}
        \includegraphics[height=7.cm,width=9.2cm,angle=0]{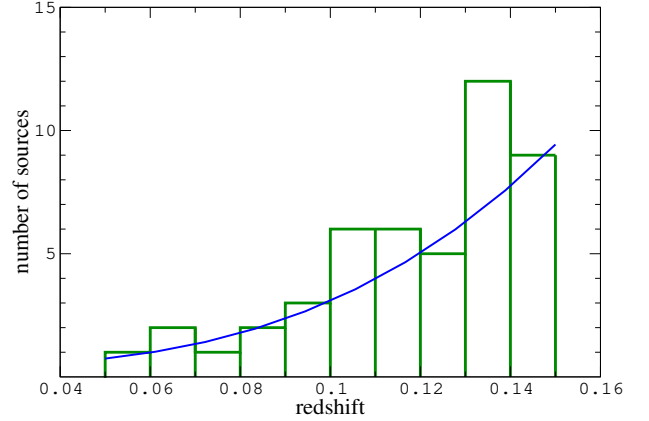}
        \caption{Distribution of the \WAT\ sources redshifts. The blue line refers to the $N \propto z^3$ relation normalized to the number of sources in the $z$ bins between 0.05 and 0.09. The number of WATs in the $z$ range between 0.13 and 0.15 appears to be consistent with this trend, indicating that the catalog is almost complete in the explored redshift range (see Section 2 for additional details).}
        \label{redfit}
\end{figure}

On the other hand, to estimate the completeness of our catalog at radio frequencies, we re-scaled the luminosities of all sources in the $z$ range between 0.05 and 0.09 computing their flux densities at 1.4 GHz and randomly assigning redshifts of sources in the last $z$ bin (see Fig. \ref{redfit}). Re-scaled flux densities all lie above the NVSS threshold. We also verified that the number of sources in the last $z$ bin is consistent with the expected number extrapolated from the $N \propto z^3$ scenario normalized to the first four $z$ bins. We found a good agreement, as shown in Fig. \ref{redfit}. Consequently, we can assert that our sample is statistically complete in the radio band within the $z$ range considered in our selection. We measured the jet sidedness values for our sample, finding  a value that is generally close to unity (as remarked in \citealt{1993ApJ...408..428O}) consistent with previous literature results.

Finally, we note that the BH12 catalog includes sources whose NVSS flux density at 1.4 GHz is higher  than 5 mJy, while our selection is mainly based on the surface brightness distribution of FIRST images.

\section{Multifrequency properties of WATs and their host galaxies}
\label{hosts}

\subsection{WAT hosts: infrared and optical properties}
\begin{figure}
        \includegraphics[height=7.cm,width=9.2cm,angle=0]{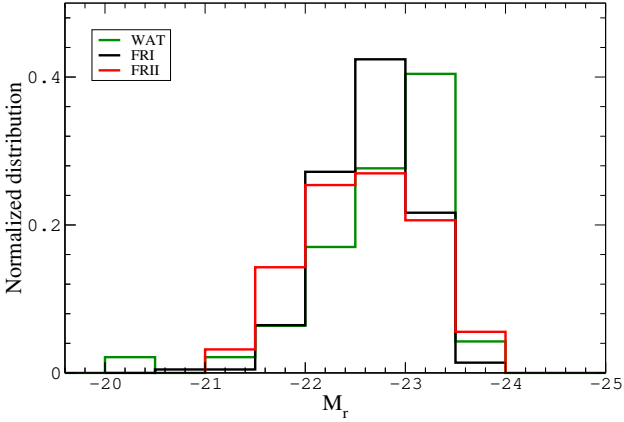}
        \caption{Comparison between the normalized distributions of the  host absolute magnitude in the $r$-band ($M_{r}$) for different classes of radio galaxies. The histograms show the  sources listed in the \FRONE\ (red) and   the \FRTWO\ (black), and  the selected \WAT\ sources (green). 
        }
        \label{mhist}
\end{figure}

All optical spectra obtained from the SDSS database were inspected. We discovered that all sources minus one listed in the \WAT\ are classified as low-excitation galaxies (LEGs), in agreement with the classification reported by \citet{best12} based on the ratios of the optical emission lines in their SDSS spectra.  The only exception is SDSS J155343.59+234825.4, being a high excitation galaxy (HEG). Given the high level of completeness of the \WAT\, this is the first time that is proven on a statistical basis for this source population that the majority of WATs is LEG. The logarithm of [OIII] line luminosities spans a range between 38.9 and 40.4 erg/s, with a mean value of 39.72 erg/s in log scale. However, it is worth noting that for four sources (SDSS J144904.27+025802.7, SDSS J141927.23+233810.2, SDSS J222455.24-002302.3, SDSS J155343.59+234825.4) we were not able to retrieve information regarding the [OIII] line luminosity because this measurement was not present in the SDSS database. Thanks to the multifrequency observations available, as done in previous radio galaxy catalogs, we used various diagnostics to provide a complete overview on the morphological and spectroscopic classification of the WAT host galaxies. The distribution of absolute magnitude of the \WAT\ host galaxies spans the range $-20.5 \gtrsim M_r \gtrsim -23.7$ with a distribution peaking at $M_r \sim -23$ (see Fig. \ref{mhist}).

The distribution of the concentration index $C_r$, defined as the ratio between the radii including 90\% and 50\% of the light in the $r$-band was obtained for each source directly from the SDSS database. The $C_r$ distribution is consistent with WATs being hosted in early-type galaxies (ETGs). In general ETGs show higher values of $C_r$ than late-type galaxies (LTGs). The thresholds  adopted to distinguish between ETGs and LTGs are the same used for the \FRONE\ and \FRTWO\ catalogs \citep[see][for more details]{strateva01,nakamura03,shen03,kauffmann03b,bell03,bernardi10}.

We compared $C_r$ values with those of the Dn(4000) spectroscopic index, defined as the ratio of the flux density measured on the red side of the Ca~II break (4000--4100 \AA) to that on the blue side (3850--3950 \AA) \citep{balogh99}. Low redshift ($z < 0.1$) red galaxies show Dn(4000)$= 1.98 \pm 0.05$, which is a value that decreases to $ 1.95 \pm 0.05$ for $0.1 < z < 0.15$ galaxies \citep{capetti15}. The presence of young stars or of nonstellar emission reduces the Dn(4000) index \citep{1985ApJ...297..371H}.
\begin{figure}
        \includegraphics[height=5.7cm,width=8.3cm,angle=0]{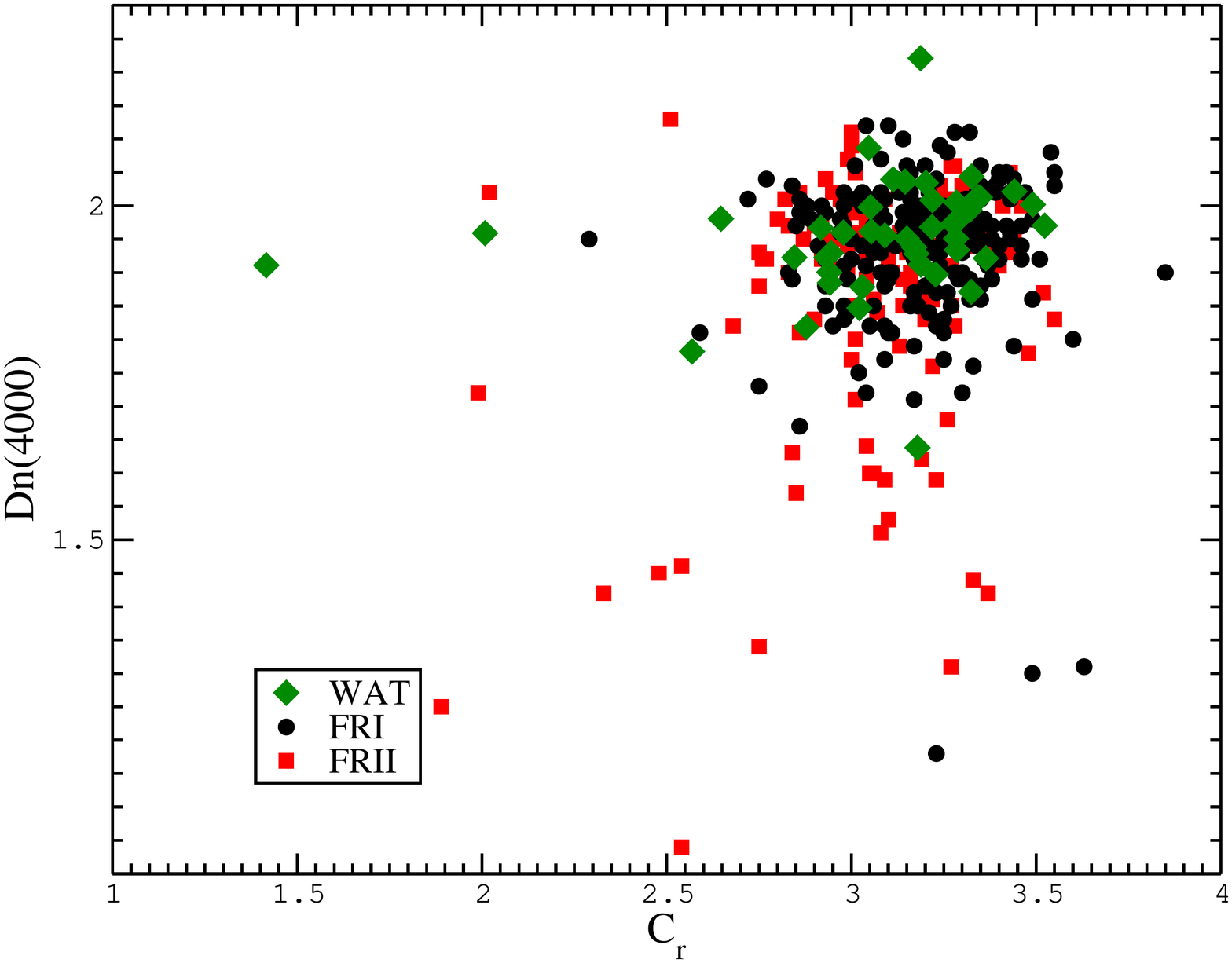}
    \includegraphics[height=7.cm,width=9cm,angle=0]{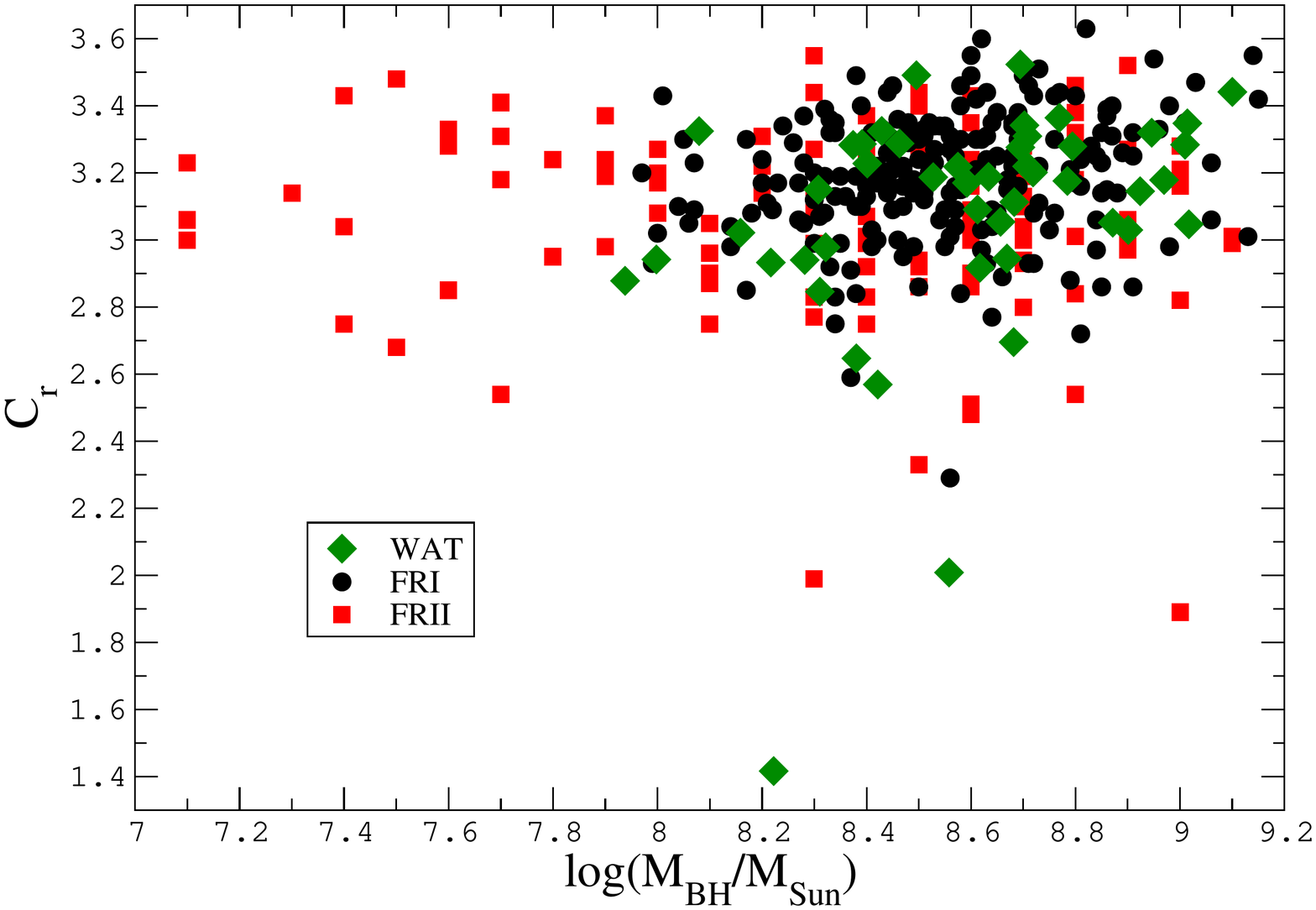}
        \caption{\textit{Upper panel} Dn(4000) spectroscopic index vs. concentration index $C_r$ for \WAT\ sources (green diamonds), \FRONE\ sources (black dots),  and \FRTWO\ sources (red squares). It appears that WATs are hosted in red early-type galaxies, as FR\,Is are, while high-excitation  FR\,IIs show lower values of Dn(4000). \textit{Lower panel} Concentration index $C_r$ vs. black hole
mass (in solar units). WATs show a similar behavior to FR\,Is, while FR\,IIs extend to  lower black hole masses with respect to the \WAT\ sample, probably due to relatively large errors particularly in the measurement of $\sigma$*, a possible uncertain identification of their spectroscopic class, or a substantial contribution from a bright nonthermal nucleus.}
        \label{crdn}
\end{figure}

\begin{figure}
        \includegraphics[height=8.cm,width=9.8cm,angle=0]{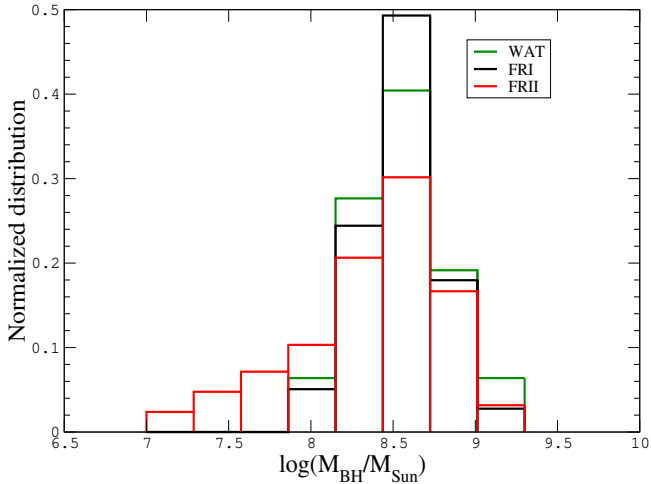}
        \caption{Comparison between black hole mass distributions for \WAT,\ \FRONE,\ and \FRTWO\ hosts.}
        \label{BHM}
\end{figure}

In Fig. \ref{crdn} we show the concentration index $C_r$ versus the Dn(4000) index (left panel) and versus M$_{\rm BH}$ (right panel) for the \WAT, \FRONE,\ and \FRTWO\ sources.  The vast majority of the WAT host galaxies lies in the region of high $C_r$ and Dn(4000) values, indicating that they are all typical red ETGs. We performed a two-dimensional Kolmogorov-Smirnov (KS) test (\citealt{1983MNRAS.202..615P,1987MNRAS.225..155F}) and we found that compairing WATs with FR\,Is and FR\,IIs we obtain (1-p-value) $\sim$ 10$^{-16}$, which implies that the two samples (WATs vs. FR\,Is and WATs vs. FR\,IIs) are not significantly different. There are only two exceptions, namely SDSS J095716.41+190651.2 and SDSS J135315.36+550648.2,  both of which have low values of  $C_r$. The distribution of black-hole masses (see Fig. \ref{BHM}) shows that WATs, FR\,Is, and FR\,II peak at the same value, but the FR\,IIs extend to lower mass values. 

\begin{figure}
                \includegraphics[height=7.cm,width=9cm,angle=0]{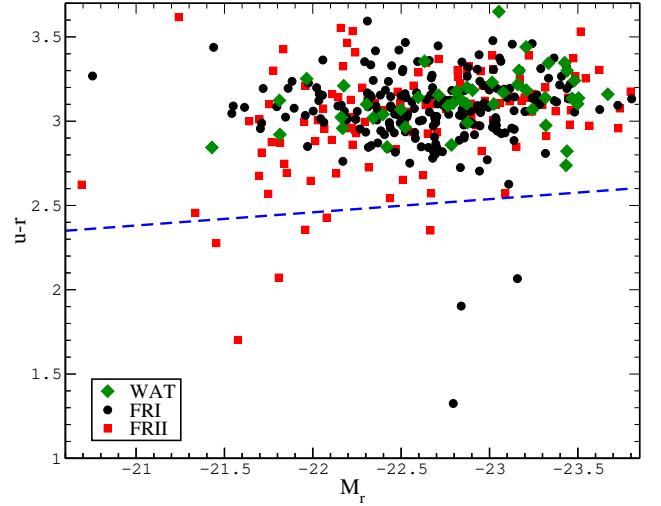}
        \caption{$u-r$ color vs. absolute $r$-band magnitude, $M_r$, for the \WAT\, \FRONE\ and \FRTWO\ hosts. Shown are   WATs (green diamonds),  FR\,Is (black dots),  and  FR\,IIs (red squares). WATs are associated with red ETGs. The dashed blue line represents the relation from \citet{schawinski09} that separates red and blue early-type galaxies.}
        \label{mur}
\end{figure}

We  also considered their $u-r$ color while studying the properties of the \WAT\ host galaxies. This allowed us to avoid uncertainties in the estimate of the Dn(4000) due to SDSS spectroscopic aperture, being  on the order of 3$\arcsec$ in diameter. In Fig. \ref{mur} we show the $u-r$ color versus the absolute $r$-band magnitude $M_{r}$ of the hosts. WATs are all located above the line separating red and blue ETGs (consistently with their $u-r$ distribution), thus following the same FR\,I trend \citep{schawinski09}. The fraction of blue ETGs decreases with increasing luminosity and these ETGs disappear for $M_{r} \lesssim -22.5$ . The lack of blue ETGs among the \WAT\ host galaxies is relevant. 

The $u-r$ distributions of WATs and FR\,IIs appear to be marginally different at a confidence level of 93\% (p-value of 0.072). The same test performed for the Dn(4000) index gave us a p-value of 0.58 for WATs and FR\,Is and 0.010 for WATs and FR II\,s. For the sake of clarity, in all our plots the uncertainties on optical parameters are not shown; however, the average uncertainties are 0.08, 0.03, and 0.004 on $C_r$, Dn(4000), and $m_r$, respectively, and $\sim$9 \kms\ on $\sigma_*$.

\begin{figure}
\includegraphics[height=7.cm,width=9cm,angle=0]{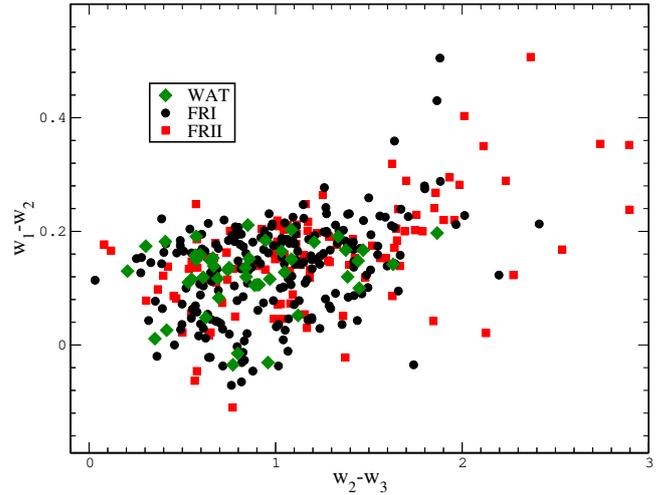}
\caption{{\em WISE} mid-IR colors of the \WAT \ hosts (green diamonds) compared to those of \FRONE\ (black dots) and \FRTWO\ (red squares). \WAT\ sources appear to have mid-IR colors mostly dominated by their host galaxies because they lie in the same color space region as elliptical galaxies and do not show contamination from the emission of their jets.}
\label{wise}
\end{figure}

As the last step, we cross-matched our \WAT\ with the latest release of the Wide-field Infrared Survey Explorer ({\em WISE}) source catalog (AllWISE) retrieving mid-IR magnitudes. The associations between the \WAT\ and the {\em WISE} catalog were computed adopting a 3\farcs3 angular separation, which corresponds to the combination of the typical positional uncertainty of the {\em WISE} all sky survey \citep{wright10} and that of the FIRST \citep{dabrusco14}. In Fig. \ref{wise} we show a comparison of the mid-IR colors of \WAT\ , \FRONE,\ and \FRTWO\ sources.
 
\WAT\ sources appear to have {\em WISE} colors mostly dominated by their host galaxies (they fall in the same region as elliptical galaxies; \citealt{wright10}) and not contaminated by the nonthermal emission of their jets, as for example occurs in blazars \citep{dabrusco12,massaro11,2016ApJ...827...67M}. Only one source (SDSS J155343.59+234825.4), classified as a  HEG, has $w_2-w_3 > 0.3$ and it is located at the onset of the sequence defined by the more luminous objects such as blazars and flat spectrum radio quasars \citep{massaro12}. Thus, the {\em WISE} infrared colors further support the idea that WATs are hosted in ``passive'' ETGs.
 
\subsection{Radio properties}
\begin{figure}
        \includegraphics[height=7.cm,width=9cm,angle=0]{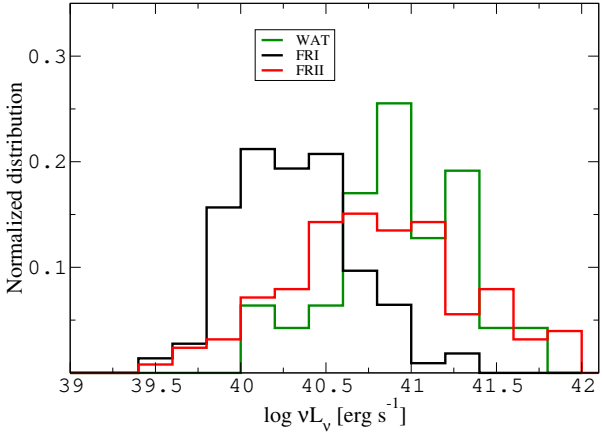}
        \caption{Normalized distribution of the NVSS luminosities at 1.4 GHz. The  histograms show the 219 \FRONE\ sources (black), the  47 \WAT\ sources (green), and the  122 \FRTWO\ sources (red). WATs are on average more powerful than FR\,Is and have radio luminosity similar to classical FR\,II radio galaxies. }
        \label{hist}
\end{figure}

In Fig. \ref{hist}, we show the distribution of the NVSS radio luminosity at 1.4 GHz ($\nu$L$_{\nu}$) for all the sources belonging to the \WAT\ in comparison with FR\,I and FR\,II listed in \citet{2017A&A...598A..49C, 2017A&A...601A..81C}. The distribution of radio luminosity at 1.4 GHz of the \WAT\ covers the range L$_{1.4}$ = $\nu_{\rm r}$L$_{\rm r}$ = $\sim 10^{40.1} - 10^{41.6}$ $\ergs$. The \citet{fanaroff74} separation between FR\,Is and FR\,IIs translates, with our adopted cosmology and by assuming a spectral index of 0.7 between 178 MHz and 1.4 GHz, into L$_{1.4} \sim 10^{41.6}$ $\ergs$. 

The distribution of the NVSS radio luminosity at 1.4 GHz clearly shows that WATs are on average more powerful than FR\,Is and have radio luminosity similar to classical FR\,II radio galaxies. According to a simple KS test we found WATs and FR\,IIs are indistinguishable, while for the comparison between WATs and FR\,Is these two populations differ within a confidence level of $\sim$10$^{-15}$.

\begin{figure}
        \includegraphics[height=7.cm,width=9cm,angle=0]{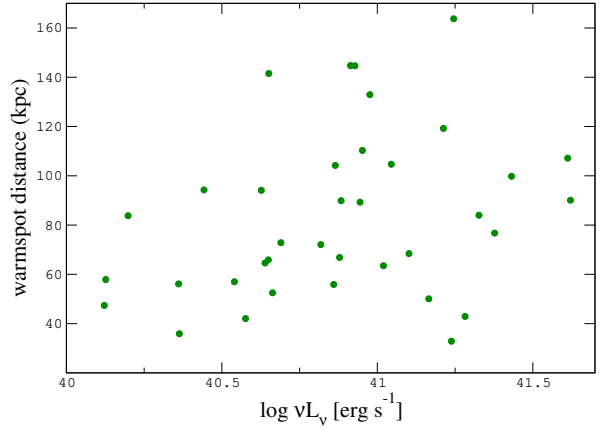}
        \caption{NVSS radio luminosity at 1.4 GHz vs.  linear size, i.e.,  physical separation between the warmspots. At large physical separations warmspots of low radio luminosity are not produced, and the region in the upper left corner appears empty.}
        \label{wd}
\end{figure}

We measured the angular separation between the peaks of surface brightness of two warmspots for each WAT on the FIRST radio maps at 1.4 GHz. Then we converted the angular separation into physical distance in kpc, thus defining their linear size: d$_{w}$. In Fig. \ref{wd} we show the relation between the d$_{w}$ and the WAT radio power. The low number of WATs, coupled with functional biases in the relations used to compute d$_{w}$ and the radio power, prevents us from evaluating a correlation between these two parameters; however, there is a hint that high radio power WATs do not produce warmspots closer to the radio core with respect to low power ones.

\section{Discussion}
Here we discuss the multifrequency properties of the \WAT\ in comparison with both the FR\,Is and FR\,IIs listed in the \FRONE\ and \FRTWO. 

The host galaxies of \WAT\ sources are remarkably uniform, being all luminous red ETGs and spectroscopically classified as LEGs, as occurs for the FR\,Is. On the other hand, their radio luminosity at 1.4 GHz is generally higher than that of FR\,Is, regardless of the value of L$_{[\rm OIII]}$. 

A KS test to compare the $r$-band absolute magnitude distributions of WATs, FR\,Is, and FR\,IIs was performed. 
According to our analysis, comparing WATs with both FR\,Is and FR\,IIs we found that they do not belong to the same parent population within a confidence level of 99\%. Our result, and its possible interpretation, will be investigated in
future works on the large-scale environment of the whole sample.

In 1974 Fanaroff \& Riley  observed that all sources in their original sample with luminosity measured at 178 MHz lower than 2$\times$10$^{25}$ W/Hz/sr (assuming a Hubble constant of 50 \kms\ Mpc$^{-1}$) were classified as FR\,Is, while more powerful radio objects  all presented the FR\,II morphology \citep{fanaroff74}. When sources belonging to the two FR classes are plotted in a radio luminosity versus optical magnitude plane where the radio power is estimated at low radio frequencies (i.e., at hundreds of MHz), this dichotomy between the two FR classes is fairly sharp (see, e.g, \citealt{ledlow96}); however, \citet{2017A&A...598A..49C,2017A&A...601A..81C}, showed that when classifying radio galaxies homogeneously on the basis of their morphology at 1.4 GHz and including low power FR\,IIs, this neat separation appears to be only a selection effect.

\begin{figure*}
\includegraphics[height=7.cm,width=9cm,angle=0]{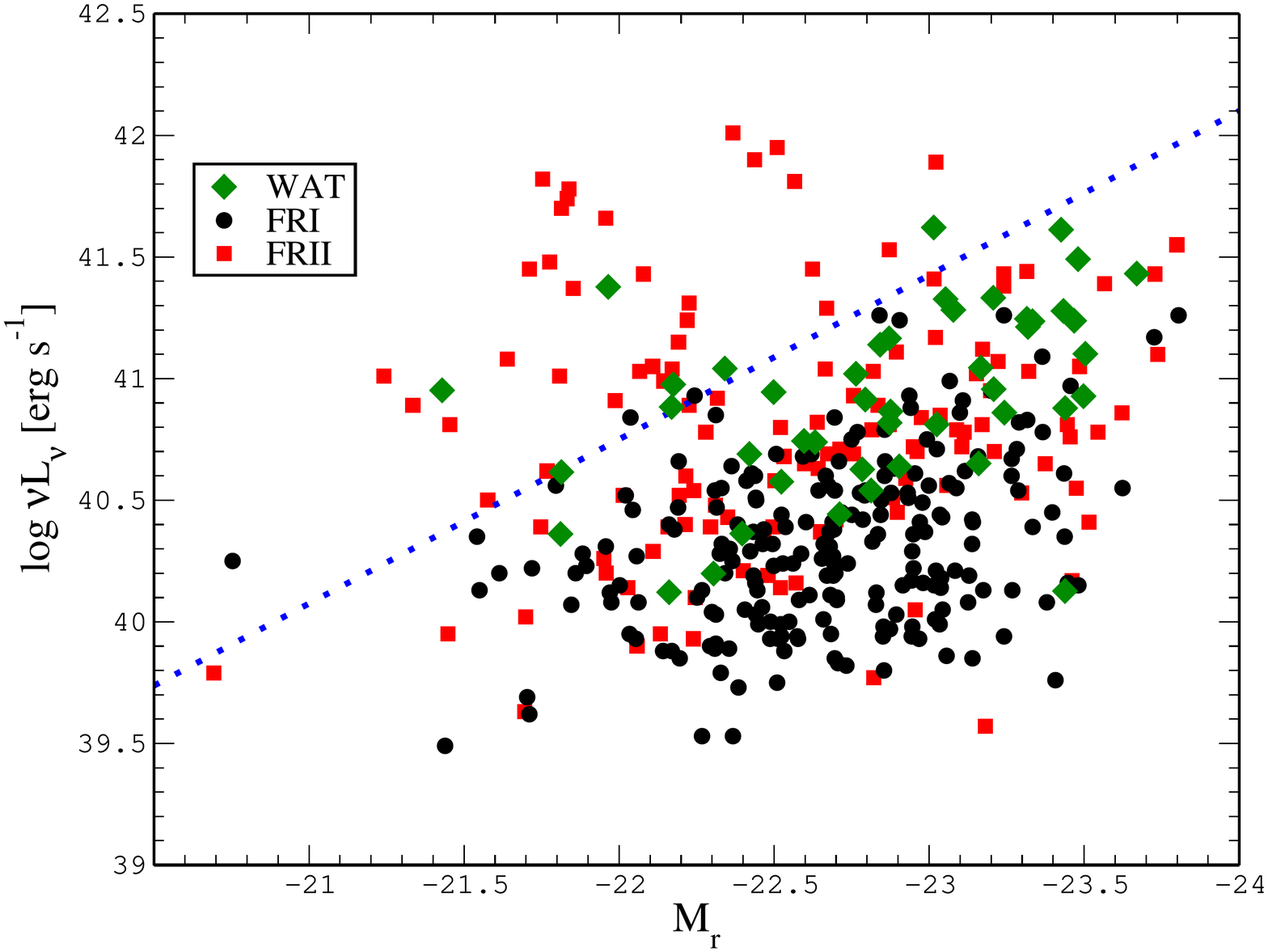}
\includegraphics[height=7.5cm,width=9.4cm,angle=0]{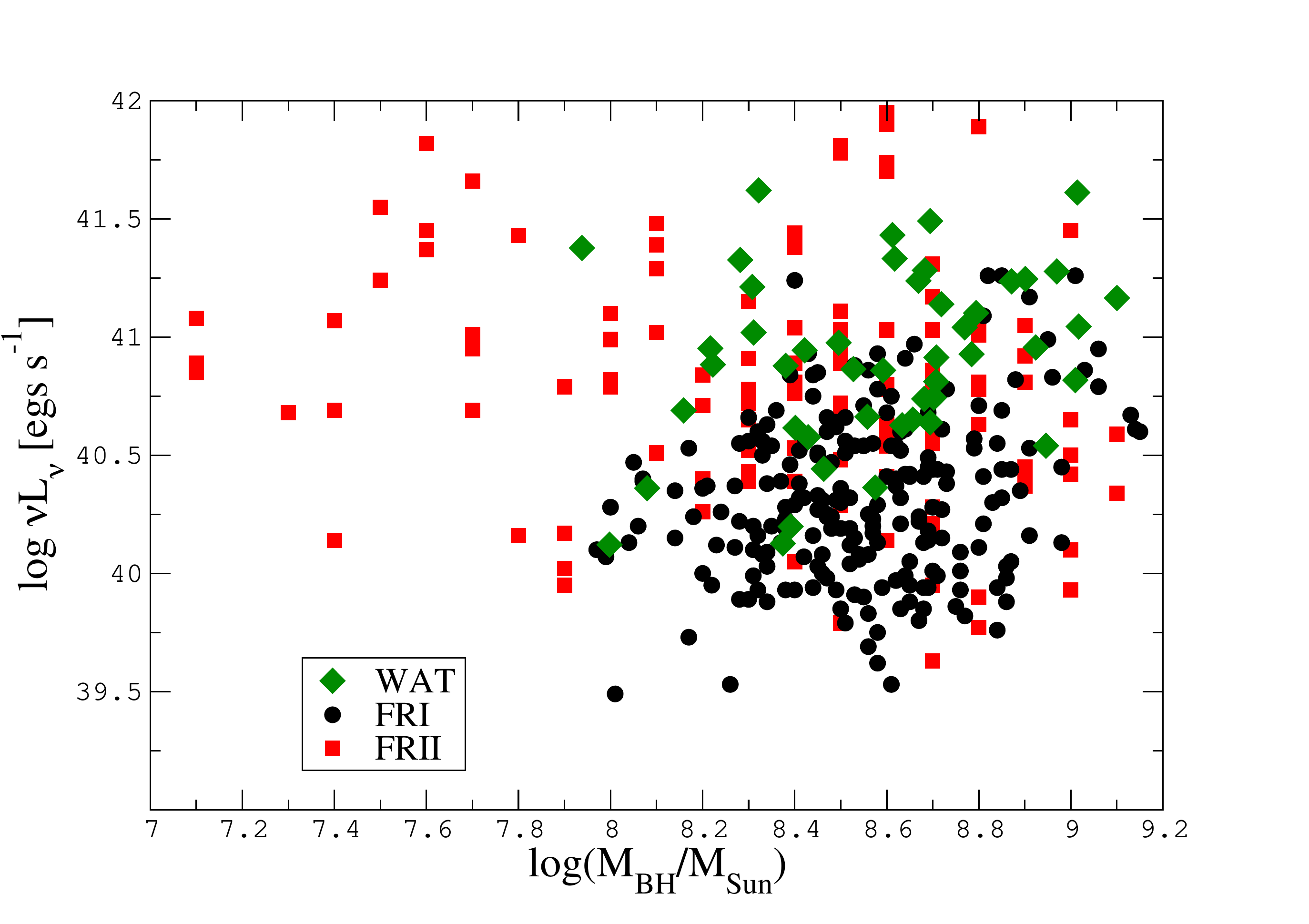}
\caption{\textit{Left panel}: Host absolute magnitude ($M_r$) vs. radio luminosity (NVSS) at 1.4 GHz, for \WAT, \FRONE,\ and \FRTWO\ sources (green diamonds, black dots, and red squares, respectively). The dashed blue line shows the separation between FR~I and FR~II, as reported by \citet{ledlow96}, to which we applied a correction of 0.34 mag to account for the different magnitude definition and the color transformation between the SDSS and Cousin systems. The bulk of the \WAT\ sources lies below the boundary between FR\,I and FR\,II radio galaxies. \textit{Right panel}: Radio luminosity (NVSS) at 1.4 GHz vs. black hole mass for \WAT, \FRONE,\ and \FRTWO\ (green diamonds, black dots, and red squares, respectively). WATs and FR\,Is show similar black holes masses, but WATs are more radio luminous.}
\label{lr}
\end{figure*}

As previously carried out for the \FRONE\ and \FRTWO\ samples, we built a re-scaled diagram plotting the radio power versus the optical magnitudes for our WATs. The bulk of \WAT\ sources lie below the boundary between FR\,I and FR\,II radio galaxies as shown in Fig.\ref{lr}, mostly lying in the region populated by FR\,I sources. We note that the dividing line between the two main classes of radio galaxies highlighted by Ledlow and Owen (1976) was appropriately shifted,  including a correction of 0.12 mag to the magnitude of the host galaxies and a correction of 0.22 mag to convert the Cousin magnitude system into the SDSS system \citep{fukugita96}. The location of WATs in this radio power versus optical magnitude confirms that the more powerful WATs tend to be  associated with more massive galaxies.

\begin{figure}
\includegraphics[height=7.cm,width=9cm,angle=0]{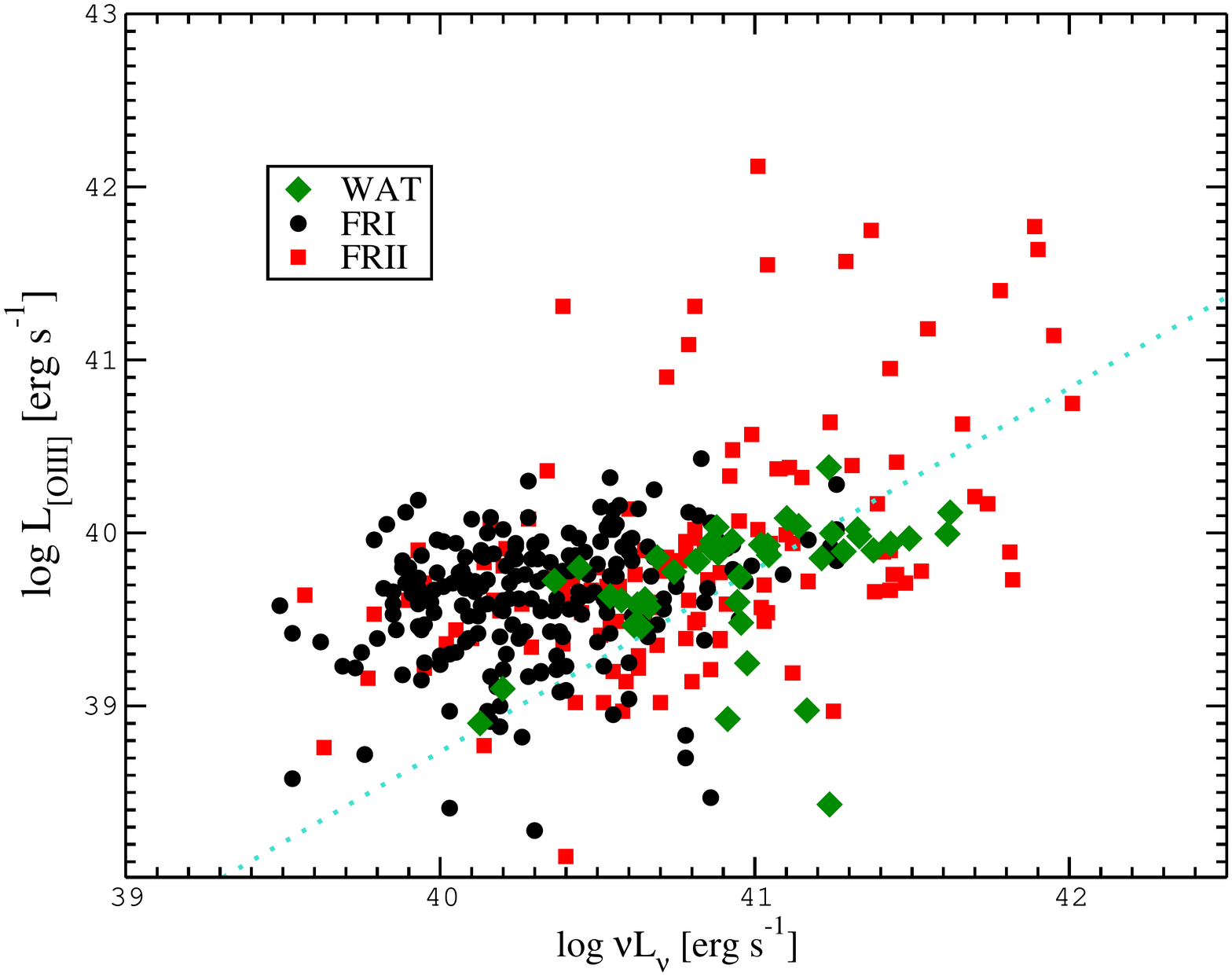}
\caption{Comparison between [OIII] line luminosity and radio luminosity (NVSS) of the \WAT\ , \FRONE,\ and \FRTWO\ samples (green diamonds, black dots, and red squares, respectively). The turquoise line shows the linear correlation between these two quantities derived from the FR~Is of the 3C sample \citep{buttiglione10}. No correlation between $\nu$ L$_{\nu}$ and L$_{\rm [OIII]}$ can be seen for  \WAT \ (as for \FRONE \ ) where, at a given line luminosity, the radio luminosities span  1-1.5 orders of magnitude.}
\label{lrlo3}
\end{figure}

As previously shown, classical FR\,Is belonging to the 3C sample show a positive trend between the [OIII] emission line and radio luminosity with a slope consistent with unity (e.g., \citealt{buttiglione10}). This indicates that a constant fraction of the AGN power, as measured from the L$_{[\rm OIII]}$, is then converted into radio power. \citet{buttiglione10} show that the same result, although with a different normalization, is found when considering the FR\,II radio galaxies again, selected from the 3C sample.

No neat trend or correlation between L$_{1.4}$ and L$_{\rm [OIII]}$ can be seen for WATs, as occurs for FR\,Is (see Figure \ref{lrlo3}), where at a given emission line luminosity radio power spans  two orders of magnitude. Similarly to 3C radio galaxies, it appears that no source has a L$_{1.4}$/L$_{\rm [OIII]}$ ratio exceeding $\sim$100, producing the scarcely populated region in the bottom right portion of this plot. On the other hand, several WATs show values of L$_{1.4}$/L$_{\rm [OIII]}$  below the 0.5 threshold. The broad distribution of the L$_{1.4}$/L$_{\rm[OIII]}$ ratio suggests that a broad range AGN bolometric power is then converted into radio emission. Furthermore, radio luminosity increases, not surprisingly, when the size of the radio source increases. Finally,  as occurred for FR\,Is, we note that the spectroscopic and host properties of the WAT host galaxies rule out the possibility of having the [OIII] emission line contaminated by  star formation, otherwise we would have been able to observe a trend, if  present, between radio power and emission line luminosity of the [OIII].

\section{Summary and conclusions}
We built a low redshift ($z\leq0.15$) catalog of 47 WATs called the WAT{\sl{CAT}}. This sample was selected out of the Best \& Heckman (2012) catalog, built combining observations available in the SDSS, NVSS, and FIRST surveys, but including a radio morphological classification. WATs are radio sources showing two-sided jets with two clear warmspots (i.e., jet knots as bright as 20\% of the nucleus) lying on the opposite side of the radio core, and having classical extended emission resembling a plume beyond them. The WAT classification is purely morphological and is based on the visual inspection of FIRST radio images at 1.4 GHz. We included sources in which the radio emission extends beyond 30 kpc from the position of their optical host galaxy. Our final catalog is also restricted to redshift  $z \leq 0.15$.

Our main results are listed below:
\begin{itemize}
    \item Using NVSS radio observations at 1.4 GHz, we find  that WATs tend to be more radio luminous than FR\,Is, and more similar to FR\,IIs;
    \item \WAT\ host galaxies are remarkably homogeneous. They do not show significant differences in their optical and infrared colors with respect to the general population of massive ETGs, and the presence of an active nucleus (and its level of activity) does not appear to affect their host galaxies;
    \item SDSS optical spectra of \WAT\ sources proved that these sources are all classified as low-excitation radio galaxies (LEGs). FR\,Is are all classified as LEGs, while FR\,IIs can be LEGs or high excitation radio galaxies (HEGs). This again shows the similar behavior between WATs and FR\,Is;
    \item $M_{r}$ versus $\nu$L$_{\nu}$ measured at 1.4 GHz showed that WATs follow the same behavior of FR\,II HEGs;
    \item using infrared data collected by {\em WISE}, WAT host galaxies appear to be similar to normal elliptical galaxies, with IR colors similar to those of FR Is, lacking significant contribution from active star formation; 
    \item values of the absolute magnitude in the $r$-band ($M_{r}$) are also in agreement that  WATs, as reported in literature, could be the brightest cluster galaxy when they reside therein;
    \item comparing the distribution of the concentration index  ($C_r$) and the dispersion index Dn(4000), we discovered that WATs, like FR\,Is, tend  not to be extremely blue and lack signatures of active star formation, which are sometimes found in FR\,IIs;
    \item from the comparison between the radio luminosity at 1.4 GHz obtained from NVSS and from the luminosity of the [OIII] emission line present in WAT optical spectra, we found that our selected WATs do not follow any trend between L$_{1.4}$ and L$_{\rm [OIII]}$, whose [OIII] emission line luminosity is even lower than expected for a given value of the radio power at 1.4 GHz. 
    
\end{itemize}

In summary, we conclude that WATs show multifrequency properties remarkably similar to FR\,I radio galaxies, being more powerful at radio frequencies and similar to typical FR\,IIs.

\begin{acknowledgements}
We thank the referee, Prof. S. Massaglia, for a careful reading of our manuscript and many helpful comments that led to improvements in the paper. This work is supported by the ``Departments of Excellence 2018 - 2022''
Grant awarded by the Italian Ministry of Education, University and
Research (MIUR) (L. 232/2016). This research has made use of resources
provided by the Compagnia di San Paolo for the grant awarded on the
BLENV project (S1618$_{}$\L1\_MASF\_01) and by the Ministry of Education,
Universities and Research for the grant MASF\_FFABR\_17\_01. This
investigation is supported by the National Aeronautics and Space
Administration (NASA) grants GO4-15096X, AR6-17012X, and GO6-17081X. F.M.
acknowledges financial contribution from the agreement ASI-INAF
n.2017-14-H.0 A.P. acknowledges financial support from the Consorzio
Interuniversitario per la fisica Spaziale (CFIS) under the agreement
related to the grant MASF\_CONTR\_FIN\_18\_02.
Part of this work is based on the NVSS (NRAO VLA Sky Survey): The National
Radio Astronomy Observatory is operated by Associated Universities, Inc.,
under contract with the National Science Foundation.
This publication makes use of data products from the Wide-field Infrared
Survey Explorer, which is a joint project of the University of California, Los
Angeles, and the Jet Propulsion Laboratory/California Institute of Technology,
funded by the National Aeronautics and Space Administration.
This research made use of the NASA/ IPAC Infrared Science Archive and
Extragalactic Database (NED), which are operated by the Jet Propulsion
Laboratory, California Institute of Technology, under contract with the
National Aeronautics and Space Administration.
Funding for SDSS-III has been provided by the Alfred P. Sloan Foundation, the
Participating Institutions, the National Science Foundation, and the
U.S. Department of Energy Office of Science. The SDSS-III web site is
http://www.sdss3.org/.  SDSS-III is managed by the Astrophysical Research
Consortium for the Participating Institutions of the SDSS-III Collaboration,
including the University of Arizona, the Brazilian Participation Group,
Brookhaven National Laboratory, University of Cambridge, Carnegie Mellon
University, University of Florida, the French Participation Group, the German
Participation Group, Harvard University, the Instituto de Astrofisica de
Canarias, the Michigan State/Notre Dame/JINA Participation Group, Johns
Hopkins University, Lawrence Berkeley National Laboratory, the Max Planck
Institute for Astrophysics, the Max Planck Institute for Extraterrestrial Physics,
New Mexico State University, New York University, The Ohio State University,
Pennsylvania State University, University of Portsmouth, Princeton University,
the Spanish Participation Group, University of Tokyo, University of Utah,
Vanderbilt University, University of Virginia, University of Washington, and
Yale University.
\end{acknowledgements}

\newpage
\onecolumn
\begin{table}[]
\begin{center}
\begin{tabular}{|l|r|r|r|r|r|r|}
\hline
  \multicolumn{1}{|c|}{Source name} &
  \multicolumn{1}{P{1.5cm}|}{Low contour level FIRST} &
  \multicolumn{1}{P{1.5cm}|}{Contour increase factor} &
  \multicolumn{1}{P{1.5cm}|}{Low contour level NVSS} &
  \multicolumn{1}{P{1.5cm}|}{Contour increase factor} &
  \multicolumn{1}{P{1.5cm}|}{Low contour level TGSS} &
  \multicolumn{1}{P{1.5cm}|}{Contour increase factor} \\
\hline
  J004312.85-103956.1 & 0.001 & x2 & 0.001 & x4 & 0.01 & x2\\
  J080101.35+134952.1 & 0.001 & +2 & 0.002 & x2 & 0.04 & +2\\
  J080337.67+105042.4 & 0.001 & +2 & 0.01 & x2 & 0.02 & +2\\
  J081803.86+543708.4 & 0.001 & x2 & 0.001 & +2 & 0.01 & x2\\
  J082718.31+463510.9 & 0.001 & +2 & 0.001 & x4 & 0.01 & x2\\
  J085116.23+082723.1 & 0.001 & +2 & 0.001 & x2 & 0.01 & x2\\
  J091337.21+031720.5 & 0.001 & x2 & 0.001 & x4 & 0.002 & x2\\
  J092428.89+141409.3 & 0.002 & x2 & 0.002 & x2 & 0.02 & x4\\
  J092539.05+362705.5 & 0.001 & +2 & 0.001 & x2 & 0.02 & x2\\
  J092612.34+324721.2 & 6.0E-4 & +2 & 0.002 & +2 & 0.01 & x2\\
  J093349.82+451957.8 & 4.0E-4 & x2 & 0.001 & x2 & 0.008 & +4\\
  J095716.40+190651.2 & 0.001 & +2 & 0.001 & x2 & 0.01 & x2\\
  J101932.33+140301.8 & 0.001 & x2 & 0.001 & x2 & 0.01 & x2\\
  J103502.61+425548.3 & 0.001 & x2 & 0.001 & x2 & 0.008 & x2\\
  J103605.76+000606.8 & 0.001 & x2 & 0.001 & x4 & 0.02 & x2\\
  J103636.24+383508.1 & 0.001 & +2 & 0.001 & +2 & 0.01 & +2\\
  J103856.37+575247.5 & 0.001 & +2 & 0.001 & x2 & 0.02 & x2\\
  J104645.86+314426.8 & 0.001 & +2 & 0.001 & x2 & 0.01 & +2\\
  J104914.08+005945.2 & 0.001 & +2 & 0.002 & x2 & 0.01 & +2\\
  J114020.22+535029.1 & 0.001 & +2 & 0.001 & +2 & 0.006 & x2\\
  J114111.81+054404.9 & 0.001 & x2 & 0.002 & x2 & 0.01 & x2\\
  J115424.56+020653.0 & 0.001 & +2 & 0.001 & +2 & 0.01 & +2\\
  J115513.65-003133.9 & 0.001 & x2 & 0.002 & +2 & 0.01 & +2\\
  J120118.19+061859.3 & 0.001 & x2 & 0.001 & x2 & 0.01 & x2\\
  J120455.01+483256.9 & 0.001 & x2 & 0.002 & x2 & 0.01 & x2\\
  J121439.53+052803.9 & 0.001 & x2 & 0.001 & x4 & 0.01 & x2\\
  J130904.46+102935.3 & 0.001 & x2 & 0.002 & x2 & 0.01 & x4\\
  J133038.38+381609.7 & 0.001 & x2 & 0.001 & x4 & 0.01 & x4\\
  J135315.36+550648.3 & 0.001 & +2 & 0.001 & x2 & 0.01 & x4\\
  J141456.58+001223.0 & 0.001 & x2 & 0.002 & x2 & 0.01 & x2\\
  J141513.98-013703.7 & 0.001 & x2 & 0.001 & x2 & 0.01 & x2\\
  J141718.94+060812.3 & 0.001 & x2 & 0.002 & x2 & 0.08 & x2\\
  J141731.27+081230.1 & 0.001 & x2 & 0.001 & x4 & 0.04 & x2\\
  J141927.23+233810.2 & 4.0E-4 & x2 & 0.001 & x2 & 0.01 & x2\\
  J143304.34+033037.6 & 0.001 & x2 & 0.001 & x2 & 0.01 & x2\\
  J143409.03+013700.9 & 0.001 & x2 & 0.001 & x4 & 0.01 & x2\\
  J144700.45+460243.5 & 0.001 & x2 & 0.001 & x2 & 0.01 & x2\\
  J144904.27+025802.7 & 0.001 & +1 & 0.001 & x2 & 0.02 & x2\\
  J150229.04+524402.1 & 0.001 & x2 & 0.001 & x2 & 0.02 & x2\\
  J151108.78+180153.2 & 0.001 & x2 & 0.002 & x4 & 0.02 & x2\\
  J154346.14+341521.6 & 4.0E-4 & x2 & 0.002 & x2 & 0.01 & x2\\
  J154729.58+145656.9 & 0.001 & x2 & 0.002 & x2 & 0.02 & x2\\
  J155343.59+234825.4 & 0.001 & x4 & 0.002 & x2 & 0.01 & x2\\
  J161828.98+295859.5 & 0.001 & x2 & 0.001 & x2 & 0.01 & x2\\
  J164527.68+272005.8 & 0.001 & x2 & 0.001 & x2 & 0.01 & x2\\
  J212546.35+005551.9 & 0.001& x2 & 0.001 & x4 & 0.01 & x2\\
  J222455.24-002302.3 & 0.001 & x2 & 0.002 & x2 & 0.01 & x2\\
\hline
\end{tabular}
\end{center}
\caption{List of \WAT\ sources with radio contour levels and increase factors, used in Fig. \ref{images} and in Appendix. Column description: (1) source name; (2) lower FIRST contour level (mJy); (3) contour increase factor (mJy); (4) lower NVSS contour level (mJy); (5) contour increase factor (mJy); (6) lower TGSS contour level (mJy); (7) contour increase factor (mJy).}
\label{tableA}
\end{table}

\twocolumn

\newpage
\onecolumn

\begin{table}[]
\begin{center}
\begin{tabular}{|l|r|r|r|r|l|r|r|}
        \hline
        \multicolumn{1}{|c|}{Source name} &
        \multicolumn{1}{c|}{$z$} &
        \multicolumn{1}{c|}{log$\nu L_{\nu}$} &
        \multicolumn{1}{c|}{M$_{r}$} &
        \multicolumn{1}{c|}{logL$_{[OIII]}$} &
        \multicolumn{1}{c|}{Dn} &
        \multicolumn{1}{c|}{C$_{r}$} &
        \multicolumn{1}{c|}{$\sigma*$} \\
        \hline
        J004312.85-103956.0 & 0.13 & 40.93 & -23.50 & 19.85 & 1.93 & 3.18 & 290.93\\
        J080101.35+134952.2 & 0.11 & 41.21 & -23.32 & 22.18 & 1.95 & 3.15 & 221.46\\
        J080337.67+105042.4 & 0.14 & 40.98 & -22.18 & 3.06 & 2.00 & 3.49 & 246.58\\
        J081803.86+543708.4 & 0.12 & 41.43 & -23.67 & 22.72 & 1.96 & 3.09 & 263.64\\
        J082718.31+463510.8 & 0.12 & 40.74 & -22.60 & 13.73 & 2.01 & 3.34 & 277.60\\
        J085116.24+082723.1 & 0.06 & 40.36 & -22.40 & 51.05 & 1.97 & 3.22 & 258.04\\
        J091337.21+031720.5 & 0.14 & 40.94 & -22.50 & 6.98 & 1.78 & 2.57 & 236.35\\
        J092428.89+141409.3 & 0.14 & 40.44 & -22.71 & 11.58 & 1.98 & 3.29 & 242.06\\
        J092539.06+362705.6 & 0.11 & 41.49 & -23.48 & 26.98 & 1.97 & 3.28 & 276.34\\
        J092612.34+324721.2 & 0.14 & 41.38 & -21.97 & 14.35 & 1.82 & 2.88 & 179.16\\
        J093349.82+451957.8 & 0.13 & 41.33 & -23.05 & 20.75 & 1.90 & 2.94 & 218.17\\
        J095716.41+190651.2 & 0.09 & 40.88 & -22.17 & 35.13 & 1.91 & 1.42 & 210.85\\
        J101932.33+140301.8 & 0.15 & 41.33 & -23.21 & 15.80 & 1.97 & 2.92 & 264.37\\
        J103502.62+425548.3 & 0.14 & 40.86 & -22.88 & 17.15 & 2.22 & 3.19 & 251.18\\
        J103605.76+000606.8 & 0.10 & 41.25 & -23.32 & 39.53 & 1.88 & 3.03 & 311.06\\
        J103636.24+383508.1 & 0.14 & 41.14 & -22.84 & 18.27 & 2.03 & 3.20 & 280.19\\
        J103856.37+575247.5 & 0.10 & 40.95 & -21.43 & 20.40 & 1.92 & 2.93 & 210.19\\
        J104645.86+314426.8 & 0.11 & 40.81 & -23.02 & 19.30 & 2.01 & 3.22 & 278.41\\
        J104914.08+005945.2 & 0.11 & 40.63 & -22.78 & 12.53 & 1.91 & 3.19 & 266.85\\
        J114020.23+535029.1 & 0.15 & 41.04 & -23.17 & 11.86 & 2.09 & 3.05 & 332.40\\
        J114111.81+054405.0 & 0.10 & 41.24 & -23.47 & 1.06 & 1.93 & 2.94 & 272.32\\
        J115424.56+020653.0 & 0.13 & 40.64 & -22.90 & 5.80 & 1.97 & 3.52 & 276.32\\
        J115513.65-003133.9 & 0.13 & 41.02 & -22.76 & 17.40 & 1.92 & 2.85 & 221.82\\
        J120118.19+061859.3 & 0.14 & 40.54 & -22.81 & 8.39 & 1.99 & 3.32 & 319.14\\
        J120455.02+483256.9 & 0.07 & 40.62 & -21.81 & 26.41 & 1.90 & 3.23 & 233.62\\
        J121439.53+052803.9 & 0.08 & 40.13 & -23.44 & 5.01 & 1.95 & 3.28 & 230.03\\
        J130904.46+102935.3 & 0.09 & 41.17 & -22.87 & 4.74 & 2.02 & 3.44 & 348.60\\
        J133038.38+381609.7 & 0.11 & 40.86 & -23.24 & 24.97 & 1.94 & 3.17 & 260.61\\
        J135315.36+550648.2 & 0.14 & 40.66 & -20.49 & 6.37 & 1.96 & 2.01 & 255.58\\
        J141456.58+001223.0 & 0.13 & 40.82 & -22.87 & 15.15 & 2.01 & 3.28 & 331.01\\
        J141513.98-013703.7 & 0.15 & 41.24 & -23.33 & 37.24 & 2.00 & 3.05 & 305.79\\
        J141718.94+060812.3 & 0.11 & 40.91 & -22.79 & 2.55 & 2.00 & 3.31 & 278.47\\
        J141731.27+081230.1 & 0.06 & 40.58 & -22.52 & 49.76 & 2.04 & 3.33 & 237.39\\
        J141927.23+233810.2 & 0.14 & 40.12 & -22.16 & -1.67 & 1.88 & 2.94 & 185.40\\
        J143304.34+033037.6 & 0.15 & 40.88 & -23.44 & 17.21 & 1.98 & 2.65 & 230.86\\
        J143409.03+013700.9 & 0.14 & 41.61 & -23.43 & 18.35 & 2.01 & 3.35 & 331.87\\
        J144700.45+460243.5 & 0.13 & 41.04 & -22.34 & 18.54 & 1.92 & 3.36 & 288.44\\
        J144904.27+025802.7 & 0.12 & 40.36 & -21.81 & -3.85 & 1.87 & 3.32 & 194.29\\
        J150229.04+524402.0 & 0.13 & 41.10 & -23.50 & 24.43 & 1.98 & 3.28 & 292.56\\
        J151108.77+180153.3 & 0.12 & 41.28 & -23.08 & 21.10 & 2.04 & 3.11 & 274.49\\
        J154346.14+341521.6 & 0.12 & 40.20 & -22.30 & 3.30 & 1.93 & 3.29 & 232.29\\
        J154729.59+145657.0 & 0.09 & 40.69 & -22.42 & 37.78 & 1.85 & 3.02 & 203.33\\
        J155343.59+234825.4 & 0.12 & 41.28 & -23.43 & 307.83 & 1.64 & 3.18 & 323.50\\
        J161828.98+295859.6 & 0.13 & 40.96 & -23.21 & 5.99 & 2.04 & 3.15 & 315.10\\
        J164527.68+272005.8 & 0.10 & 40.65 & -23.16 & 14.75 & 1.96 & 3.05 & 270.40\\
        J212546.35+005551.8 & 0.14 & 41.62 & -23.02 & 25.49 & 1.96 & 2.98 & 223.22\\
        J222455.24-002302.3 & 0.14 & 40.74 & -22.63 &  &  & 2.7 & 274.29\\
        \hline
\end{tabular}
\end{center}
\caption{Main properties of the \WAT\ sources. Column description: (1) source name; (2)
redshift; (3) logarithm of the radio luminosity [erg s$^{-1}$]; (4) SDSS DR7 $r$-band AB magnitude; (5) logarithm
of the [O~III] line luminosity [erg s$^{-1}$] (6) Dn(4000) spectroscopic index; (7) concentration index $C_r$; (8) stellar velocity dispersion [\kms].}
\label{table2}
\end{table}

\twocolumn

\begin{appendix}
\onecolumn
\section{FIRST radio contours of the 47 \WAT\ sources.}

In this appendix we show FIRST radio contours for the 47 sources in the \WAT.\

\begin{table}[]
\begin{center}
\begin{tabular}{|r|l|}
\hline
  \multicolumn{1}{|c|}{Best ID} &
  \multicolumn{1}{c|}{SDSS name} \\
\hline
  84 & J103605.76+000606.8\\
  101 & J104914.08+005945.2\\
  166 & J115513.65-003133.9\\
  383 & J141456.58+001223.0\\
  1916 & J115424.56+020653.0\\
  2105 & J143409.03+013700.9\\
  2132 & J144904.27+025802.7\\
  2241 & J082718.31+463510.8\\
  2412 & J091337.21+031720.5\\
  2630 & J143304.34+033037.6\\
  3222 & J004312.85-103956.0\\
  4363 & J114111.81+054405.0\\
  4409 & J121439.53+052803.9\\
  5113 & J141513.98-013703.7\\
  5428 & J103856.37+575247.5\\
  5754 & J212546.35+005551.8\\
  6044 & J114020.23+535029.1\\
  6456 & J150229.04+524402.0\\
  6781 & J093349.82+451957.8\\
  7294 & J092539.06+362705.6\\
  7770 & J135315.36+550648.2\\
  8653 & J154346.14+341521.6\\
  8775 & J161828.98+295859.6\\
  8847 & J103502.62+425548.3\\
  8860 & J103636.24+383508.1\\
  9093 & J120455.02+483256.9\\
  9641 & J092612.34+324721.2\\
  9939 & J120118.19+061859.3\\
  10259 & J144700.45+460243.5\\
  10402 & J164527.68+272005.8\\
  11011 & J101932.33+140301.8\\
  11140 & J085116.24+082723.1\\
  11351 & J081803.86+543708.4\\
  11515 & J130904.46+102935.3\\
  11671 & J141731.27+081230.1\\
  11817 & J141718.94+060812.3\\
  13055 & J133038.38+381609.7\\
  13211 & J104645.86+314426.8\\
  13950 & J141927.23+233810.2\\
  14994 & J080101.35+134952.2\\
  15652 & J095716.41+190651.2\\
  15725 & J080337.67+105042.4\\
  15953 & J092428.89+141409.3\\
  16471 & J154729.59+145657.0\\
  858 & J222455.24-002302.3\\
  12051 & J155343.59+234825.4\\
  17907 & J151108.77+180153.3\\
\hline
\end{tabular}
\caption{\WAT\ sources names as listed in \citet{best12}, used in Appendix, and their SDSS name. Column description: (1) Best ID, from BH12; (2) SDSS DR9 source name.}
\end{center}
\end{table}

\begin{figure*}
\includegraphics[scale=0.5]{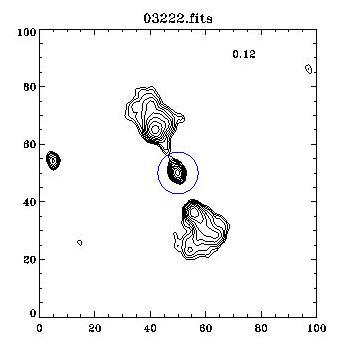}
\includegraphics[scale=0.5]{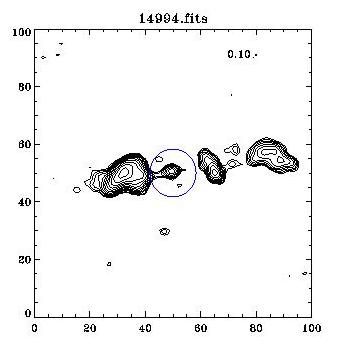}
\includegraphics[scale=0.5]{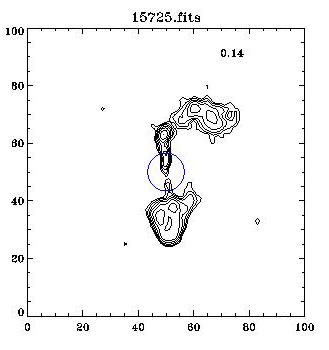}
\includegraphics[scale=0.5]{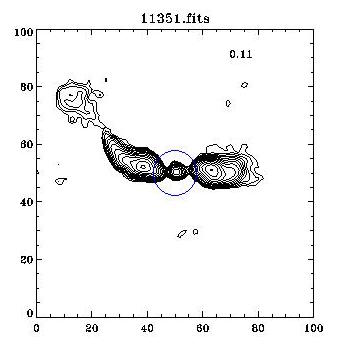}
\includegraphics[scale=0.5]{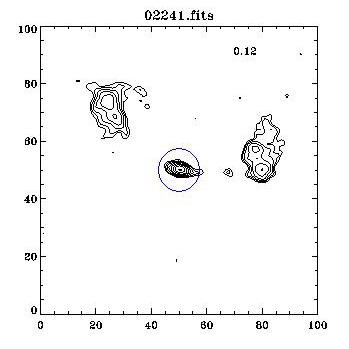}
\includegraphics[scale=0.5]{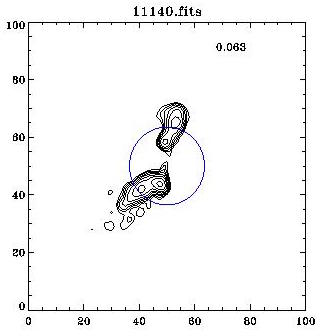}
\includegraphics[scale=0.5]{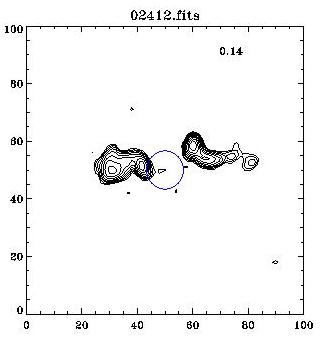}
\includegraphics[scale=0.5]{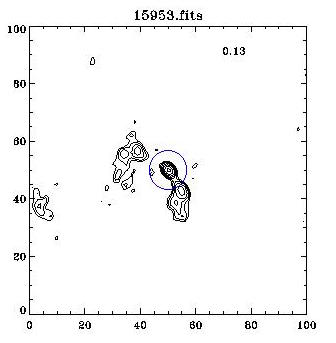}
\includegraphics[scale=0.5]{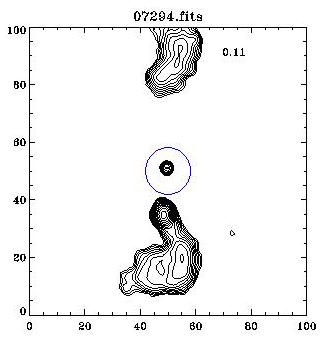}
\includegraphics[scale=0.5]{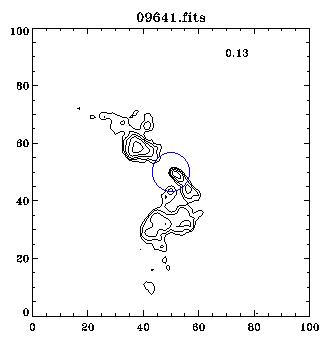}
\includegraphics[scale=0.5]{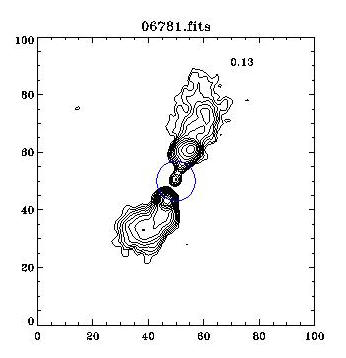}
\includegraphics[scale=0.5]{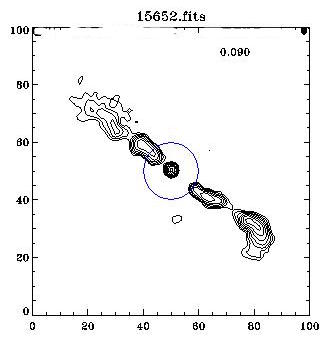}
\caption{Images of the \WAT\ sources, ordered by right ascension. Contours, corrected for cosmological effects to a redshift of $z$ = 0.15, are drawn starting from 0.45 mJy/beam and increase with a geometric progression with a common ratio of $\sqrt{2}$. The field of view is 3' x 3'. The black
circle is centered on the host galaxy and has a radius of 30 kpc. The source ID and redshift are shown in the upper corners.}
\end{figure*}

\addtocounter{figure}{-1}
\begin{figure*}
\includegraphics[scale=0.5]{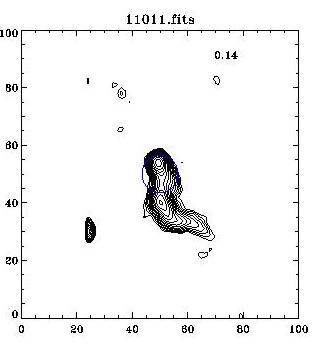}
\includegraphics[scale=0.5]{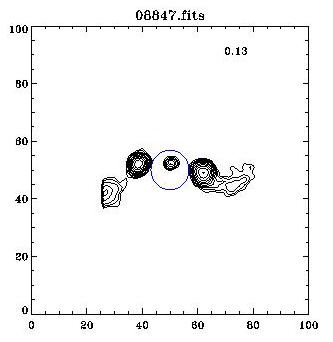}
\includegraphics[scale=0.5]{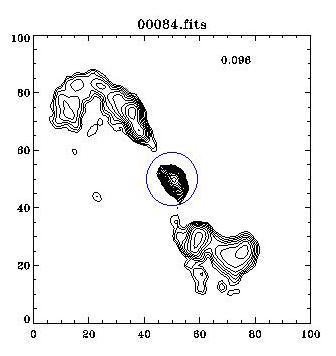}
\includegraphics[scale=0.5]{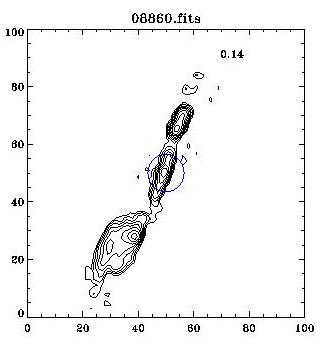}
\includegraphics[scale=0.5]{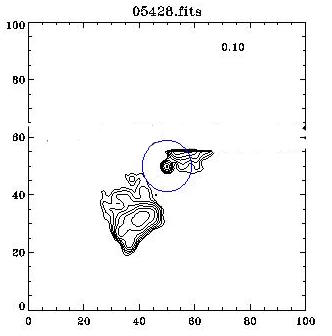}
\includegraphics[scale=0.5]{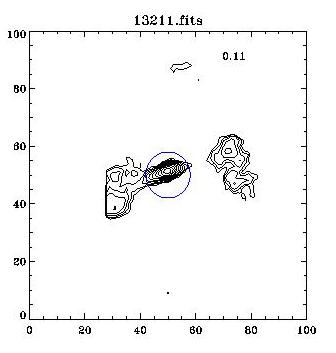}
\includegraphics[scale=0.5]{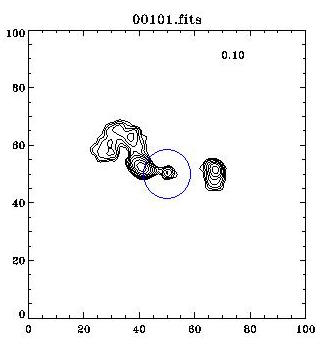}
\includegraphics[scale=0.5]{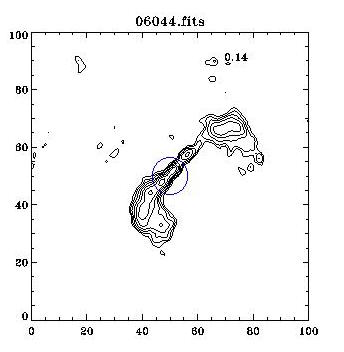}
\includegraphics[scale=0.5]{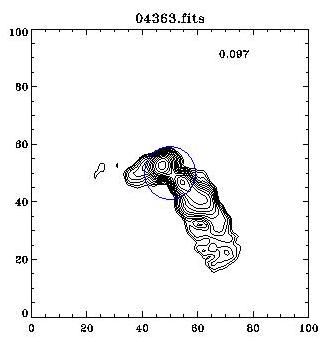}
\includegraphics[scale=0.5]{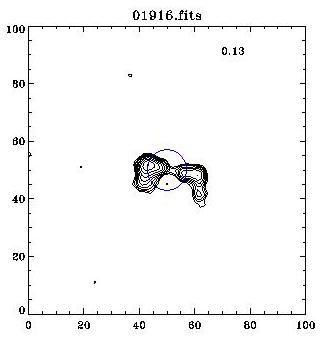}
\includegraphics[scale=0.52]{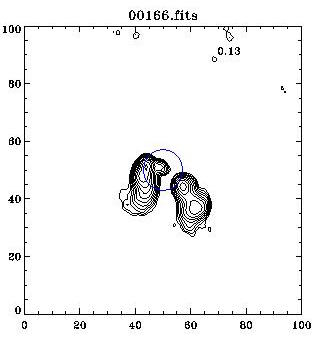}
\includegraphics[scale=0.53]{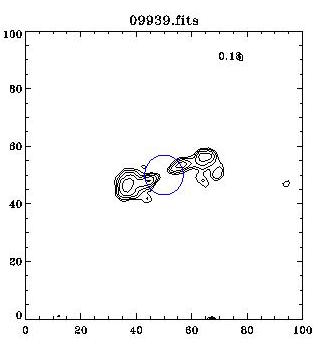}
\caption{(continued)}
\end{figure*}

\addtocounter{figure}{-1}
\begin{figure*}

\includegraphics[scale=0.5]{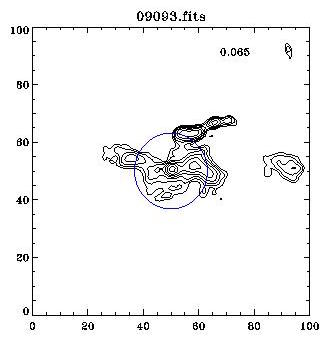}
\includegraphics[scale=0.5]{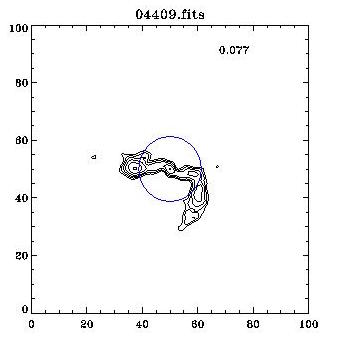}
\includegraphics[scale=0.5]{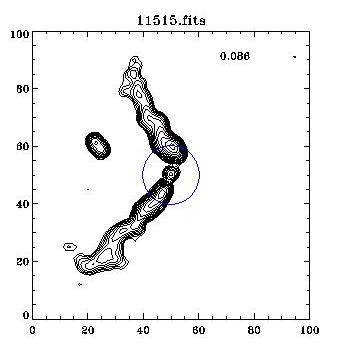}
\includegraphics[scale=0.5]{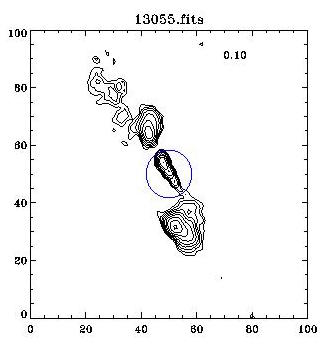}
\includegraphics[scale=0.5]{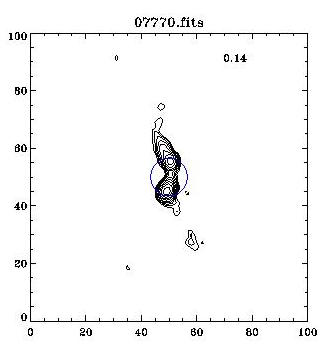}
\includegraphics[scale=0.5]{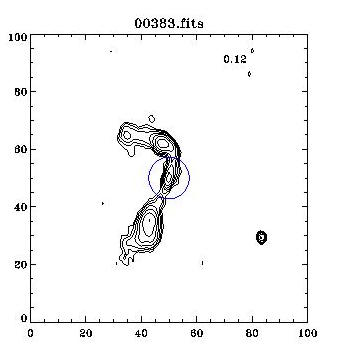}
\includegraphics[scale=0.5]{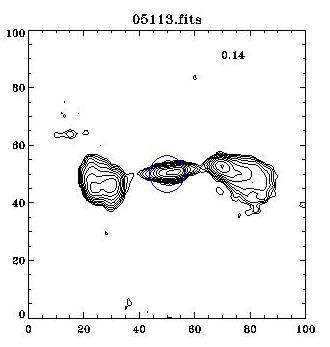}
\includegraphics[scale=0.5]{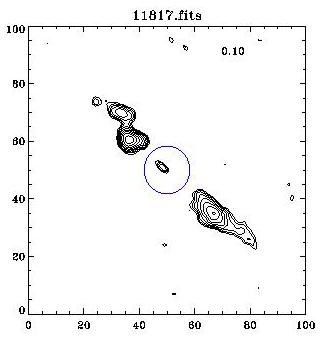}
\includegraphics[scale=0.5]{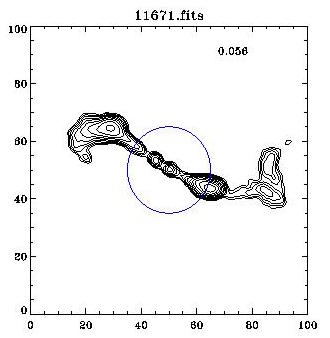}
\includegraphics[scale=0.5]{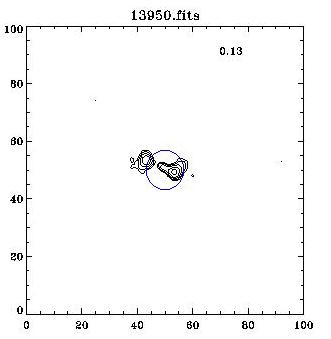}
\includegraphics[scale=0.53]{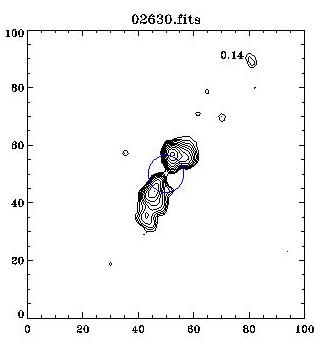}
\includegraphics[scale=0.55]{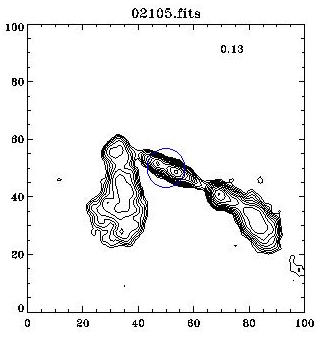}
\caption{(continued)}
\end{figure*}

\addtocounter{figure}{-1}
\begin{figure*}

\includegraphics[scale=0.5]{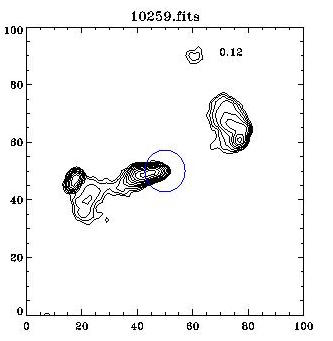}
\includegraphics[scale=0.5]{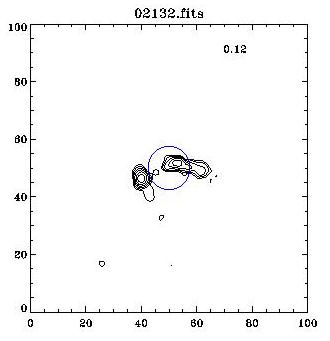}
\includegraphics[scale=0.5]{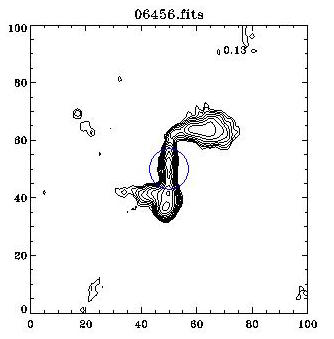}
\includegraphics[scale=0.5]{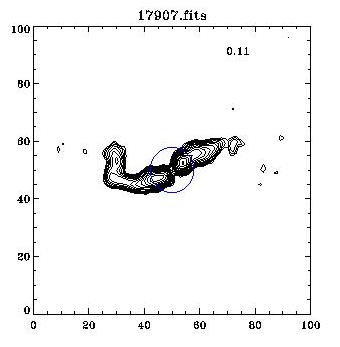}
\includegraphics[scale=0.5]{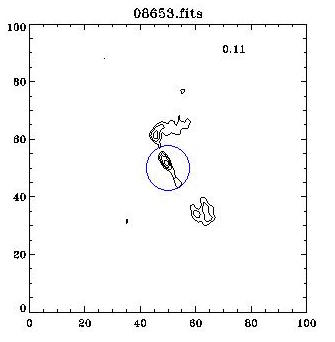}
\includegraphics[scale=0.5]{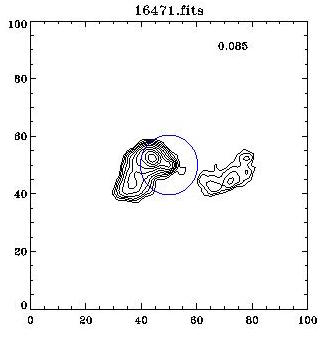}
\includegraphics[scale=0.5]{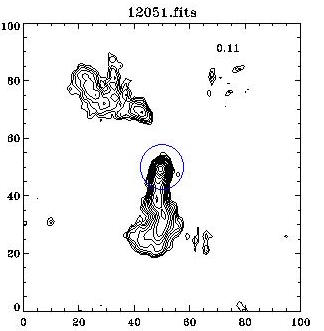}
\includegraphics[scale=0.5]{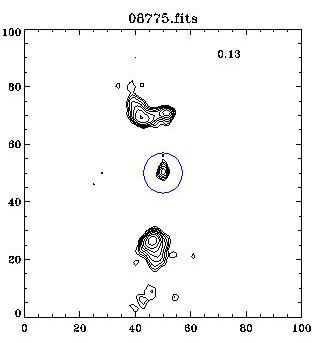}
\includegraphics[scale=0.5]{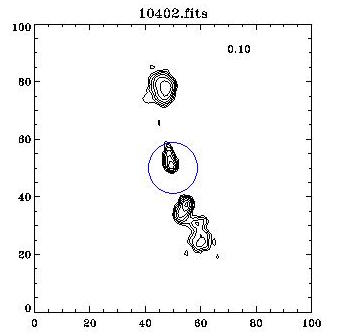}
\includegraphics[scale=0.5]{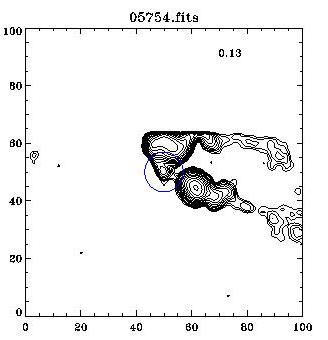}
\includegraphics[scale=0.5]{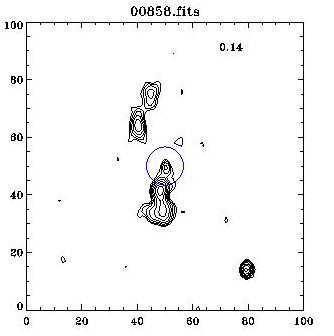}
\caption{(continued)}
\end{figure*}

\end{appendix}

\begin{appendix}
\onecolumn
\section{FIRST, NVSS, and TGSS radio contours of the 47 \WAT\ sources.} 
In this appendix we show FIRST, NVSS and TGSS radio contours for the 47 sources in the \WAT.\
\onecolumn

\begin{figure*}
\includegraphics[height=5.cm,width=6.cm,angle=0]{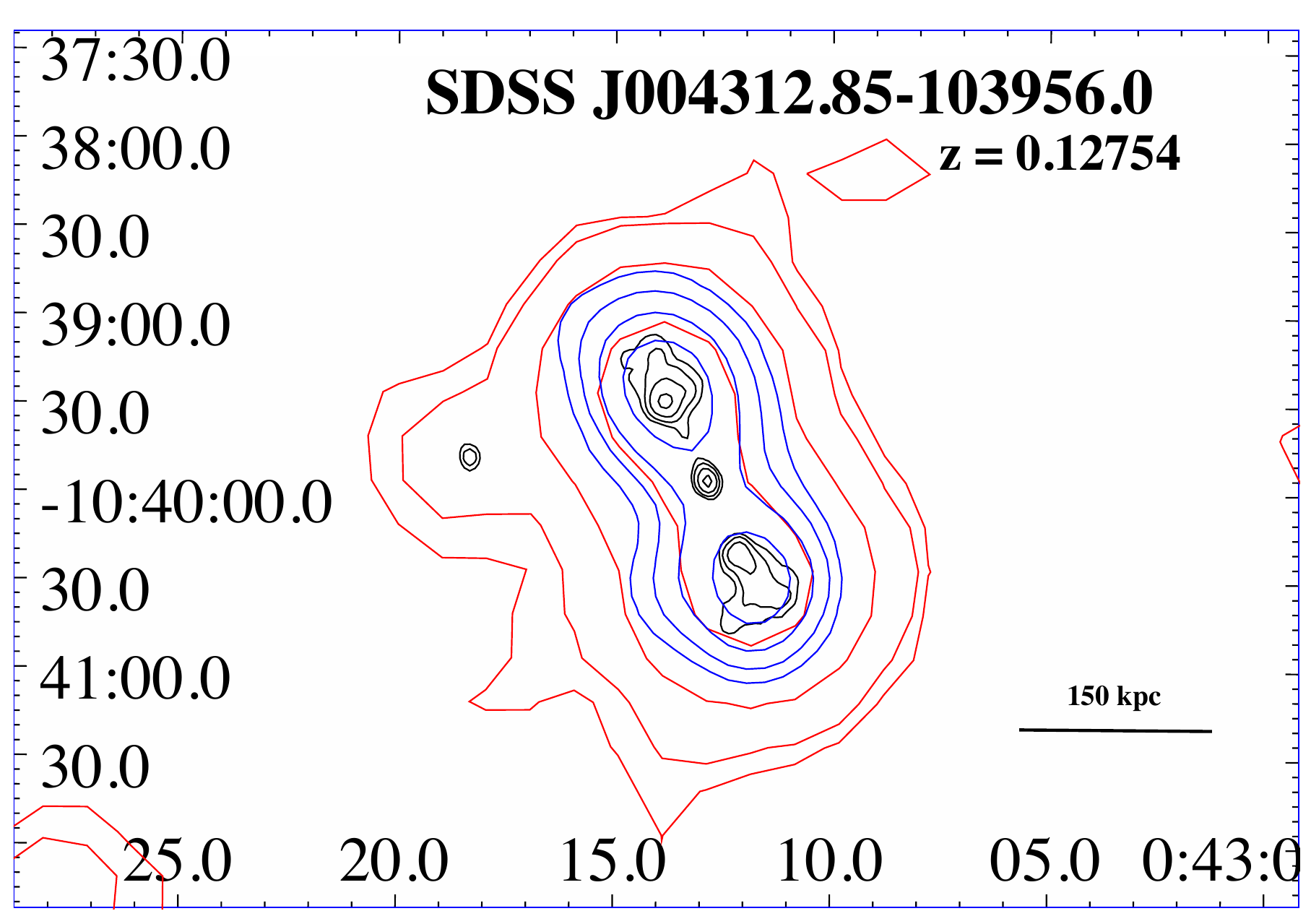}
\includegraphics[height=4.9cm,width=6.cm,angle=0]{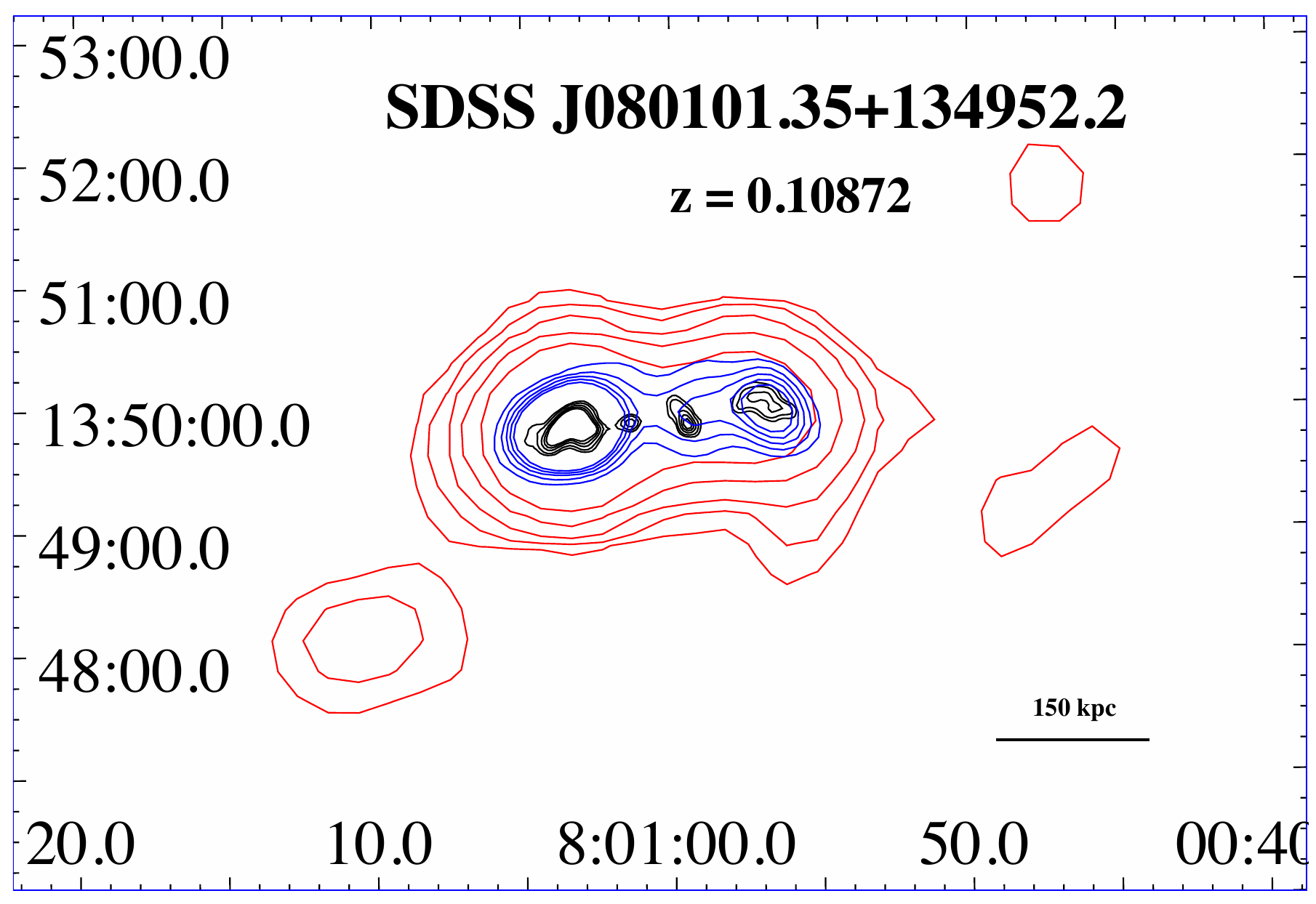} 
\includegraphics[height=5.01cm,width=6.cm,angle=0]{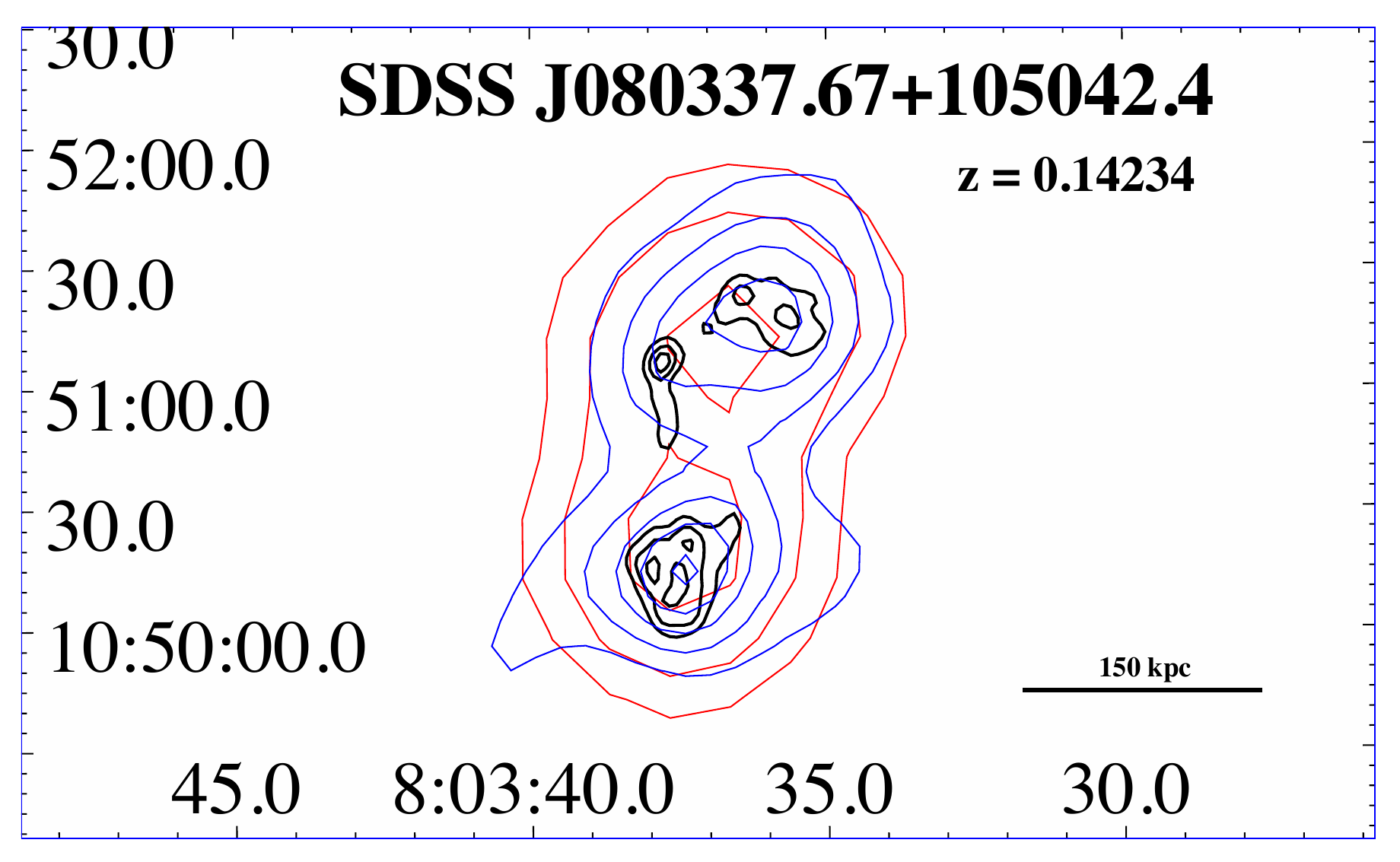} 
\includegraphics[height=5.cm,width=6.cm,angle=0]{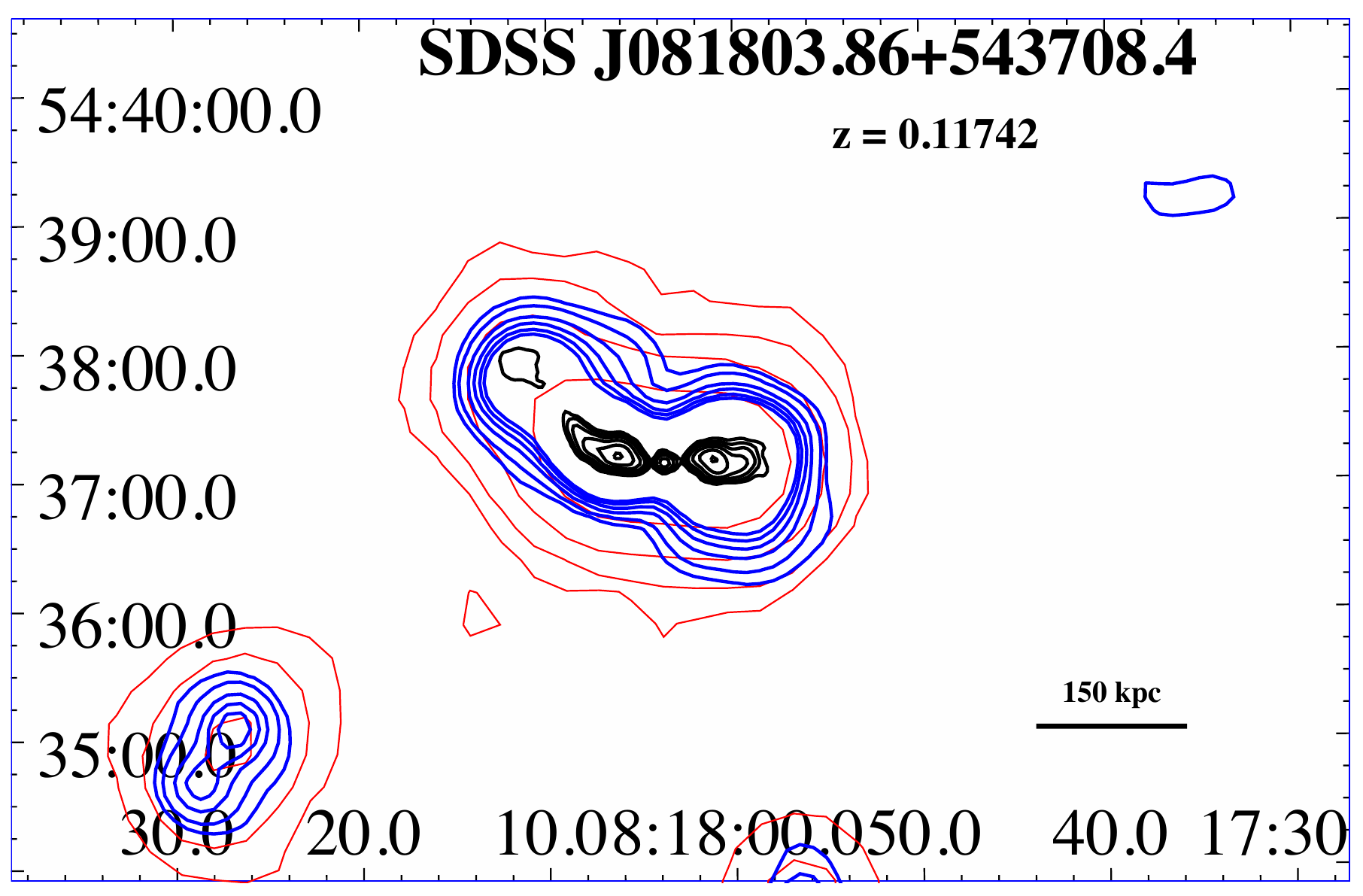} 
\includegraphics[height=5.cm,width=6.cm,angle=0]{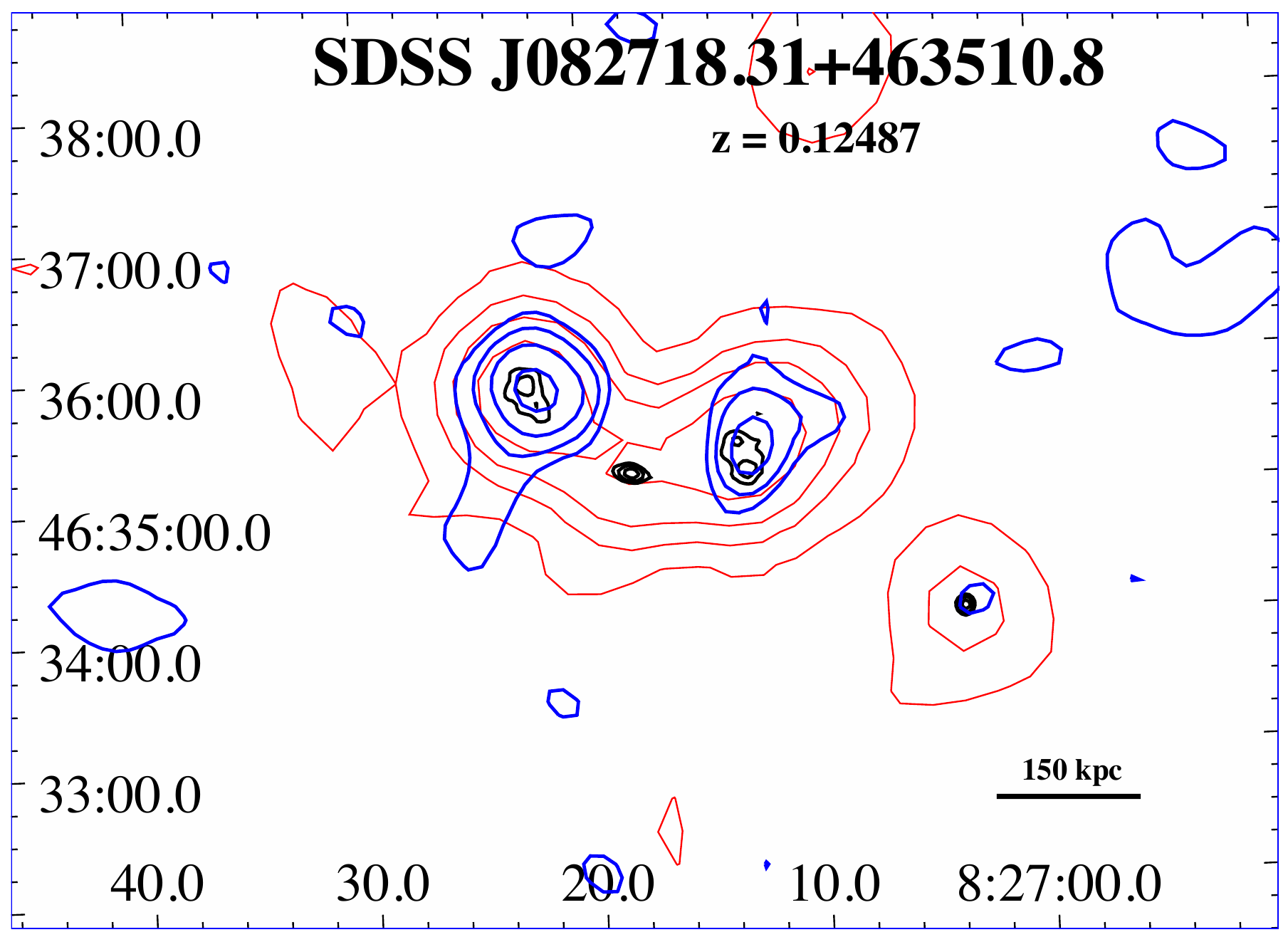} 
\includegraphics[height=5.cm,width=6.cm,angle=0]{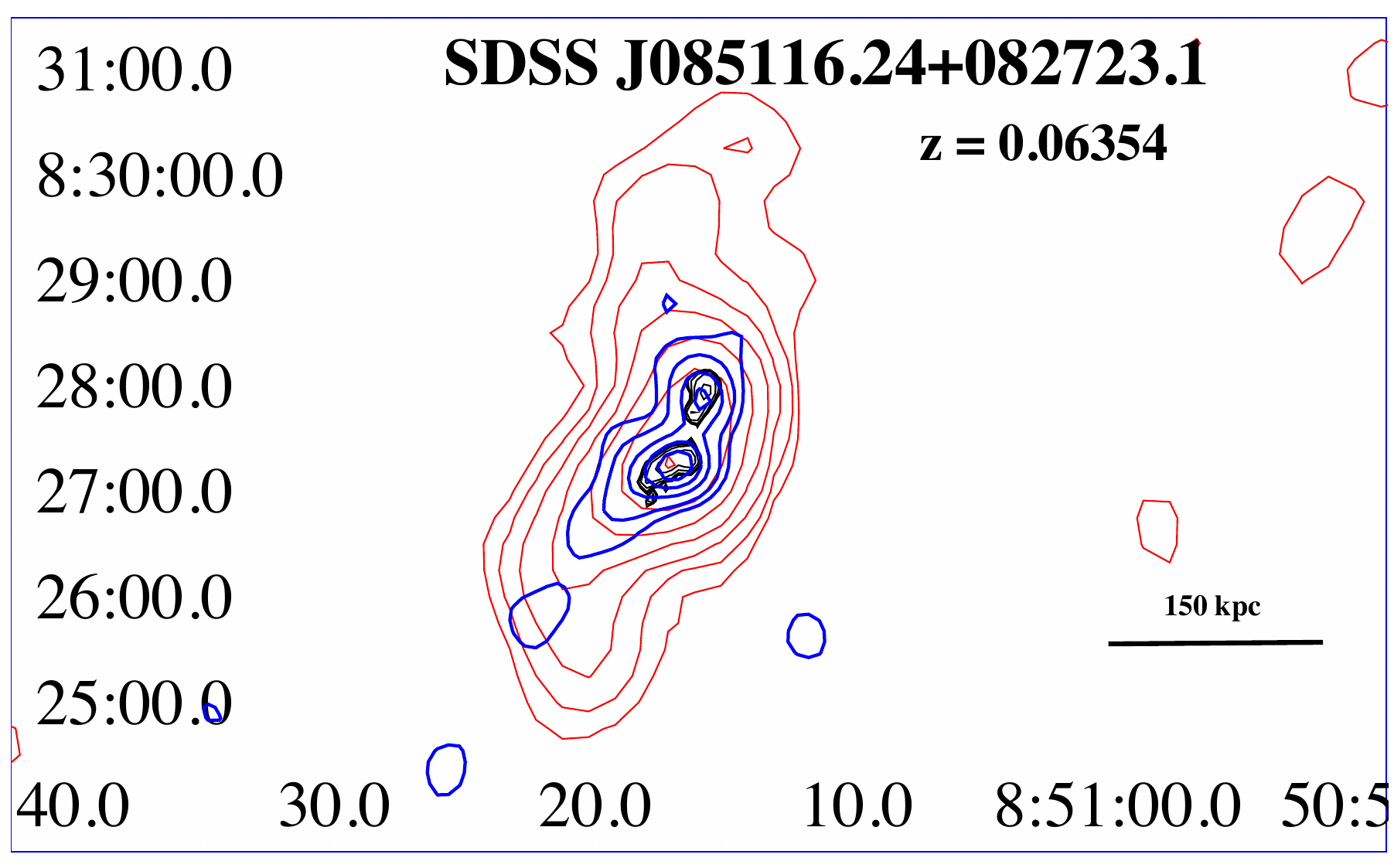} 
\includegraphics[height=5cm,width=6.cm,angle=0]{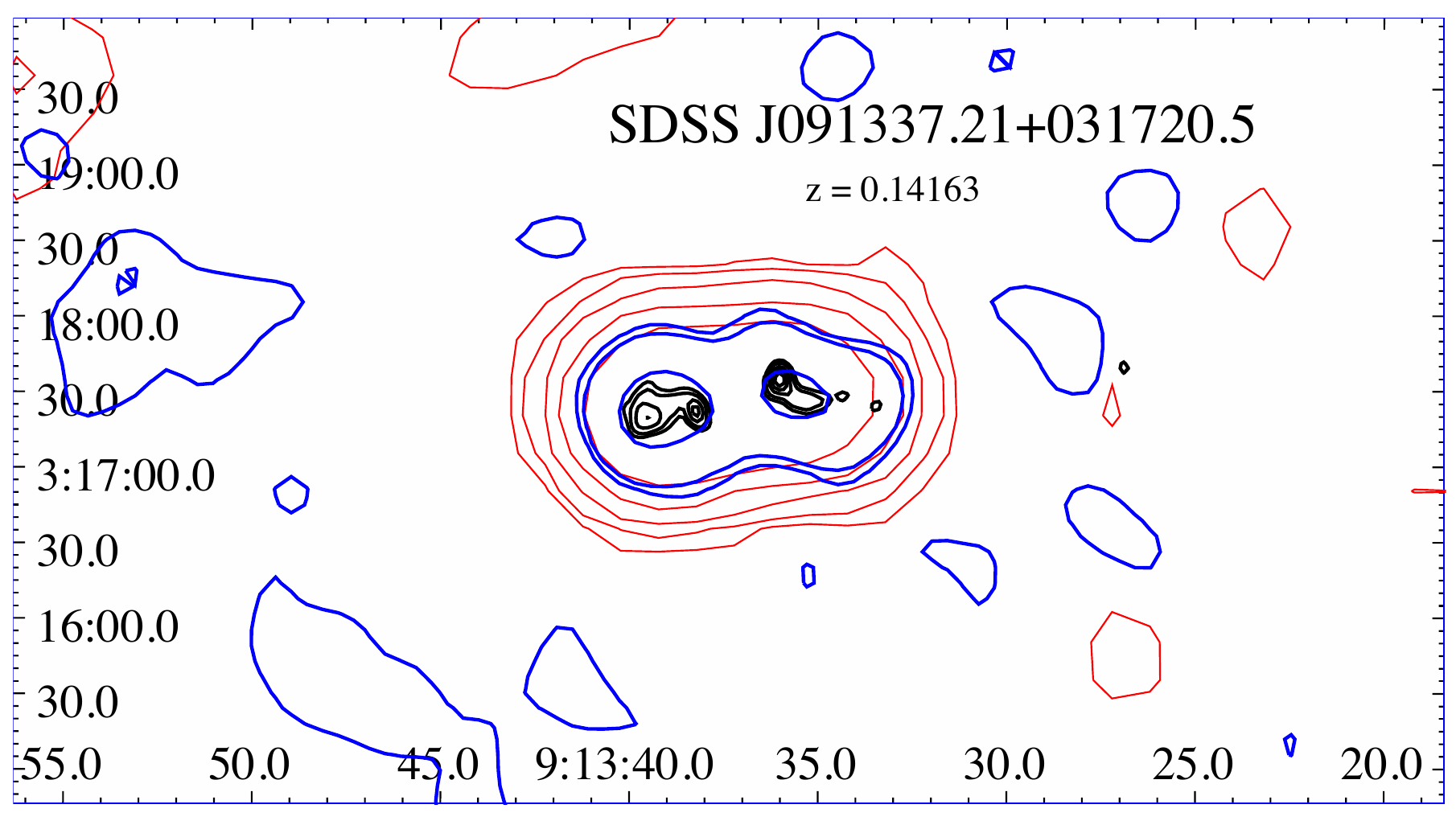} 
\includegraphics[height=5.cm,width=6.cm,angle=0]{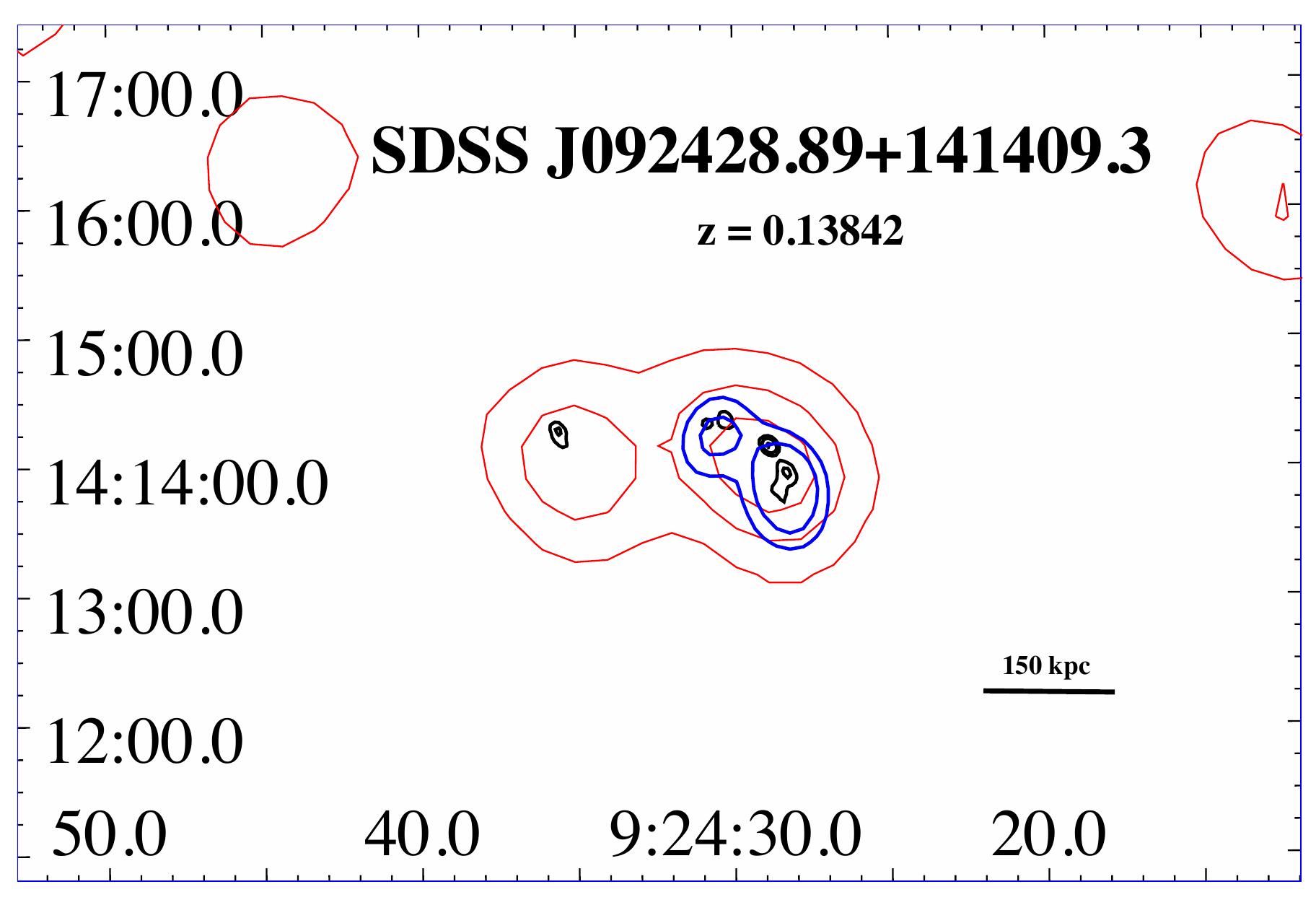} 
\includegraphics[height=5.cm,width=6.cm,angle=0]{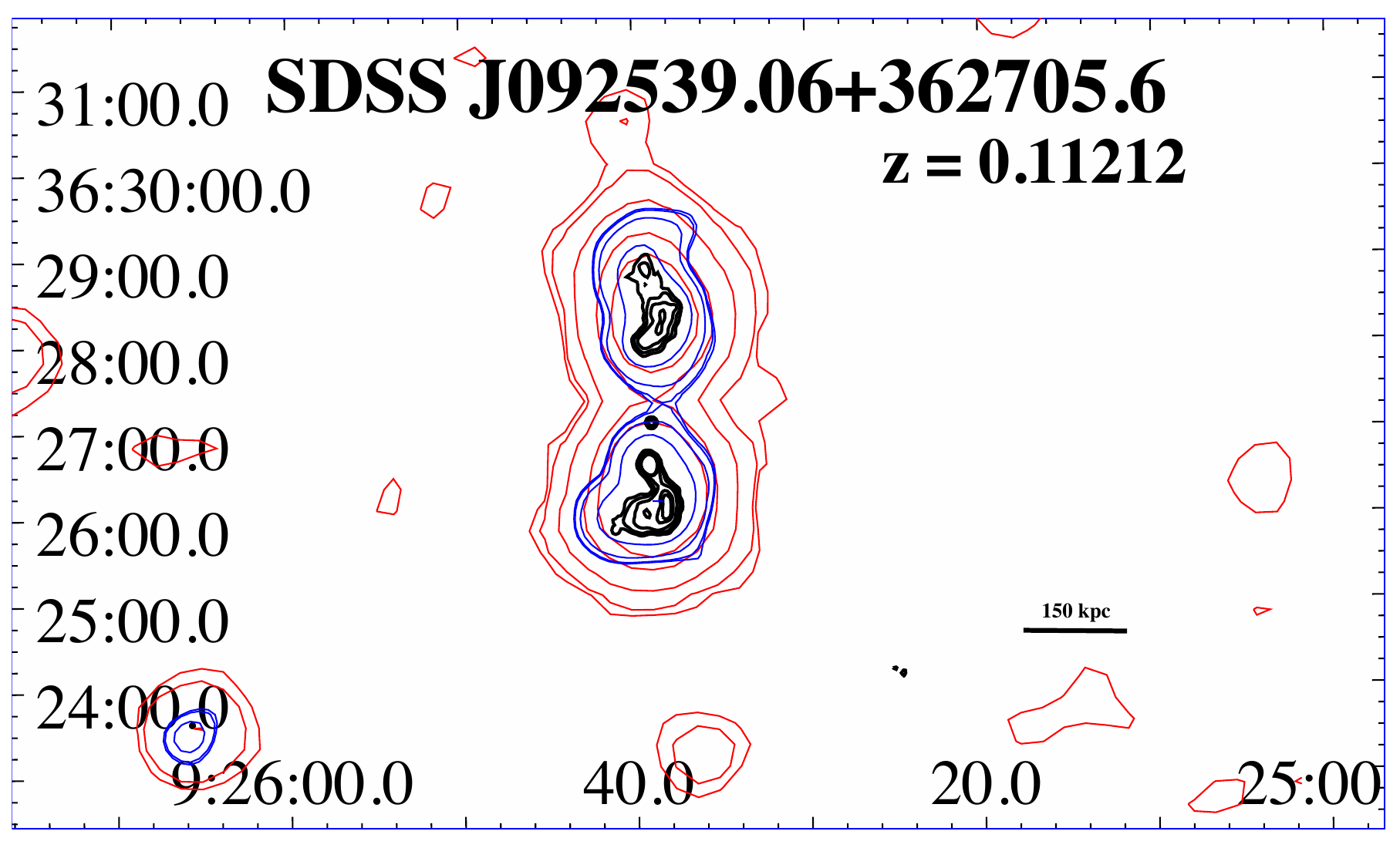} 
\includegraphics[height=5.cm,width=6.cm,angle=0]{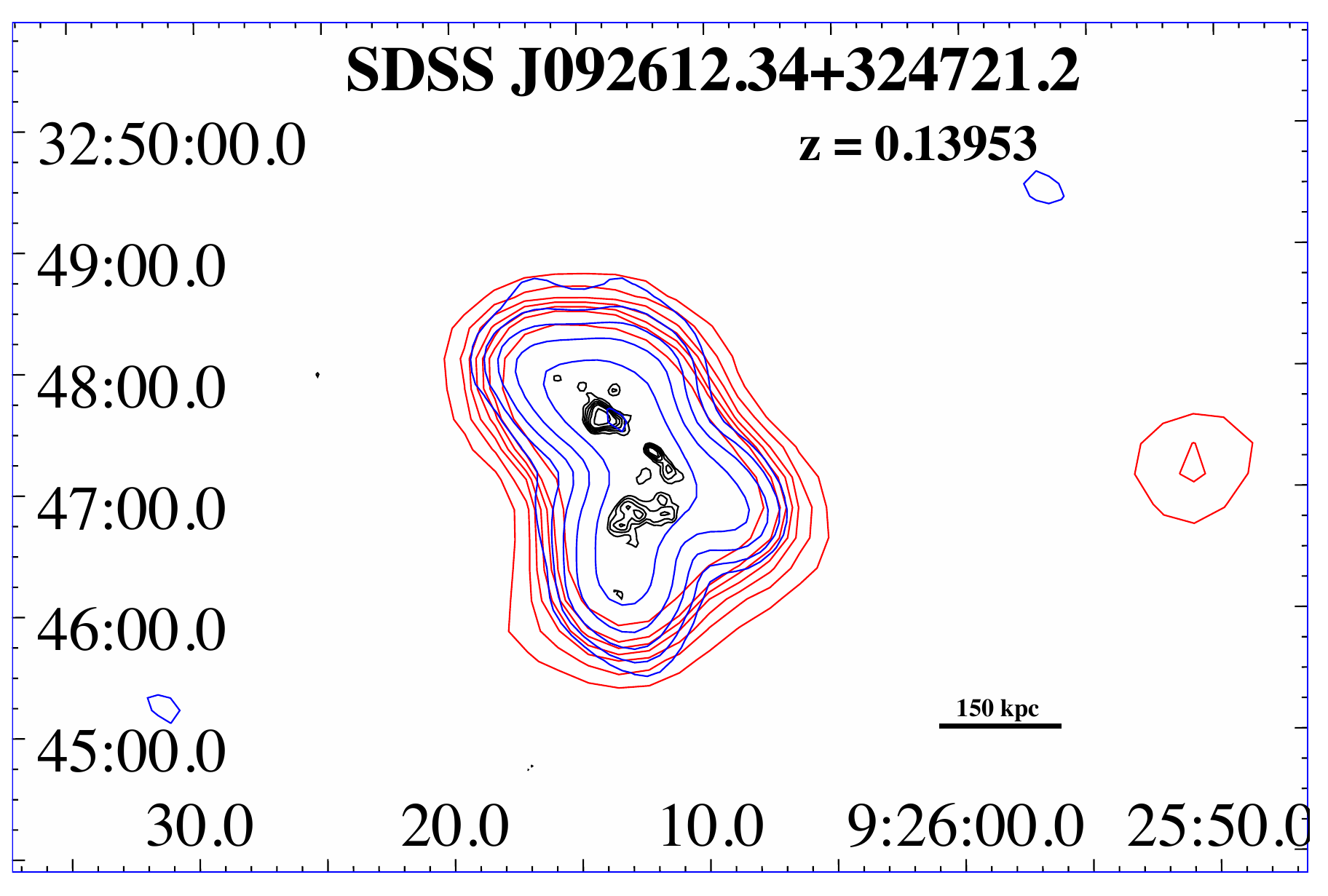} 
\includegraphics[height=5.cm,width=6.cm,angle=0]{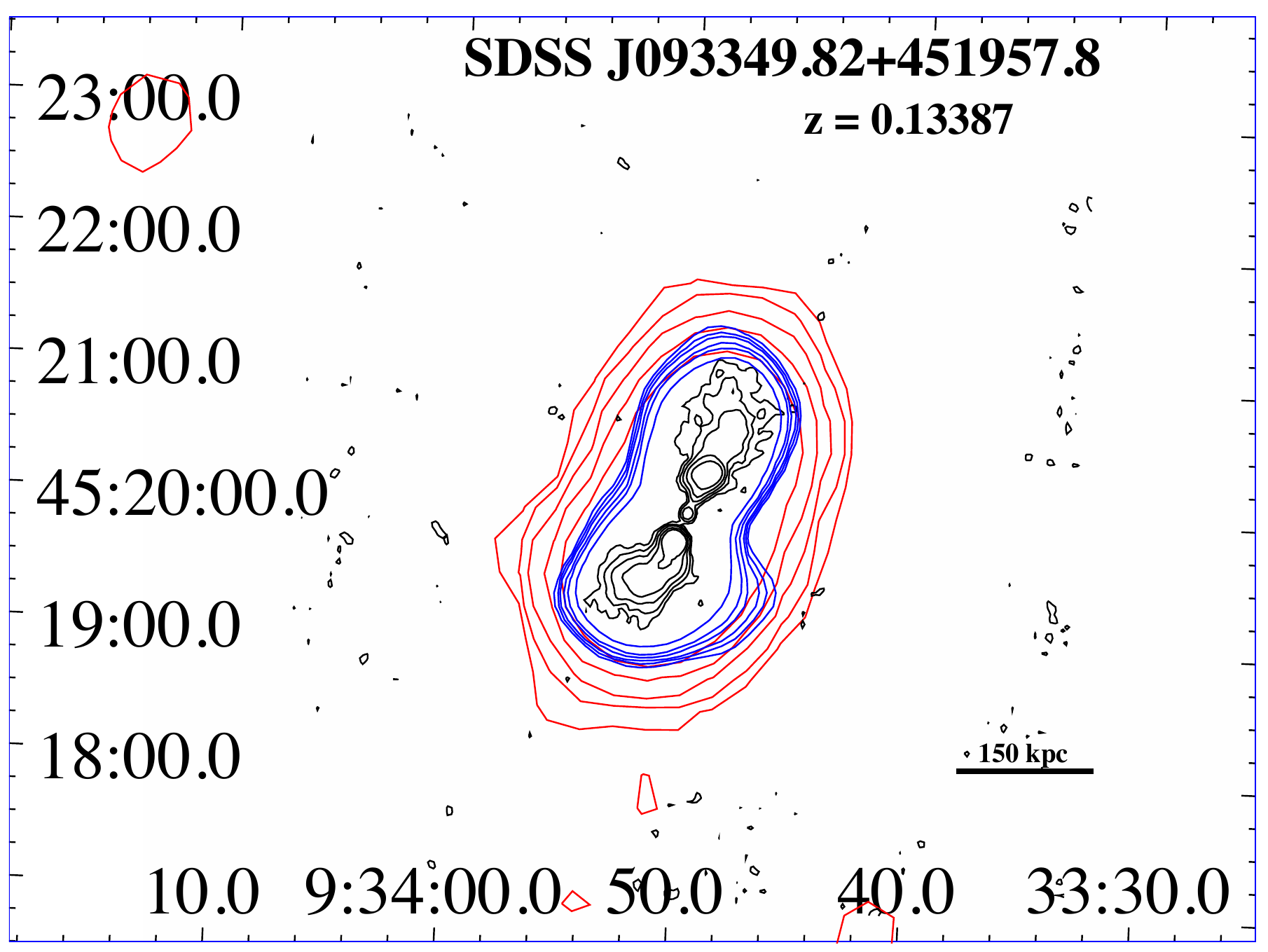} 
\includegraphics[height=5.cm,width=6.cm,angle=0]{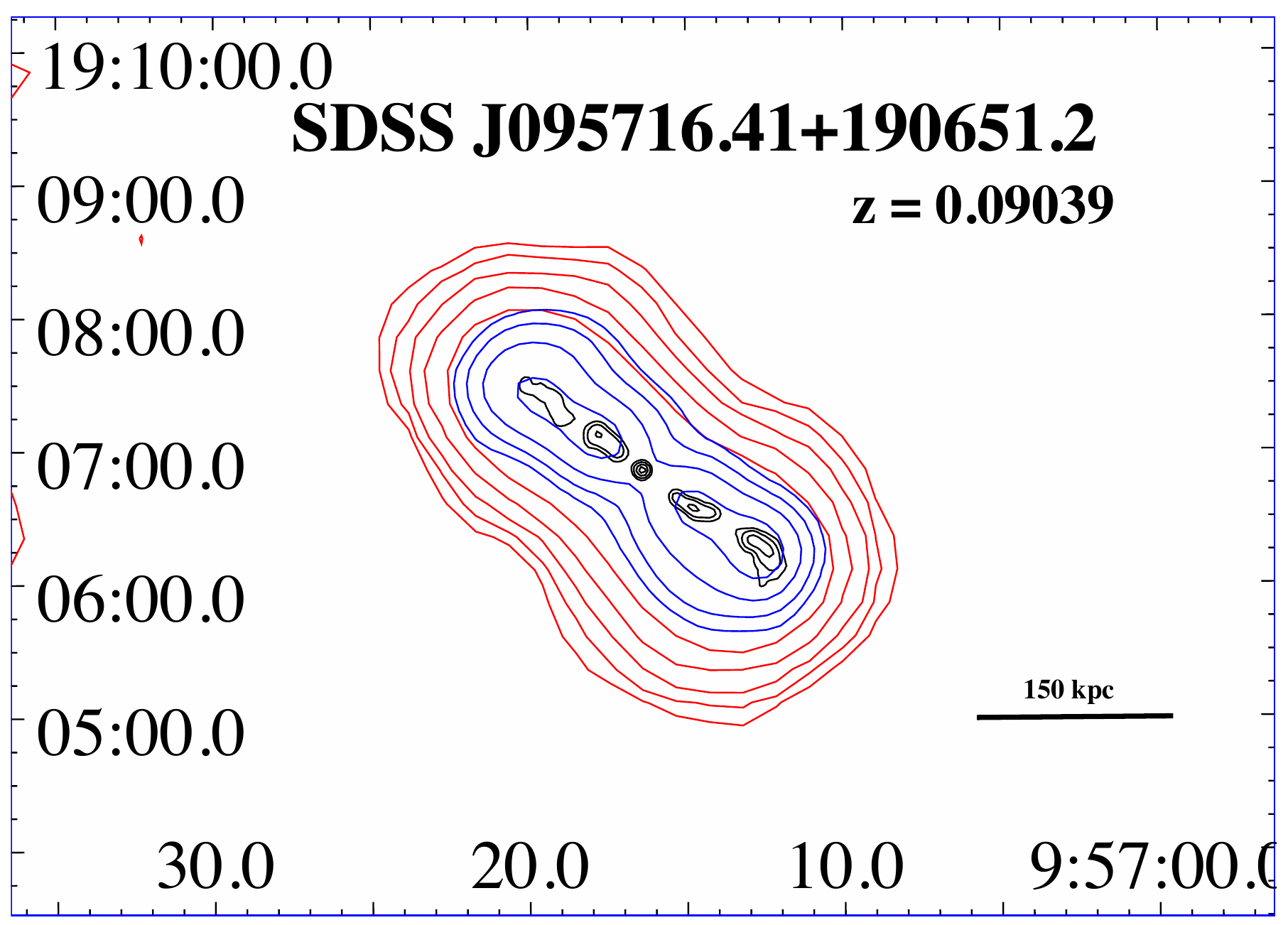} 
\caption{Images of the \WAT\ sources, ordered by right ascension. See Table \label{table1} for contour levels. The field of view is 3'$\times$3'; the black tick at the
  bottom is 150 kpc long. The source name and redshift are shown in the upper right corner.
}
\end{figure*}

\addtocounter{figure}{-1}
\begin{figure*}
\includegraphics[height=5.cm,width=6.cm,angle=0]{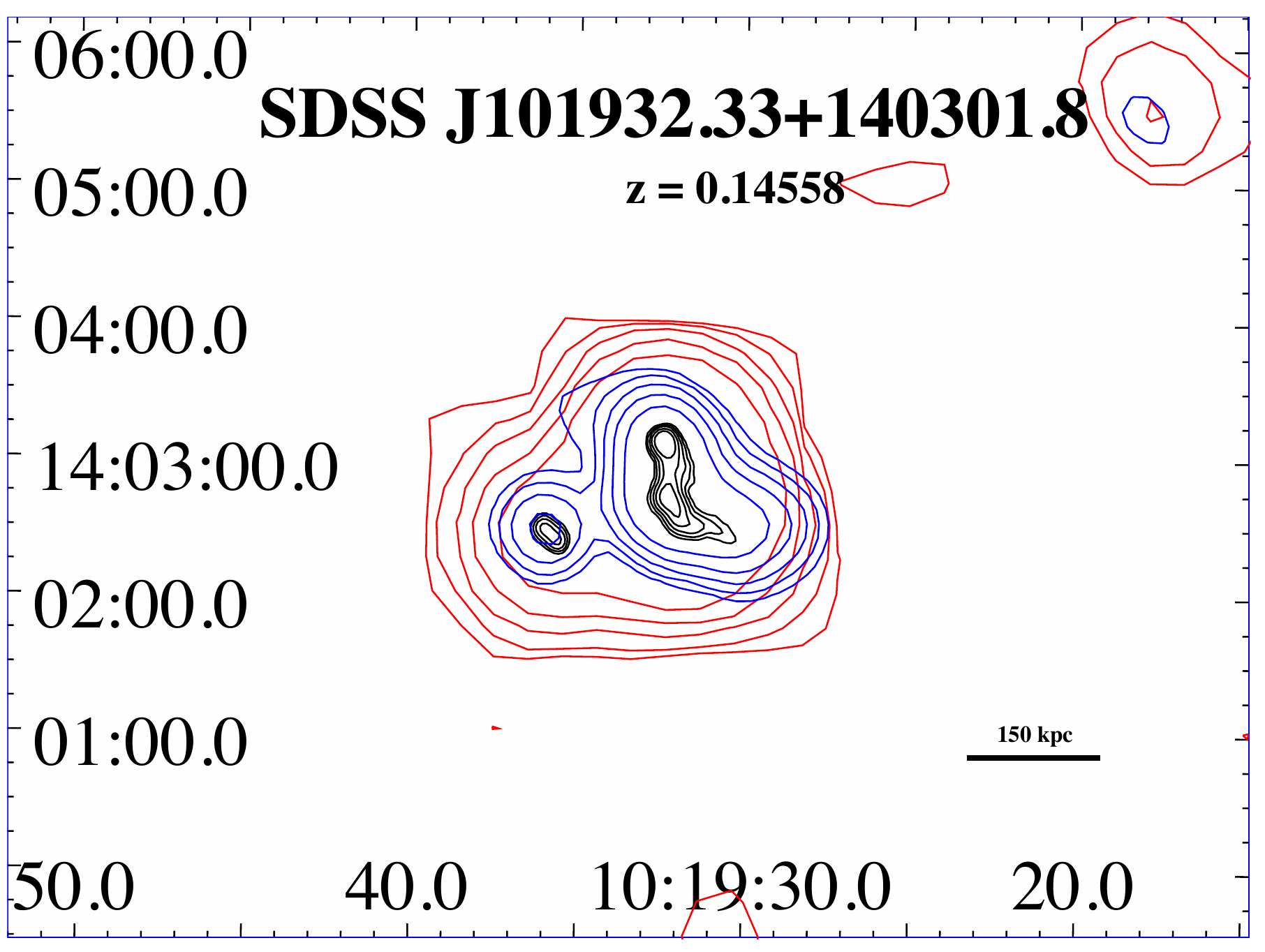}
\includegraphics[height=5.cm,width=6.cm,angle=0]{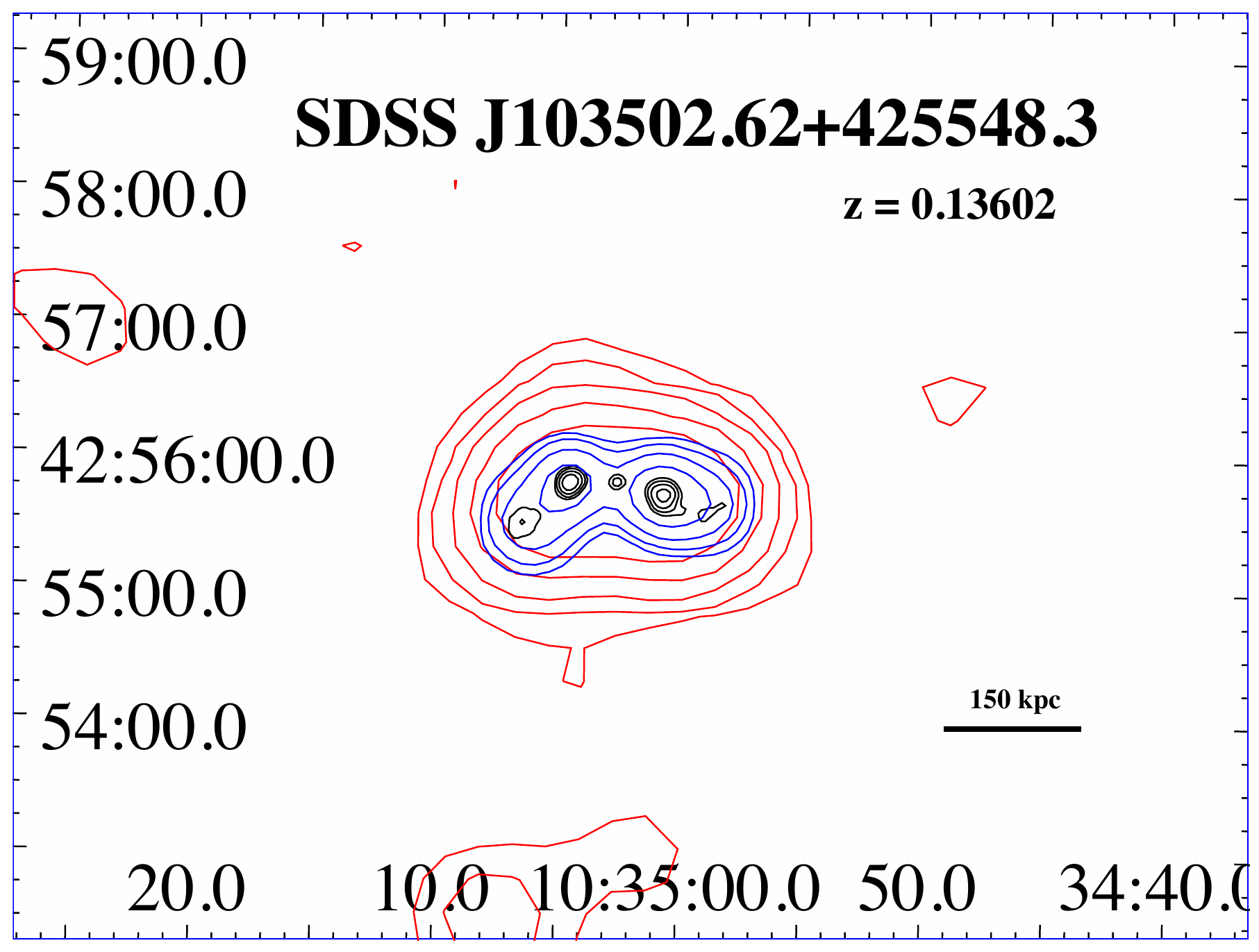} 
\includegraphics[height=5.cm,width=6.cm,angle=0]{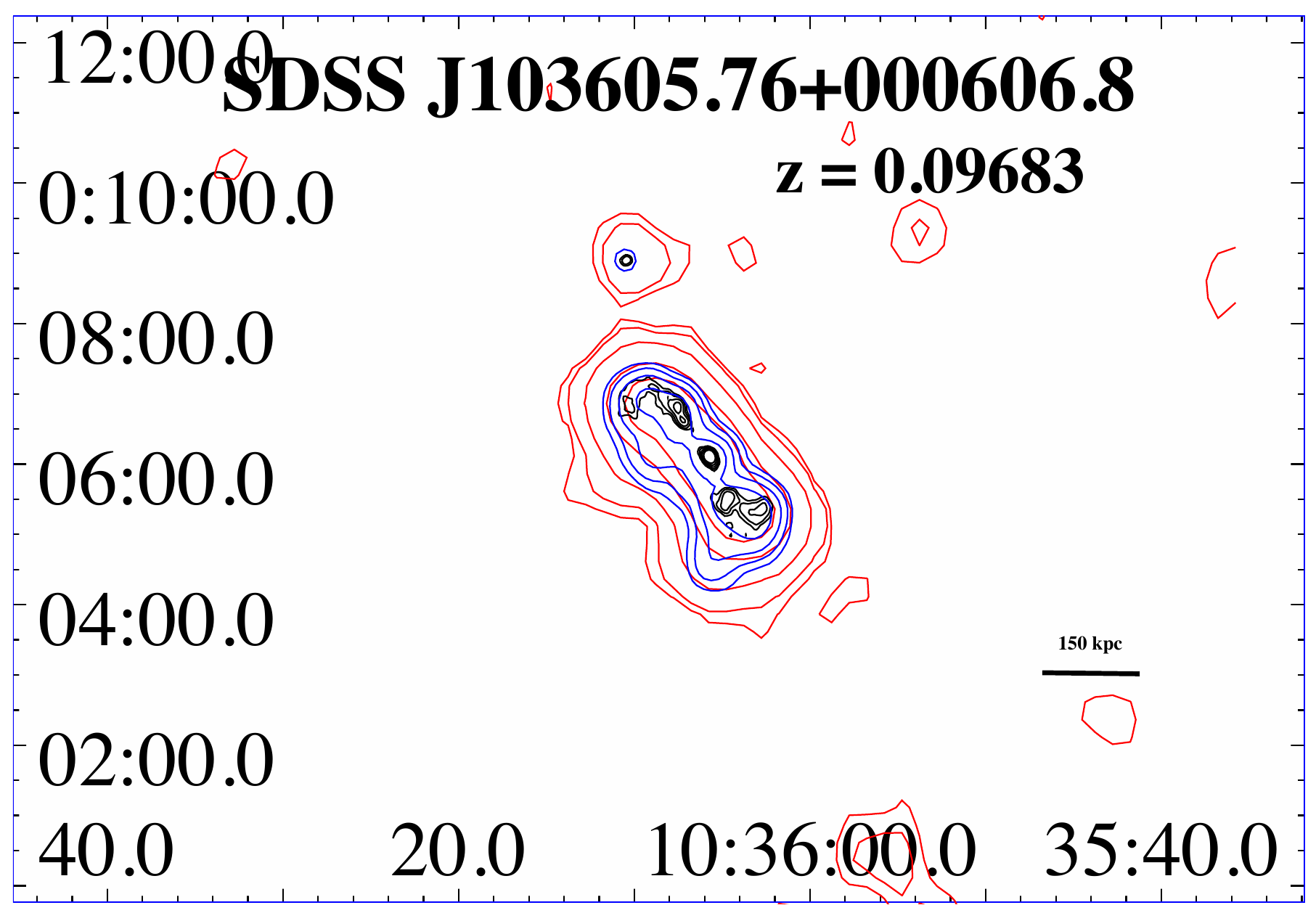} 
\includegraphics[height=5.cm,width=6.cm,angle=0]{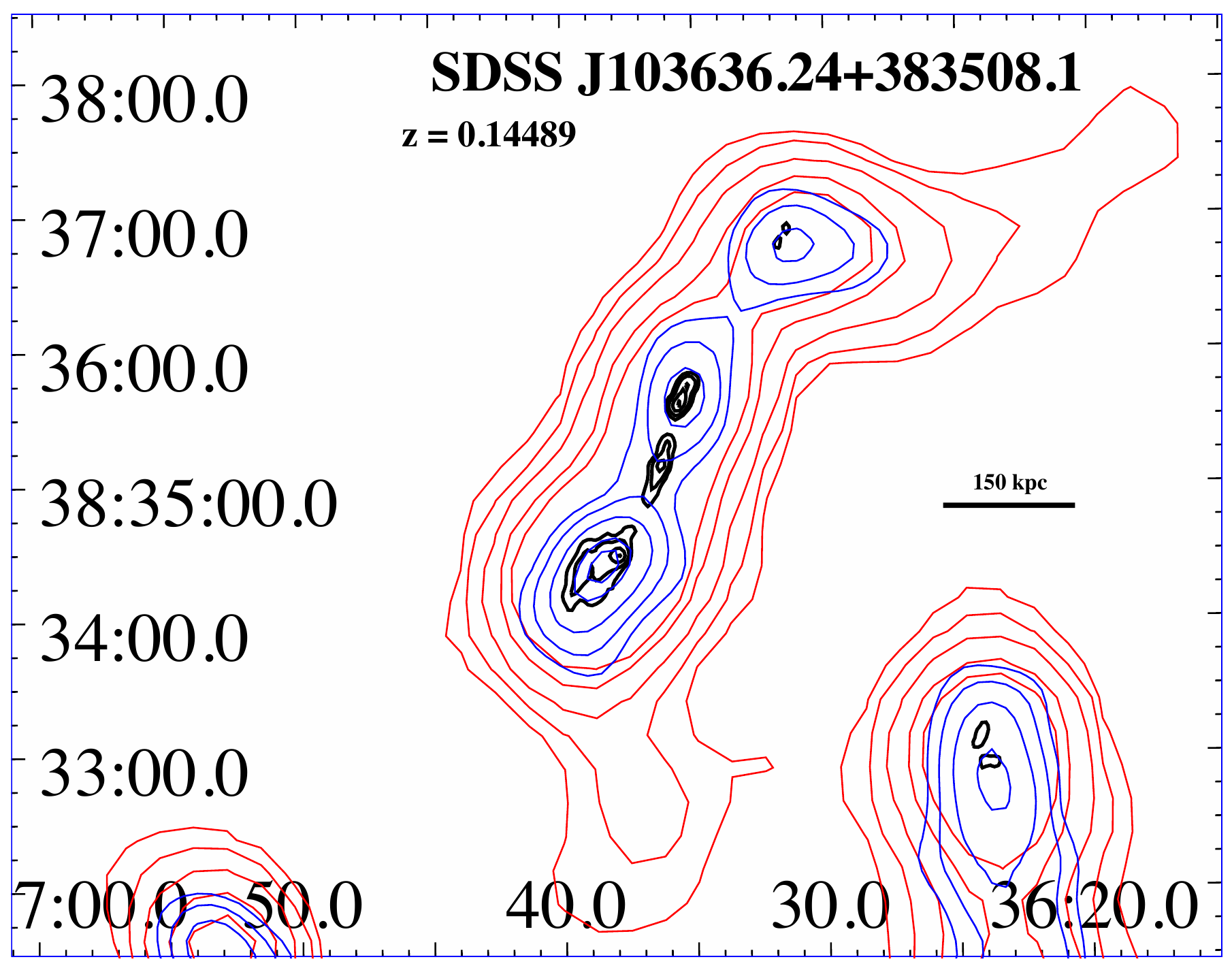} 
\includegraphics[height=5.cm,width=6.cm,angle=0]{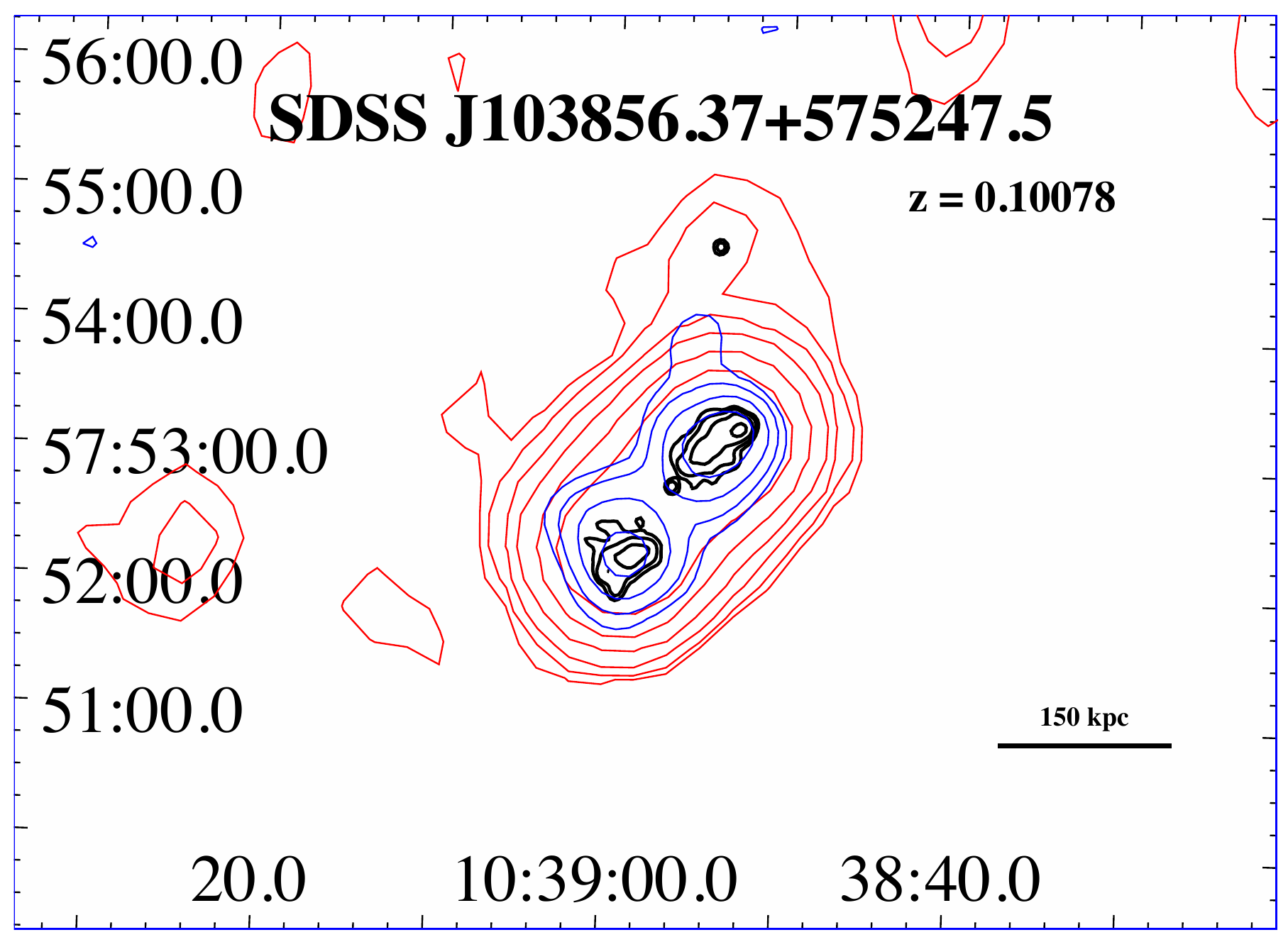} 
\includegraphics[height=5.cm,width=6.cm,angle=0]{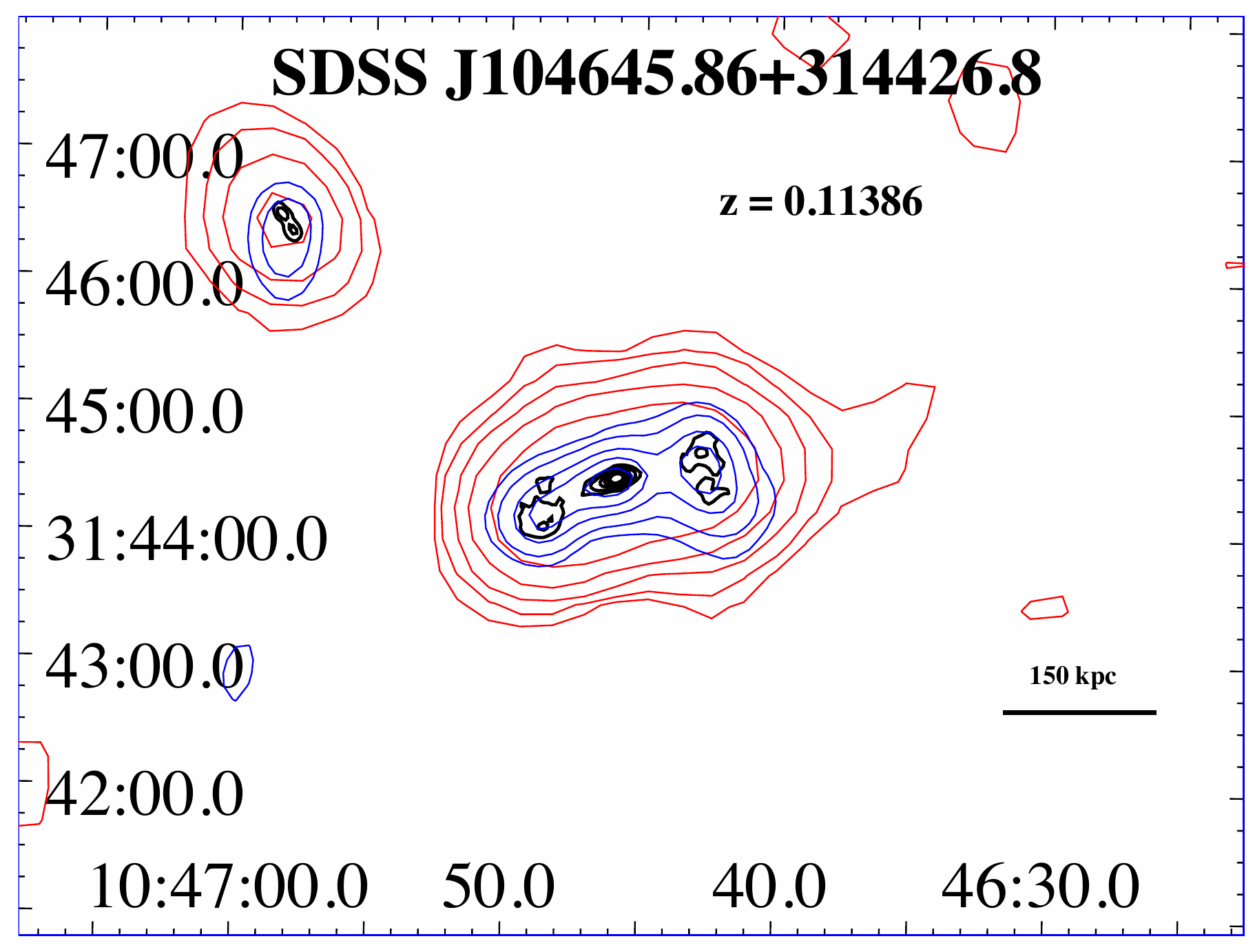} 
\includegraphics[height=5.cm,width=6.cm,angle=0]{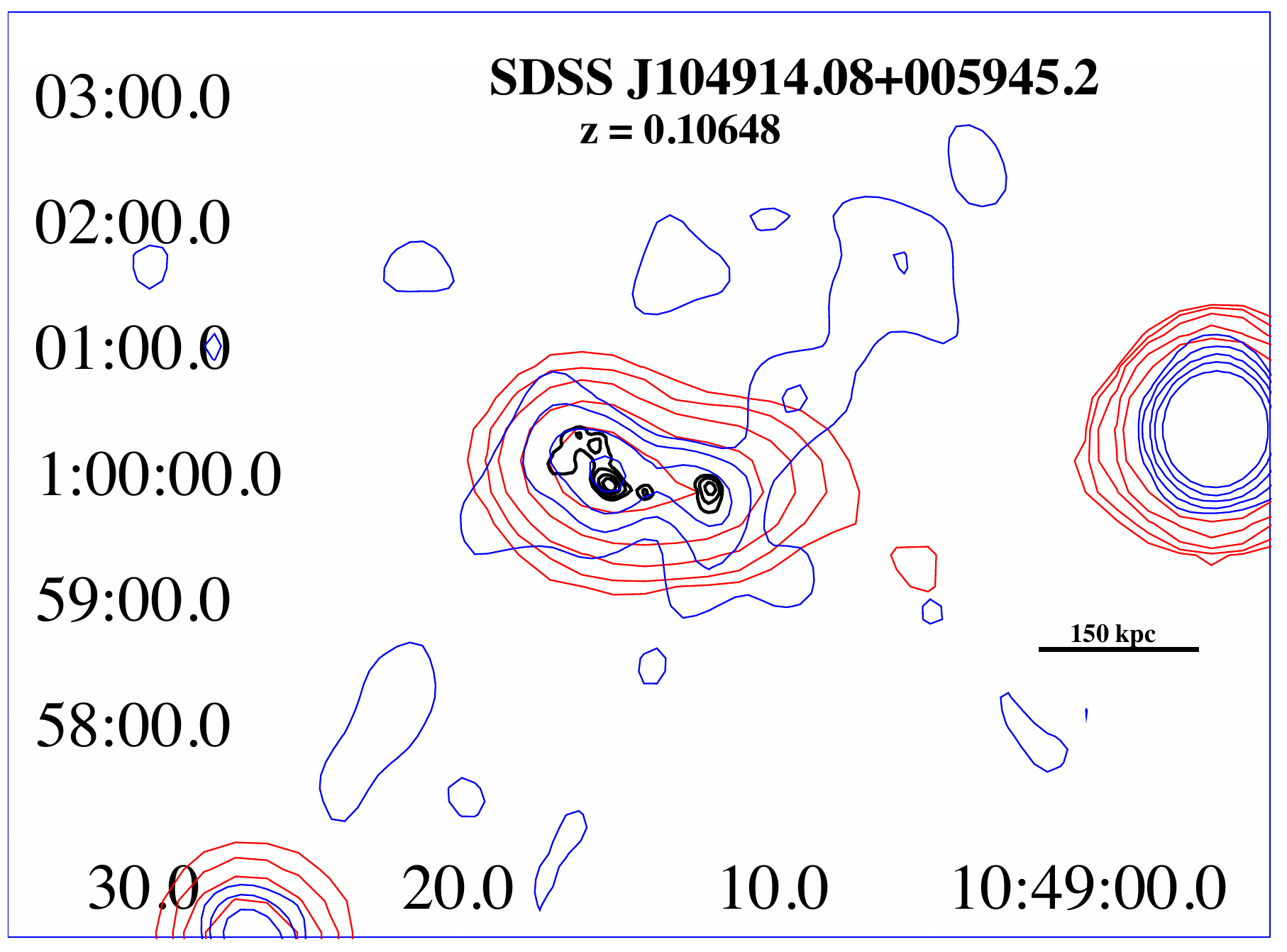} 
\includegraphics[height=5.cm,width=6.cm,angle=0]{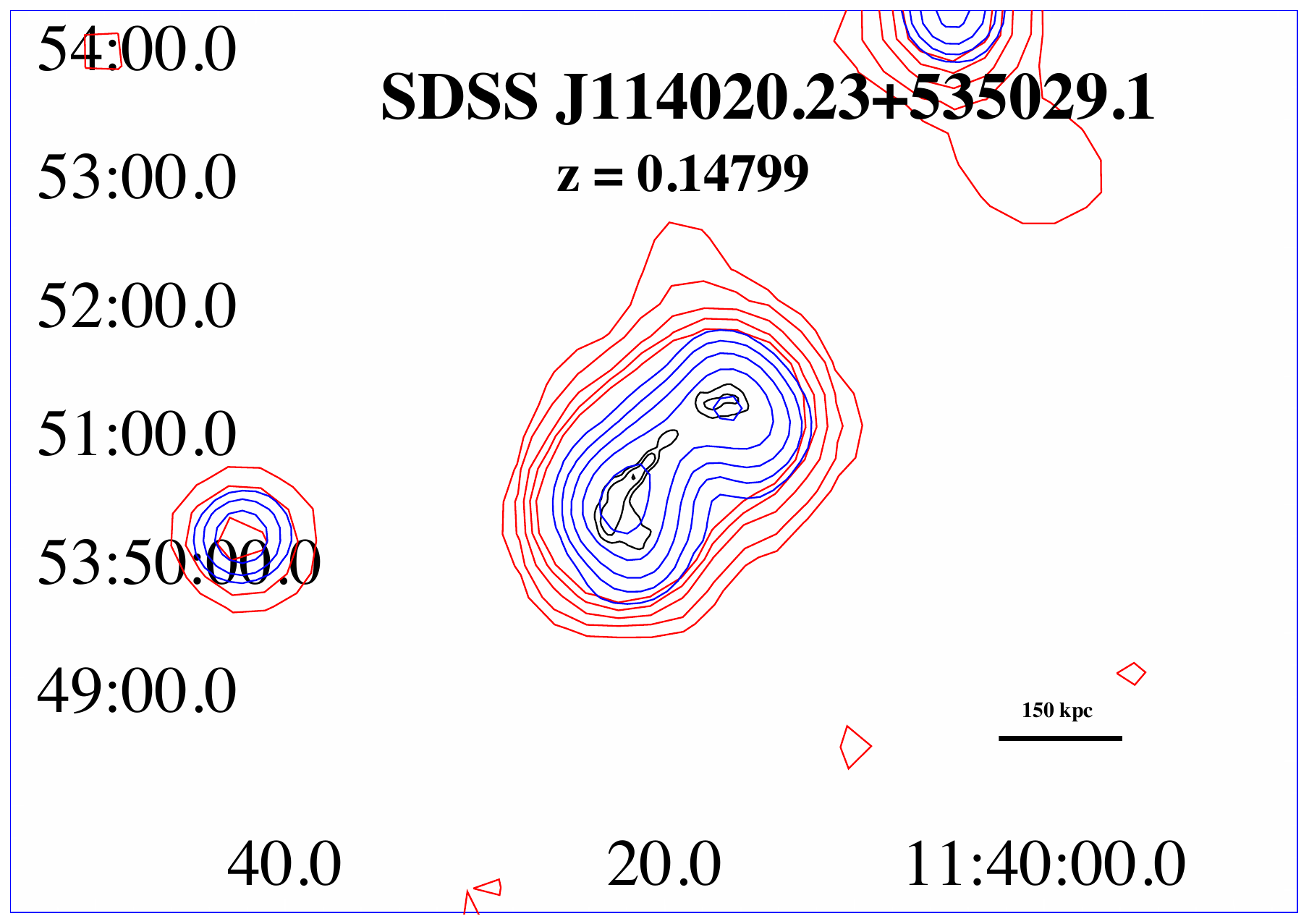} 
\includegraphics[height=5.cm,width=6.cm,angle=0]{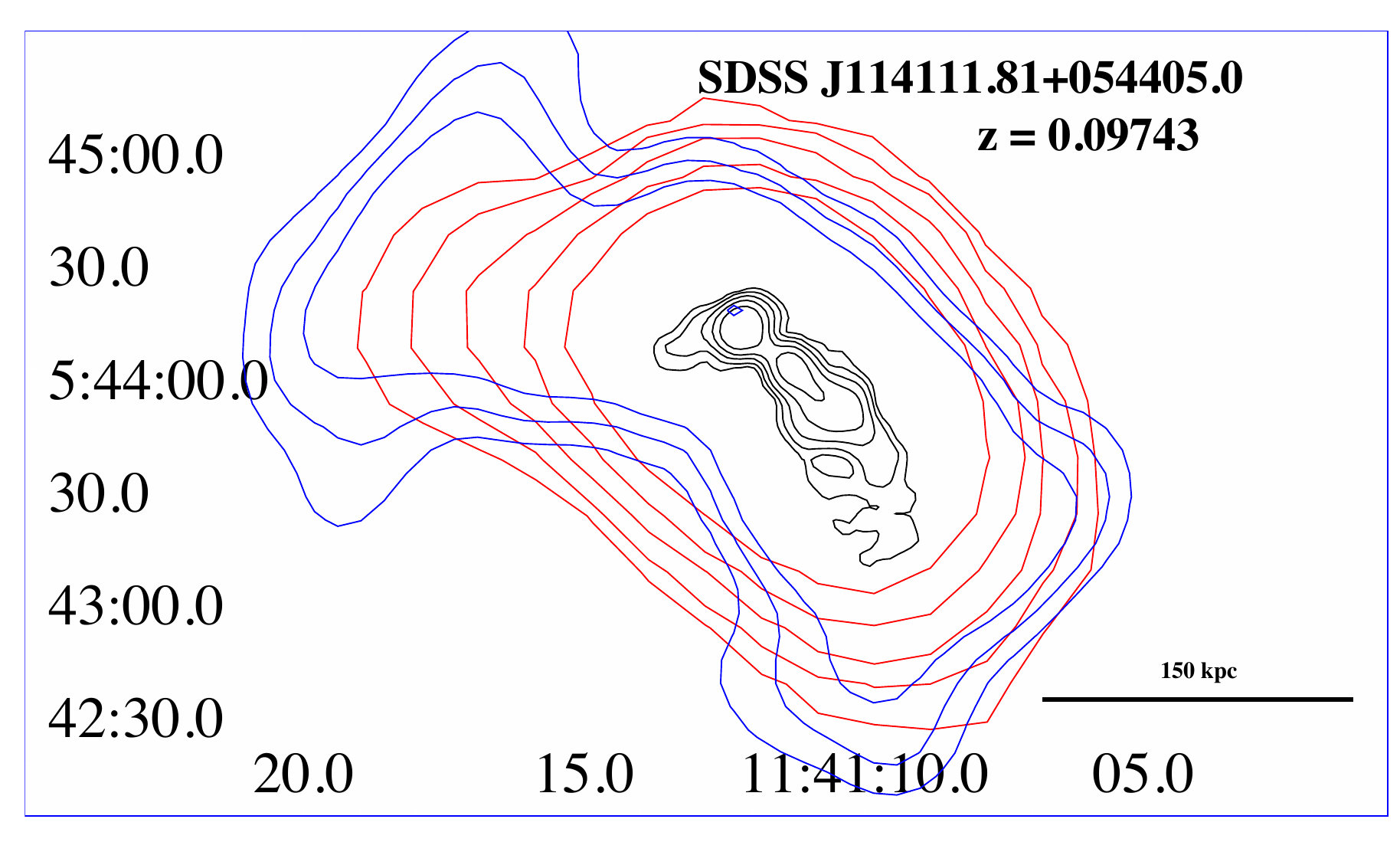} 
\includegraphics[height=5.cm,width=6.cm,angle=0]{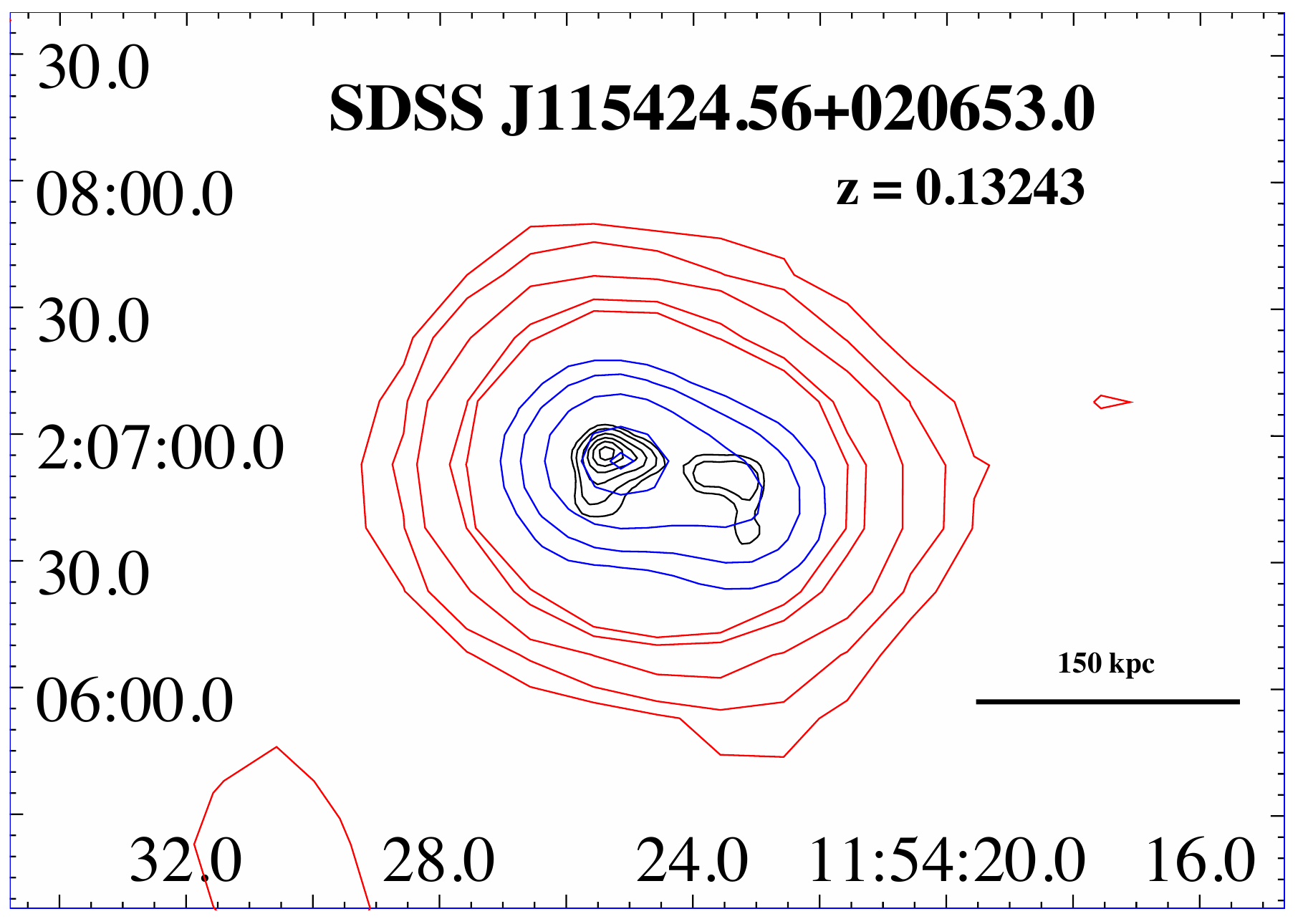} 
\includegraphics[height=5.cm,width=6.cm,angle=0]{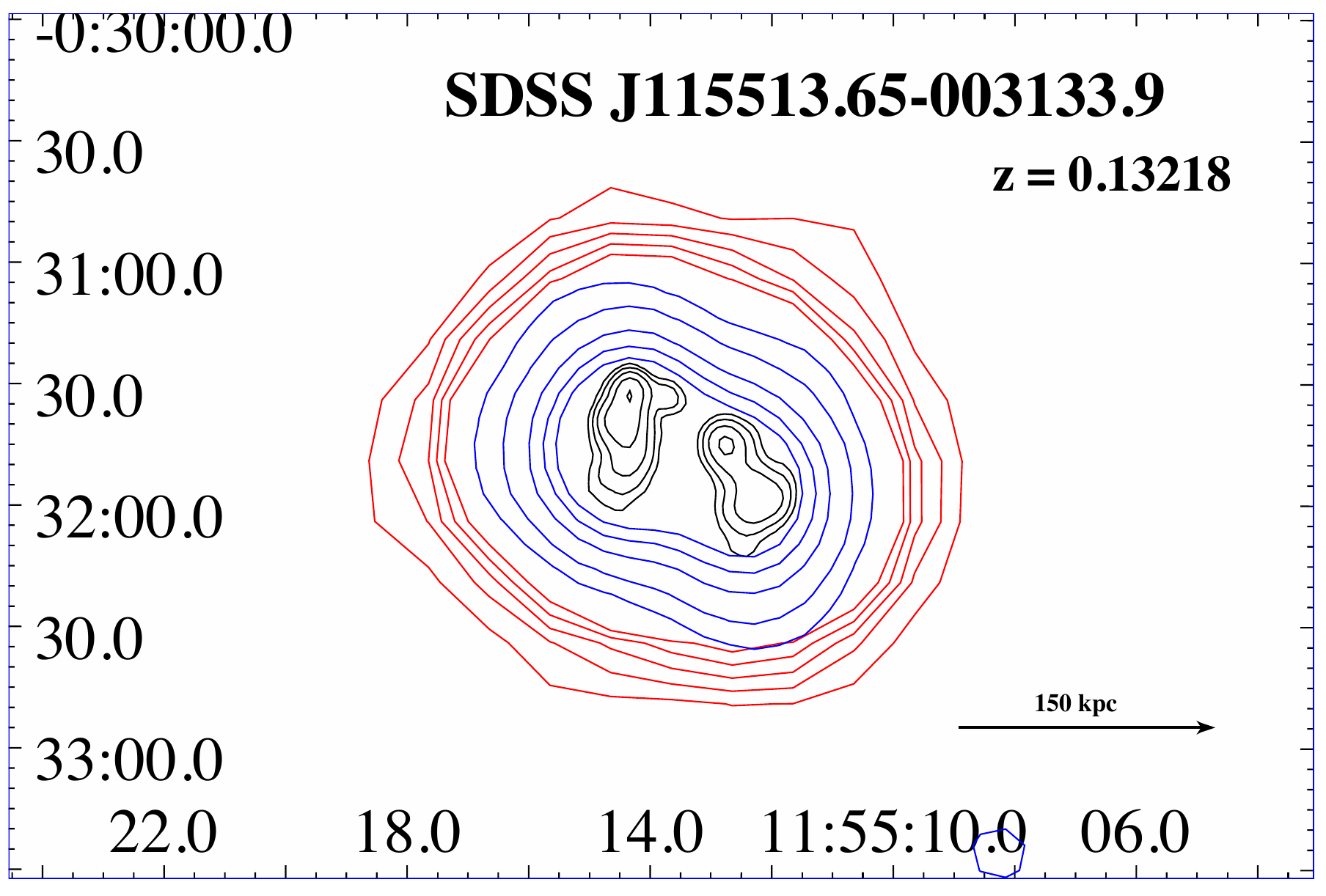} 
\includegraphics[height=5.cm,width=6.cm,angle=0]{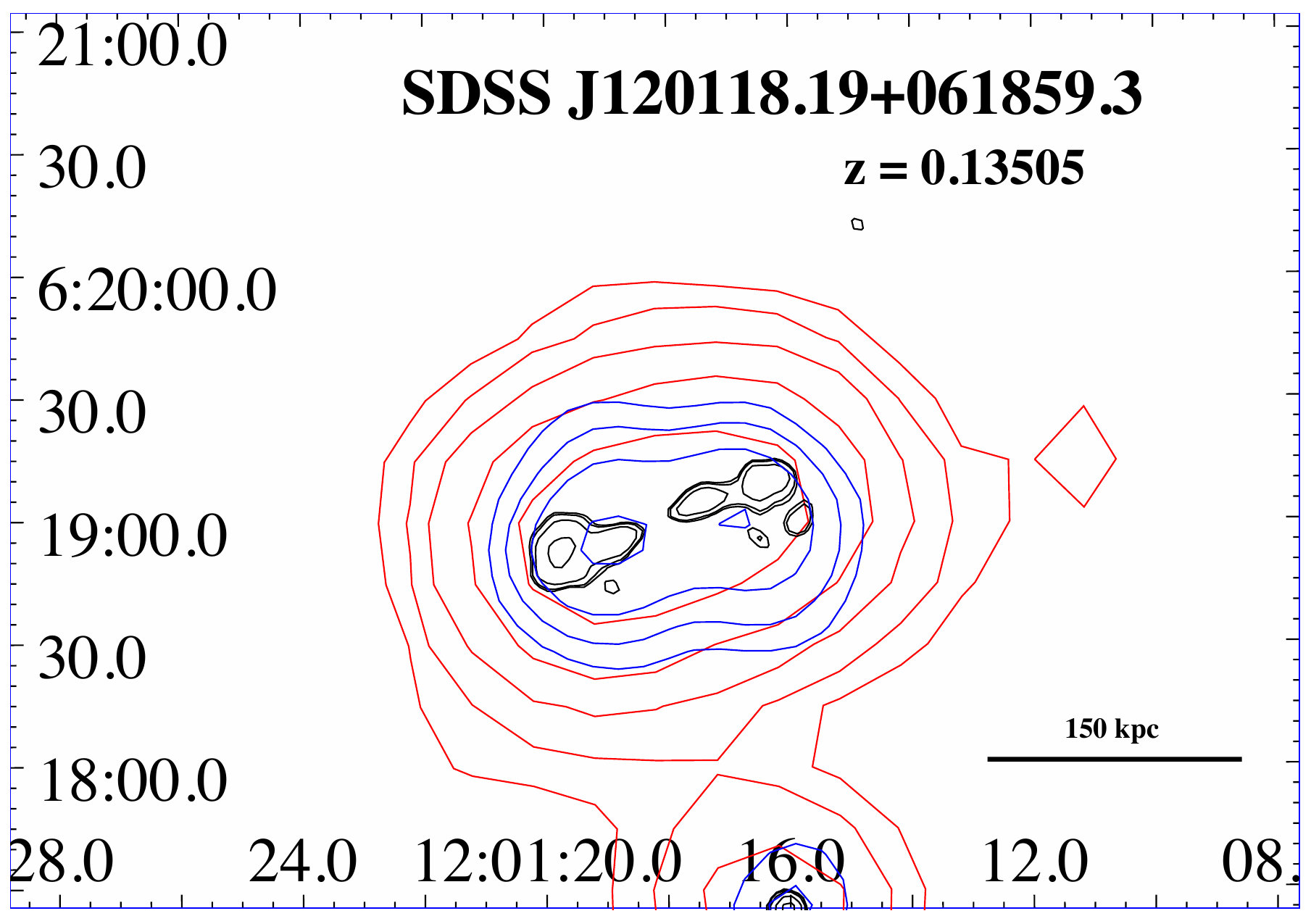} 
\caption{(continued)}
\end{figure*}

\addtocounter{figure}{-1}
\begin{figure*}
\includegraphics[height=5.cm,width=6.cm,angle=0]{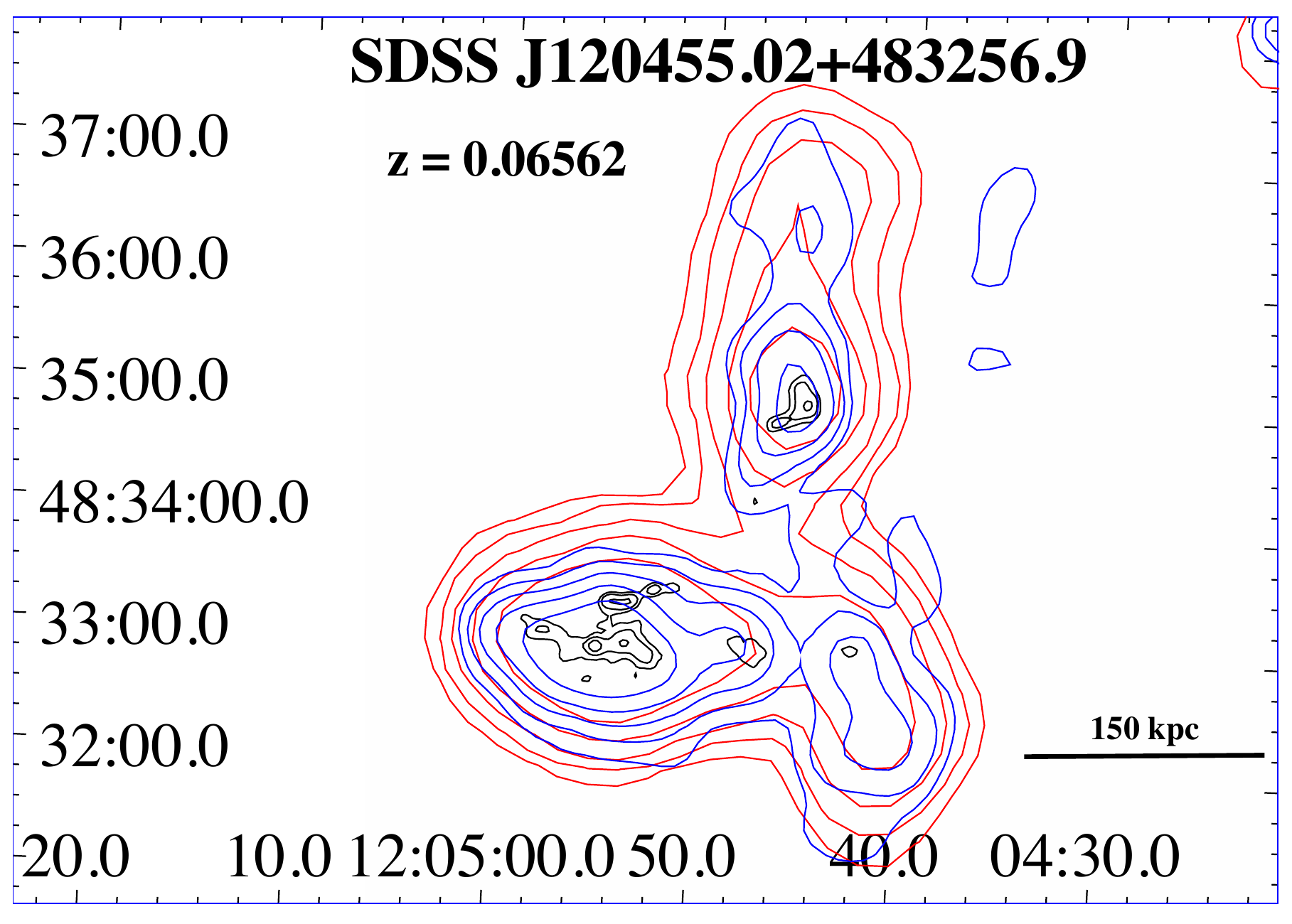}
\includegraphics[height=5.cm,width=6.cm,angle=0]{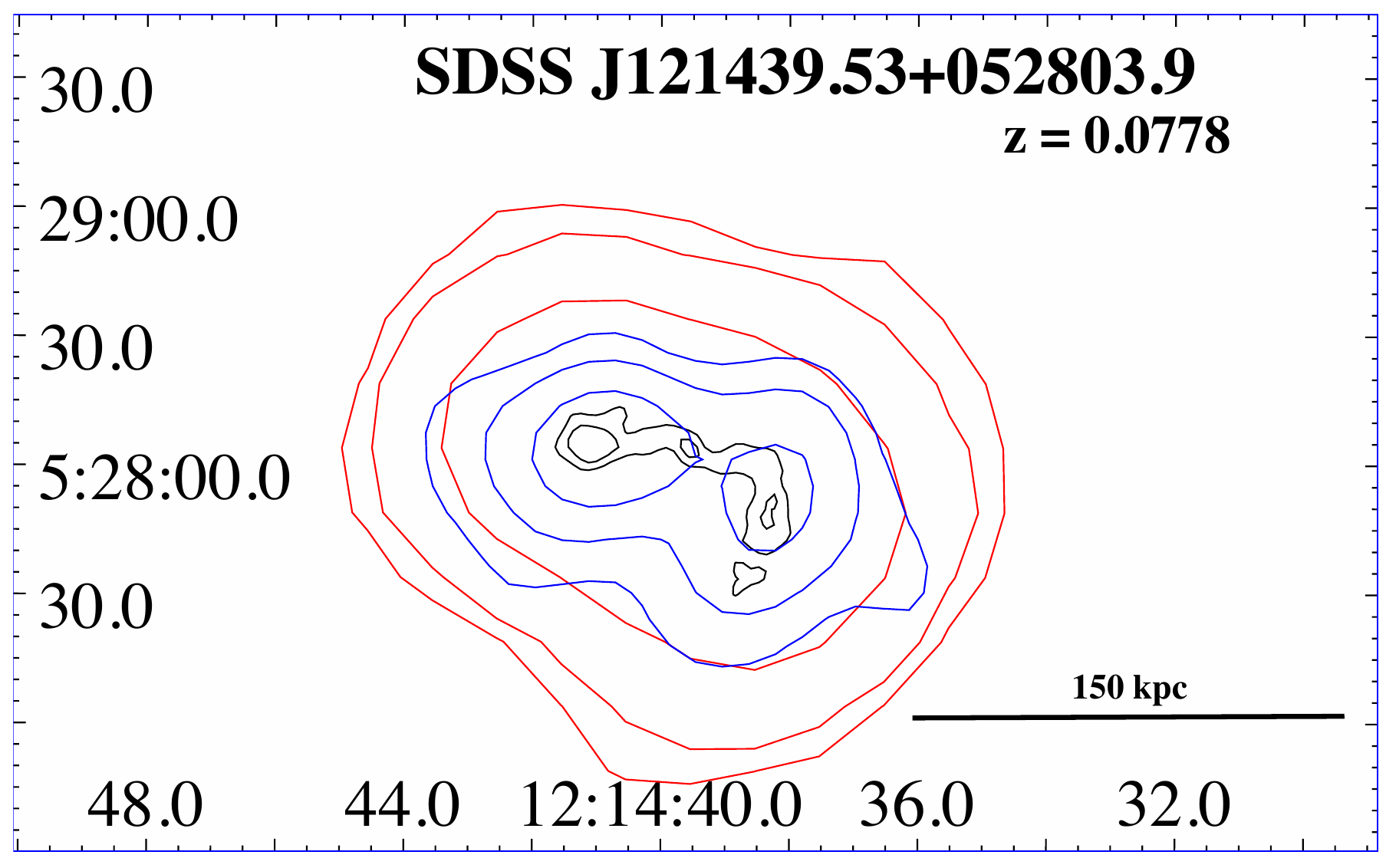}
\includegraphics[height=5.cm,width=6.cm,angle=0]{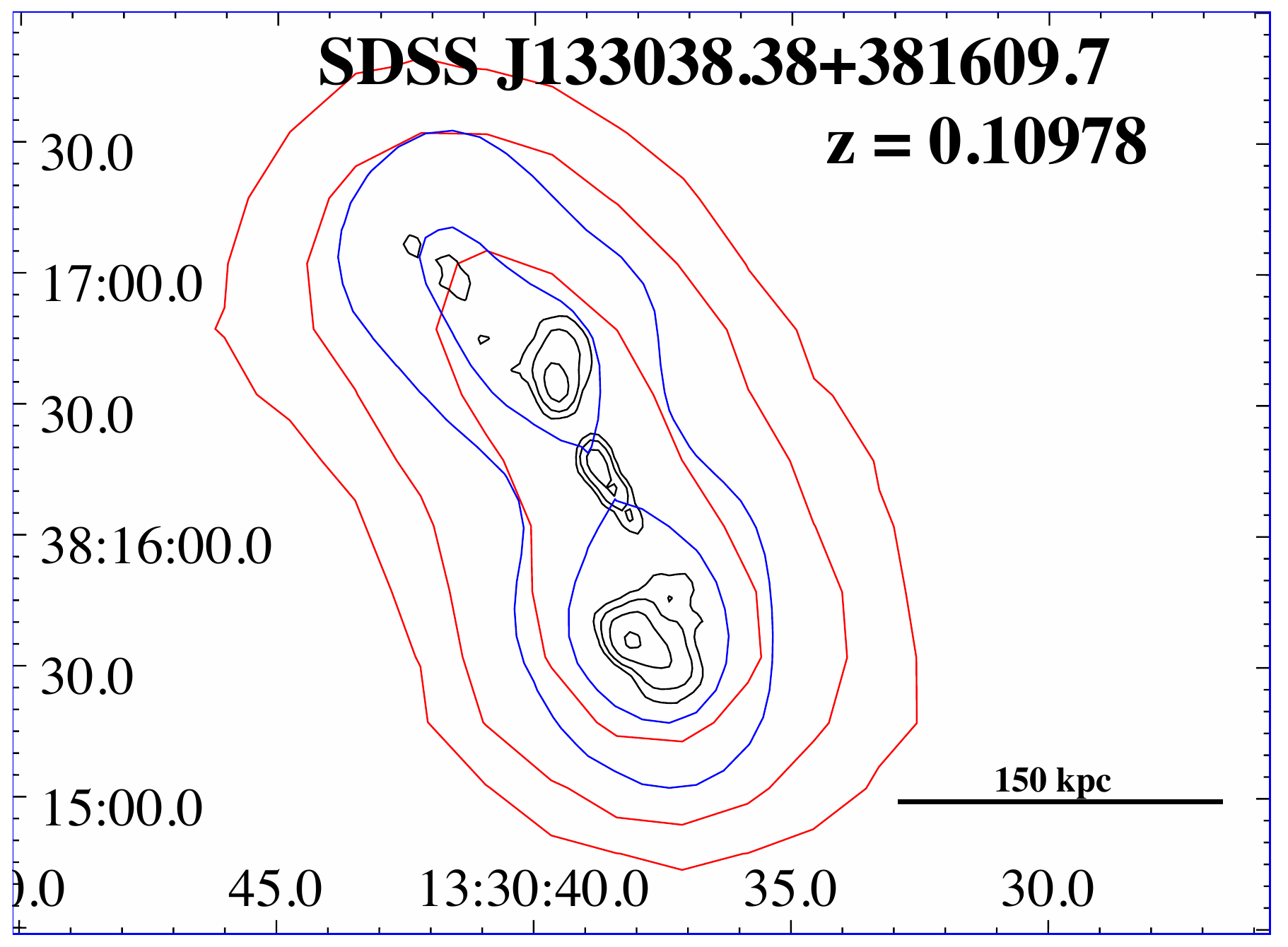} 
\includegraphics[height=5.cm,width=6.cm,angle=0]{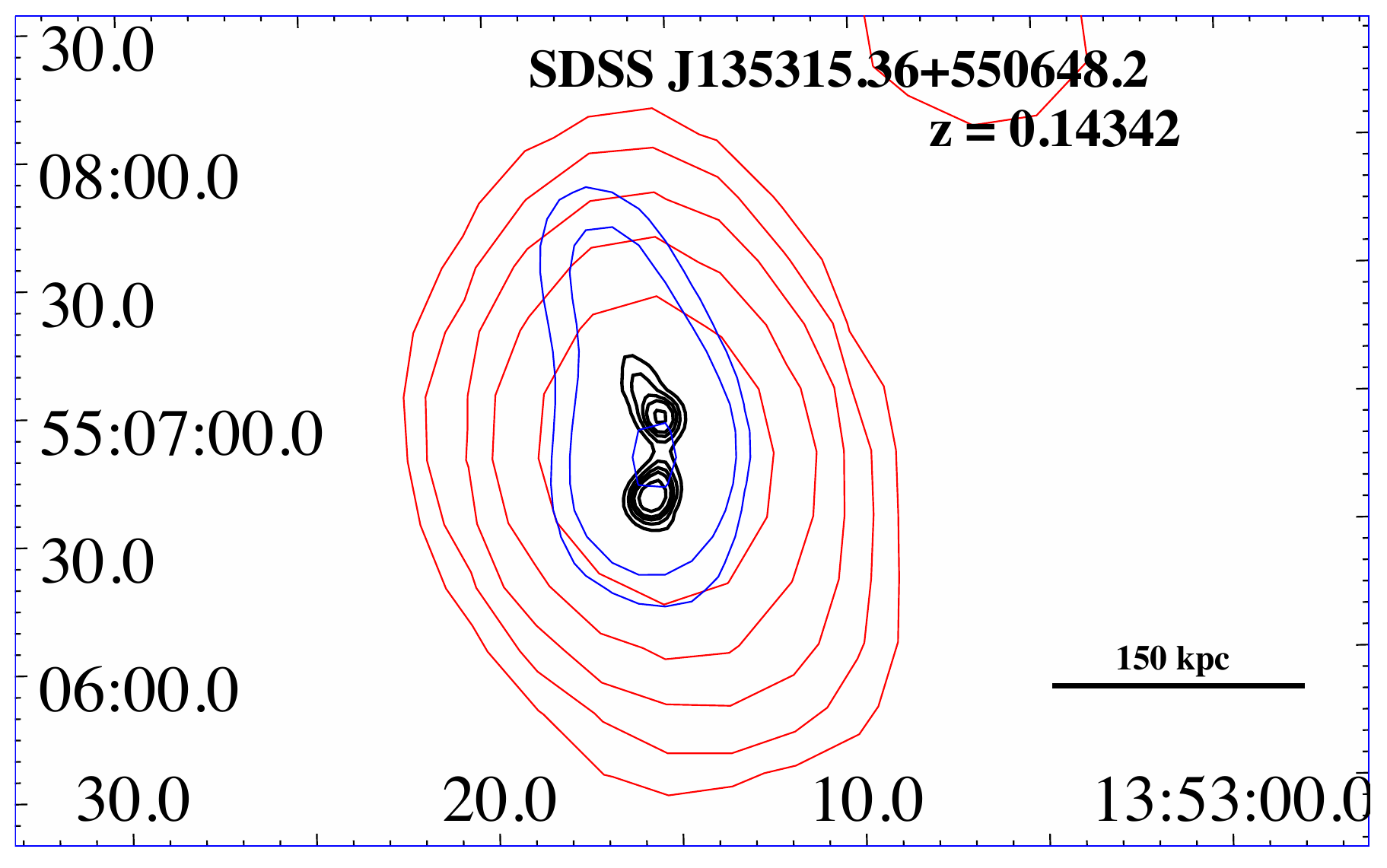} 
\includegraphics[height=5.cm,width=6.cm,angle=0]{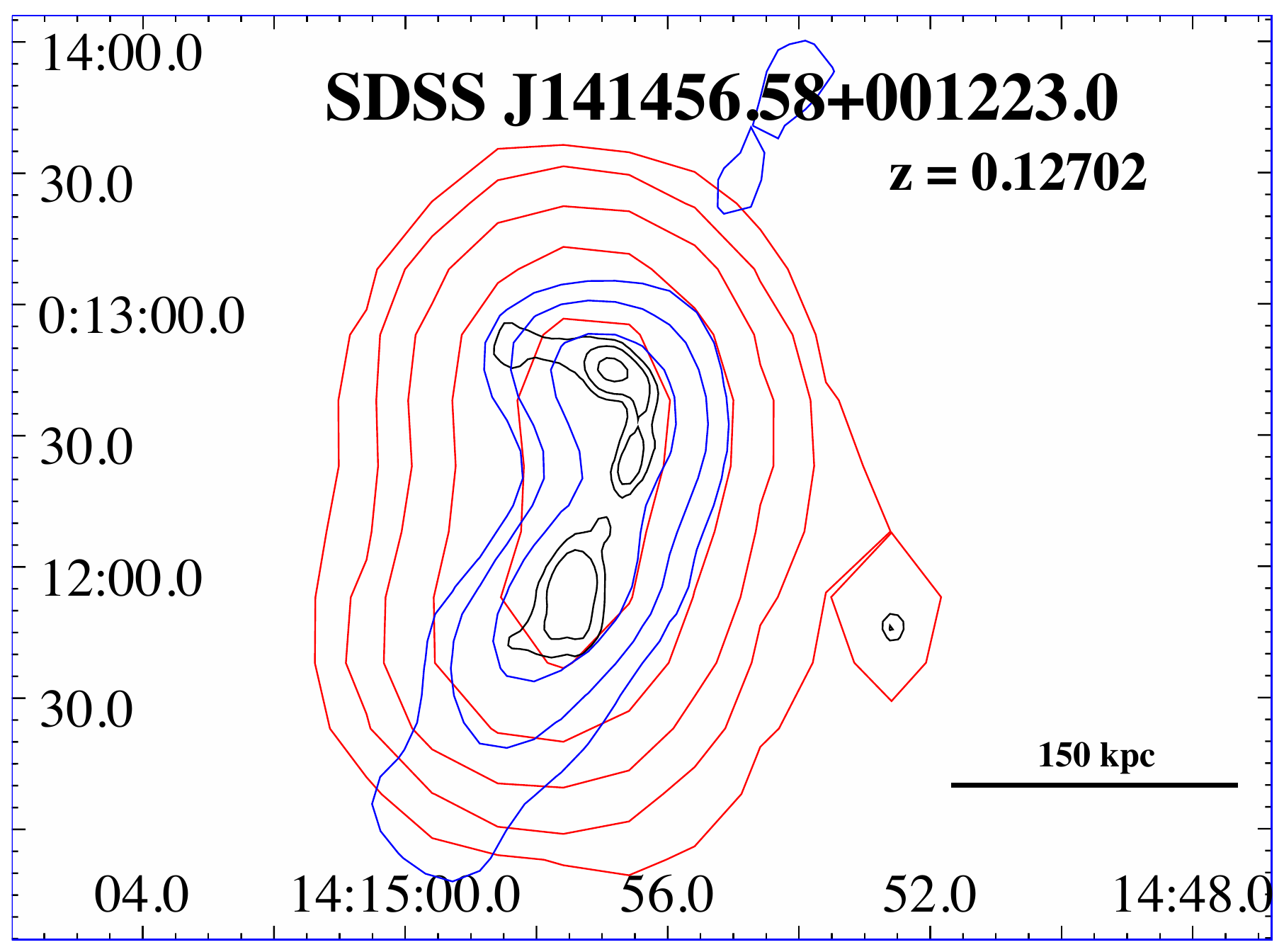} 
\includegraphics[height=5.cm,width=6.cm,angle=0]{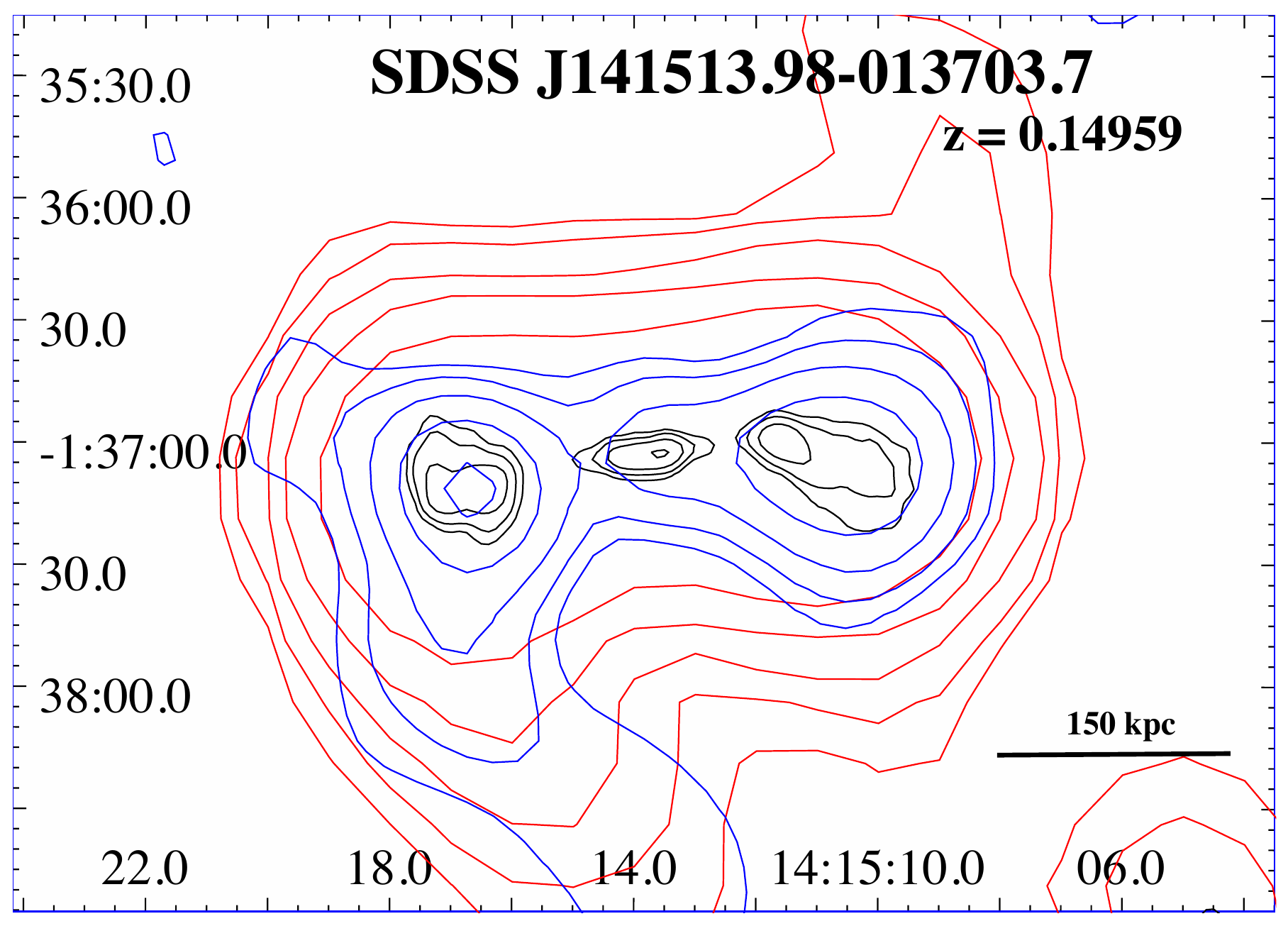} 
\includegraphics[height=5.cm,width=6.cm,angle=0]{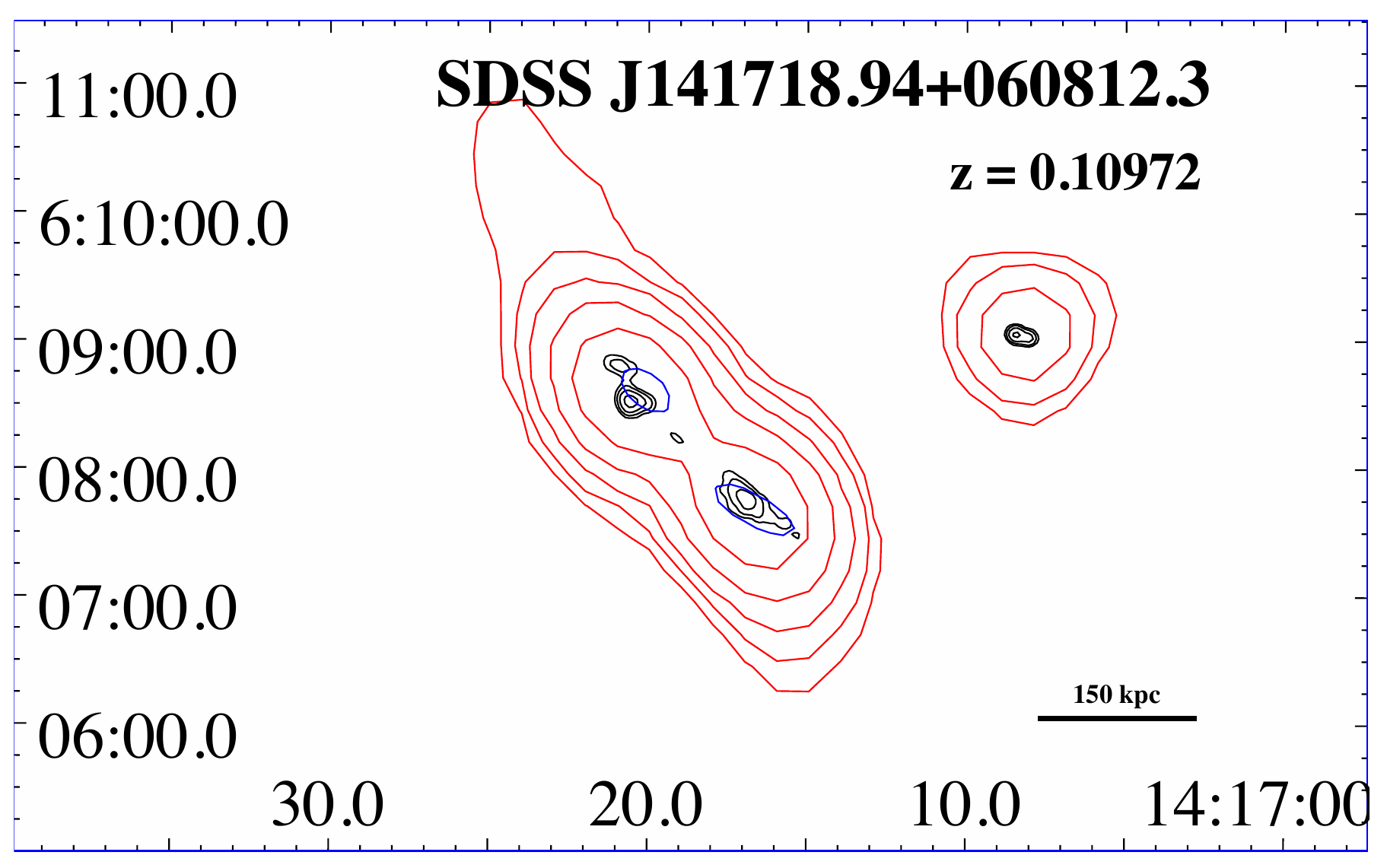} 
\includegraphics[height=5.cm,width=6.cm,angle=0]{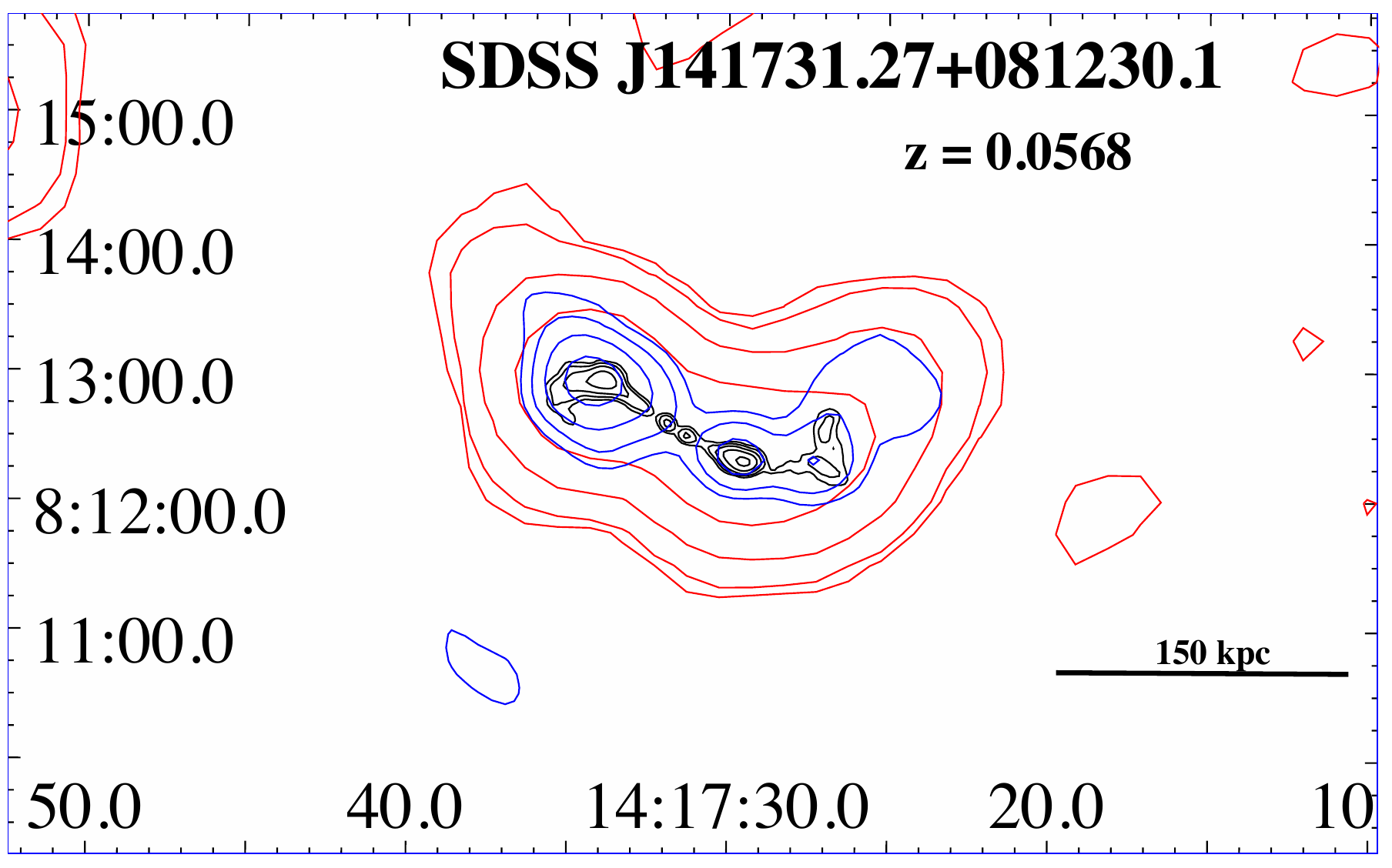} 
\includegraphics[height=5.cm,width=6.cm,angle=0]{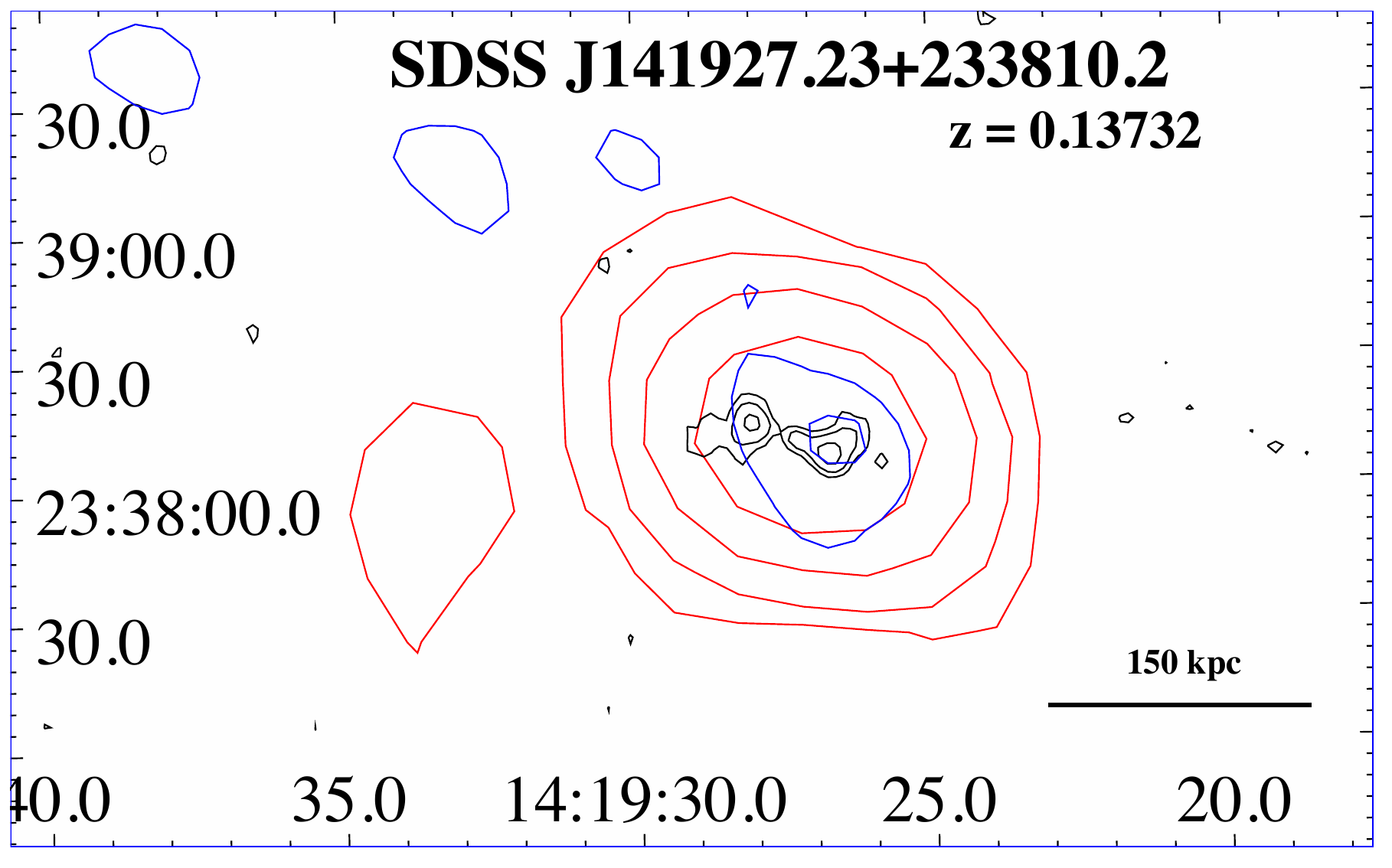} 
\includegraphics[height=5.cm,width=6.cm,angle=0]{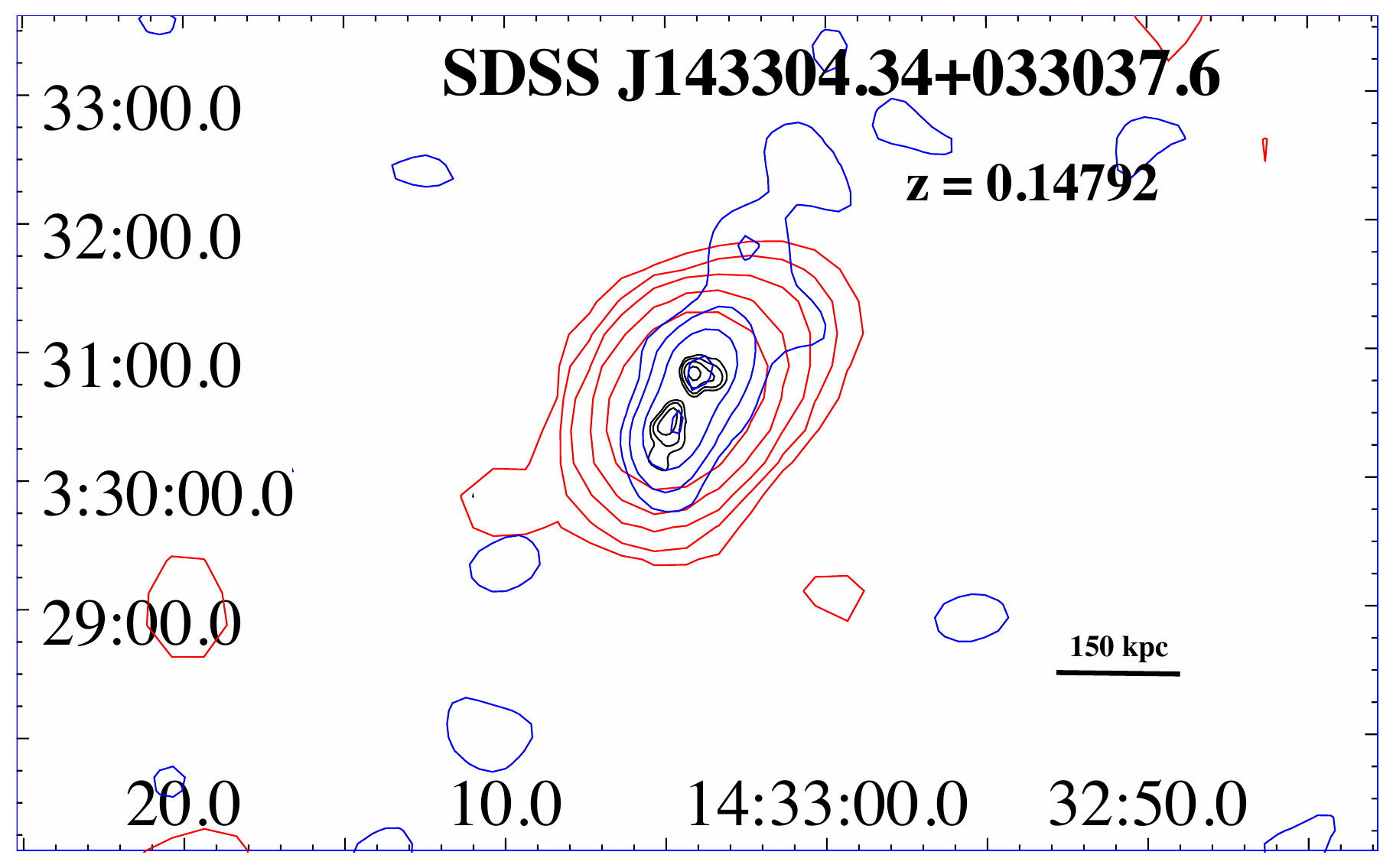} 
\includegraphics[height=5.cm,width=6.cm,angle=0]{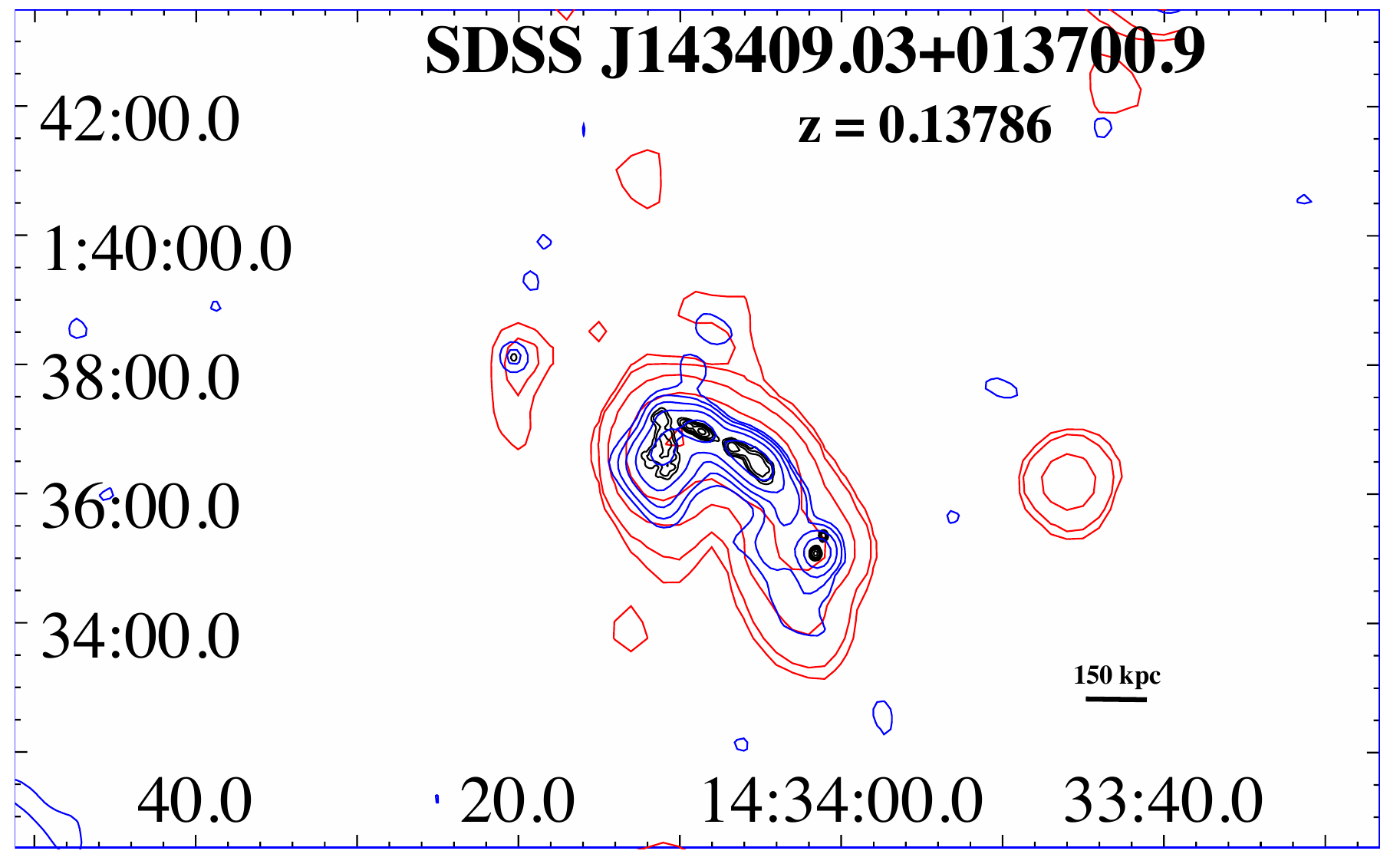} 
\includegraphics[height=5.cm,width=6.cm,angle=0]{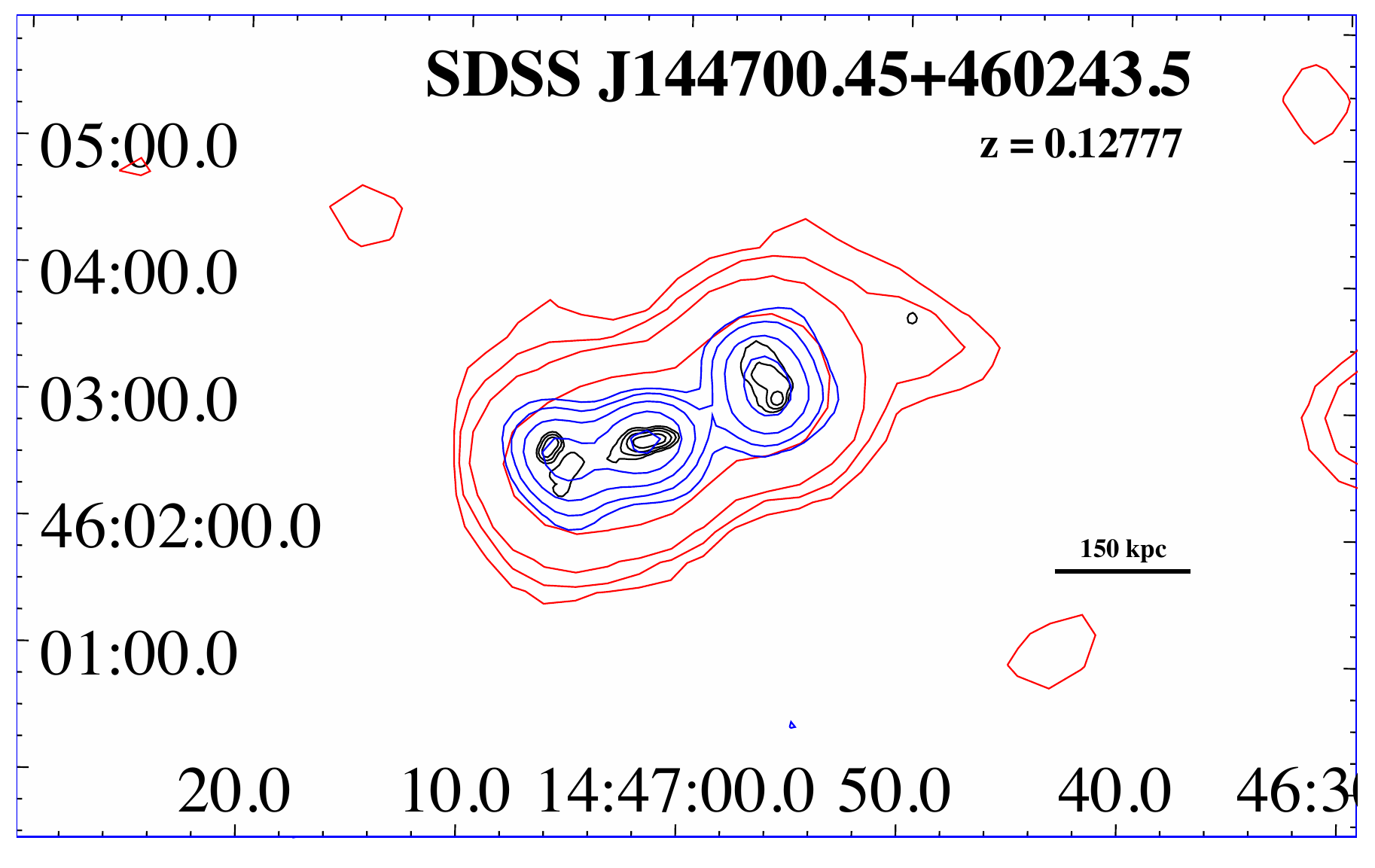}
\includegraphics[height=5.cm,width=6.cm,angle=0]{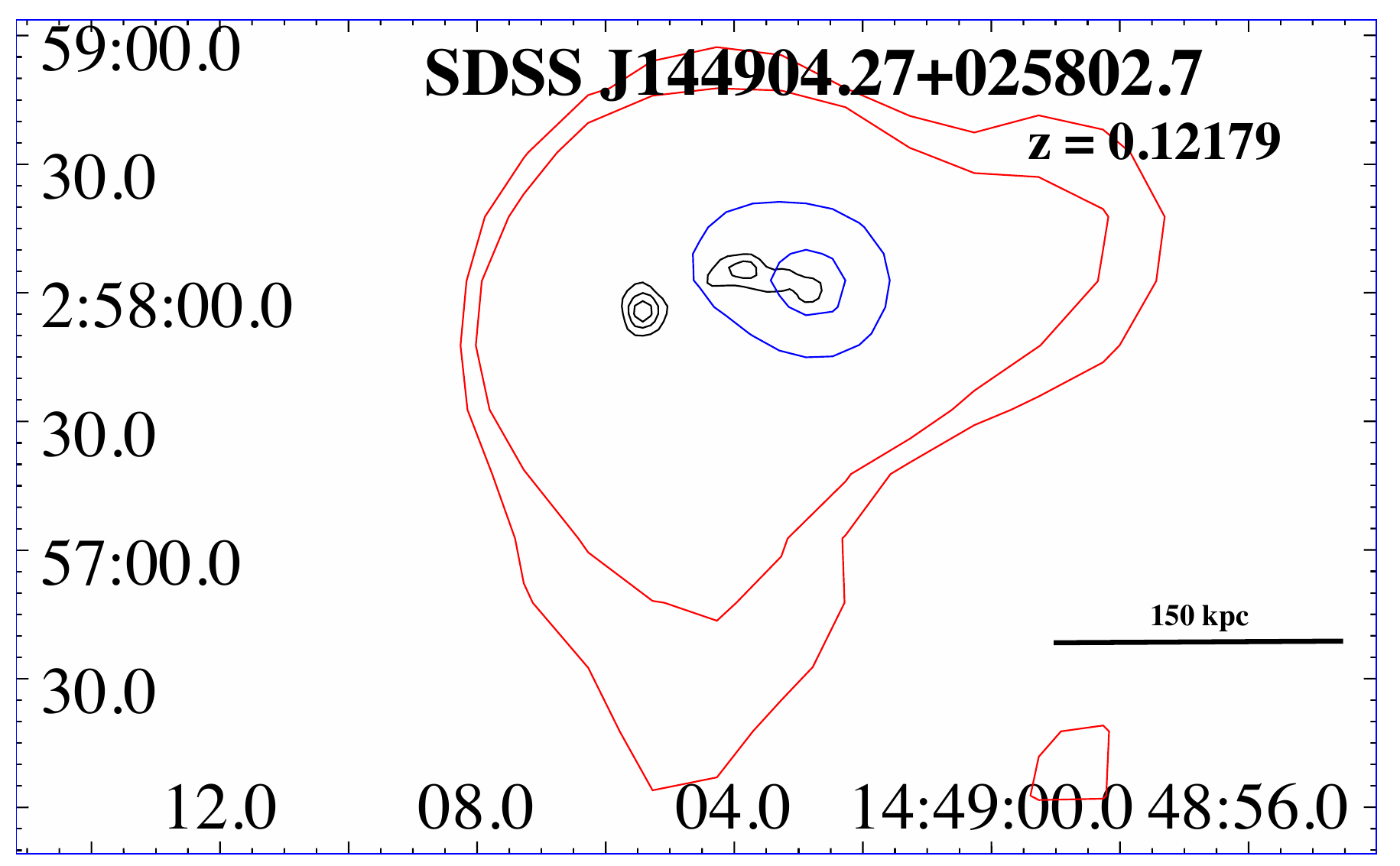}
\includegraphics[height=5.cm,width=6.cm,angle=0]{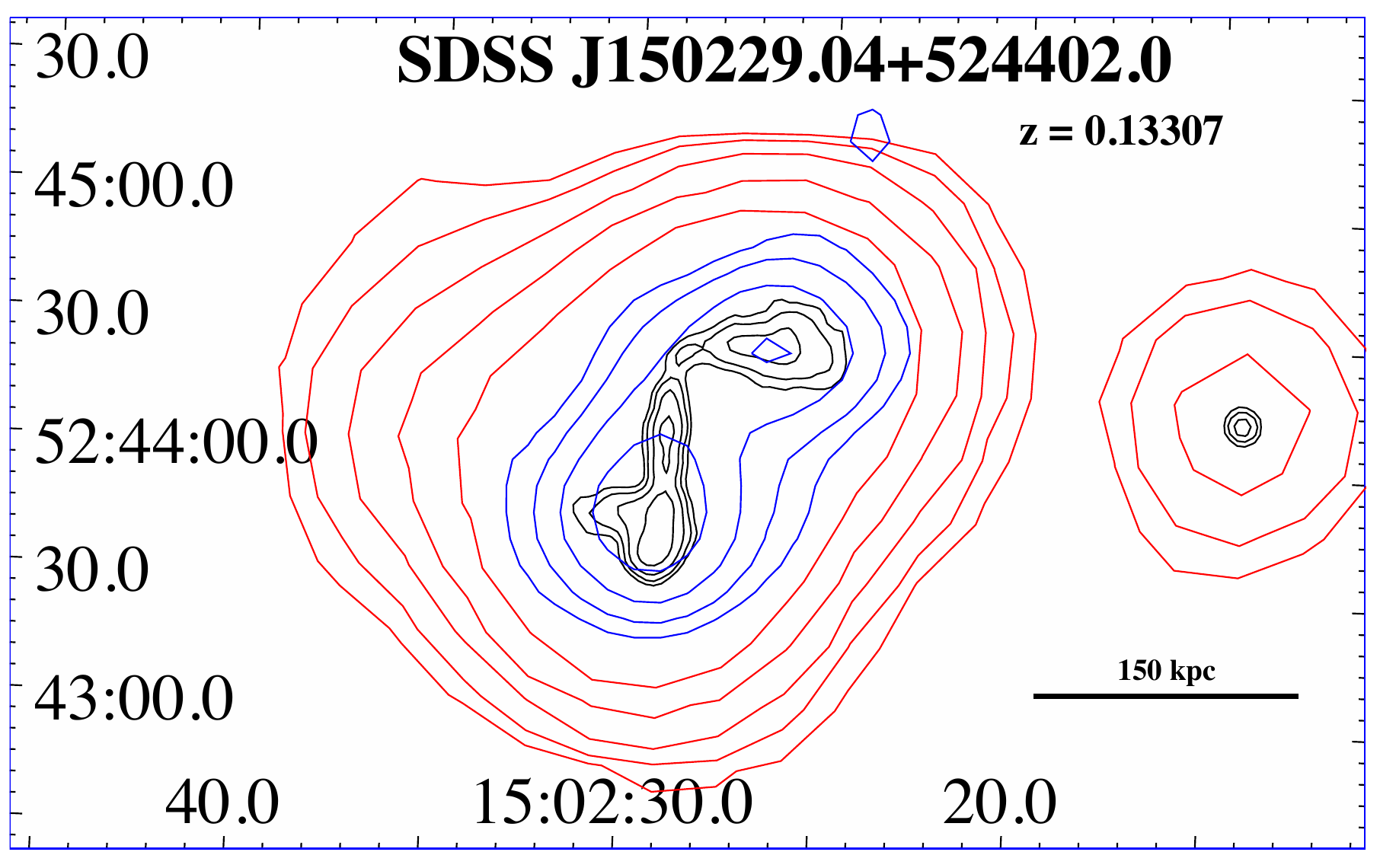} 
\includegraphics[height=5.cm,width=6.cm,angle=0]{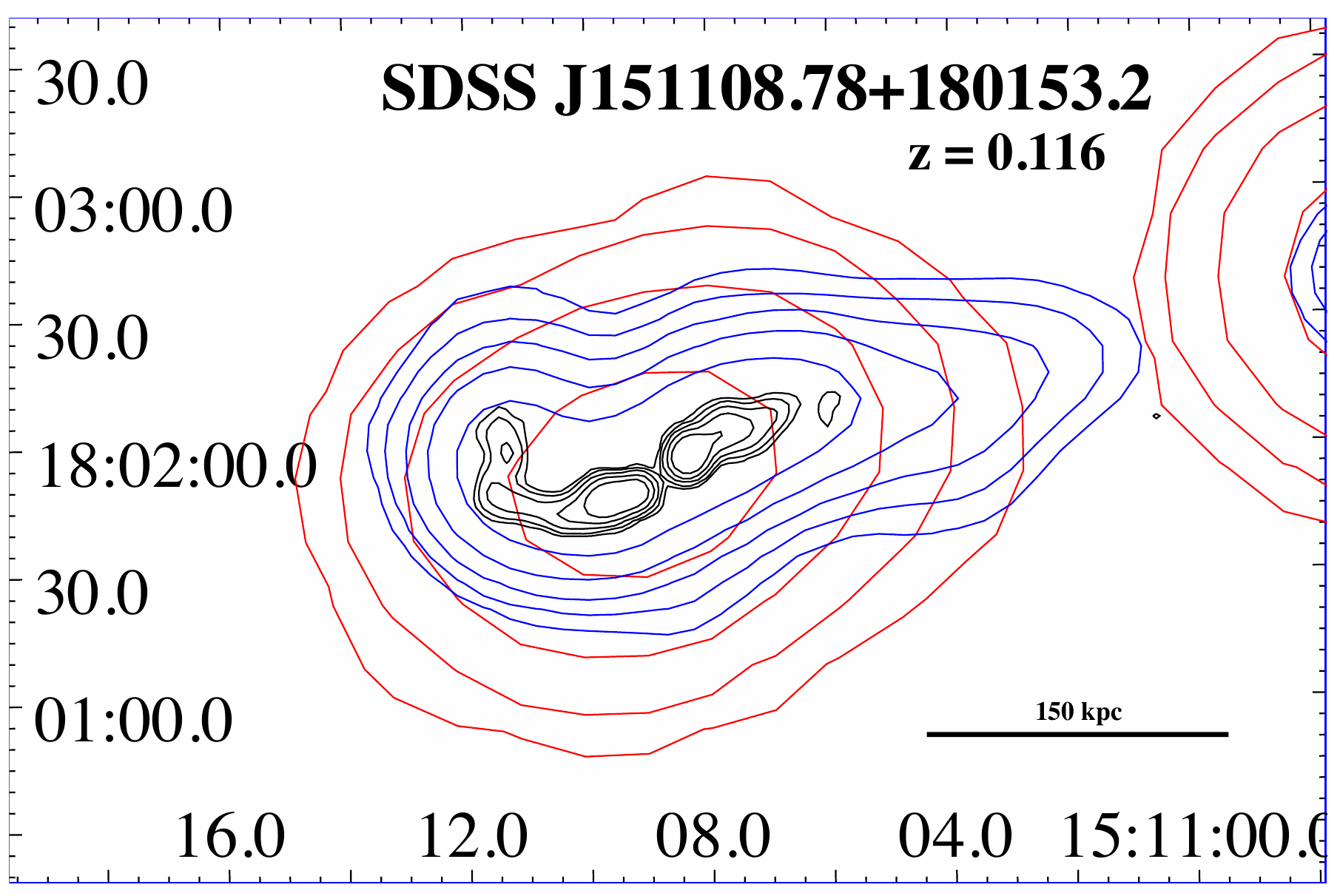}  
\caption{(continued)}
\end{figure*}

\addtocounter{figure}{-1}
\begin{figure*}
        
        \includegraphics[height=5.cm,width=6.cm,angle=0]{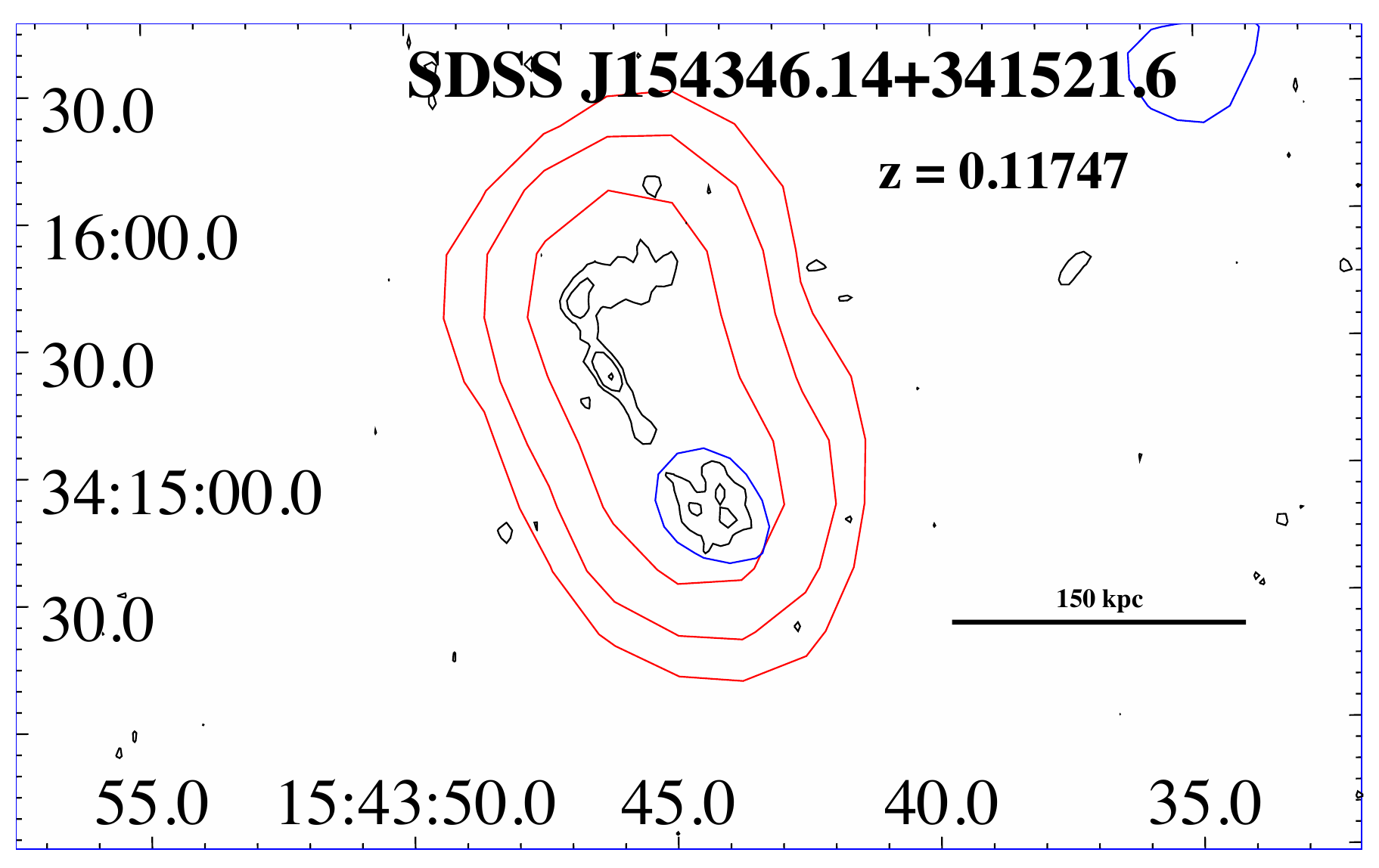} 
        \includegraphics[height=5.cm,width=6.cm,angle=0]{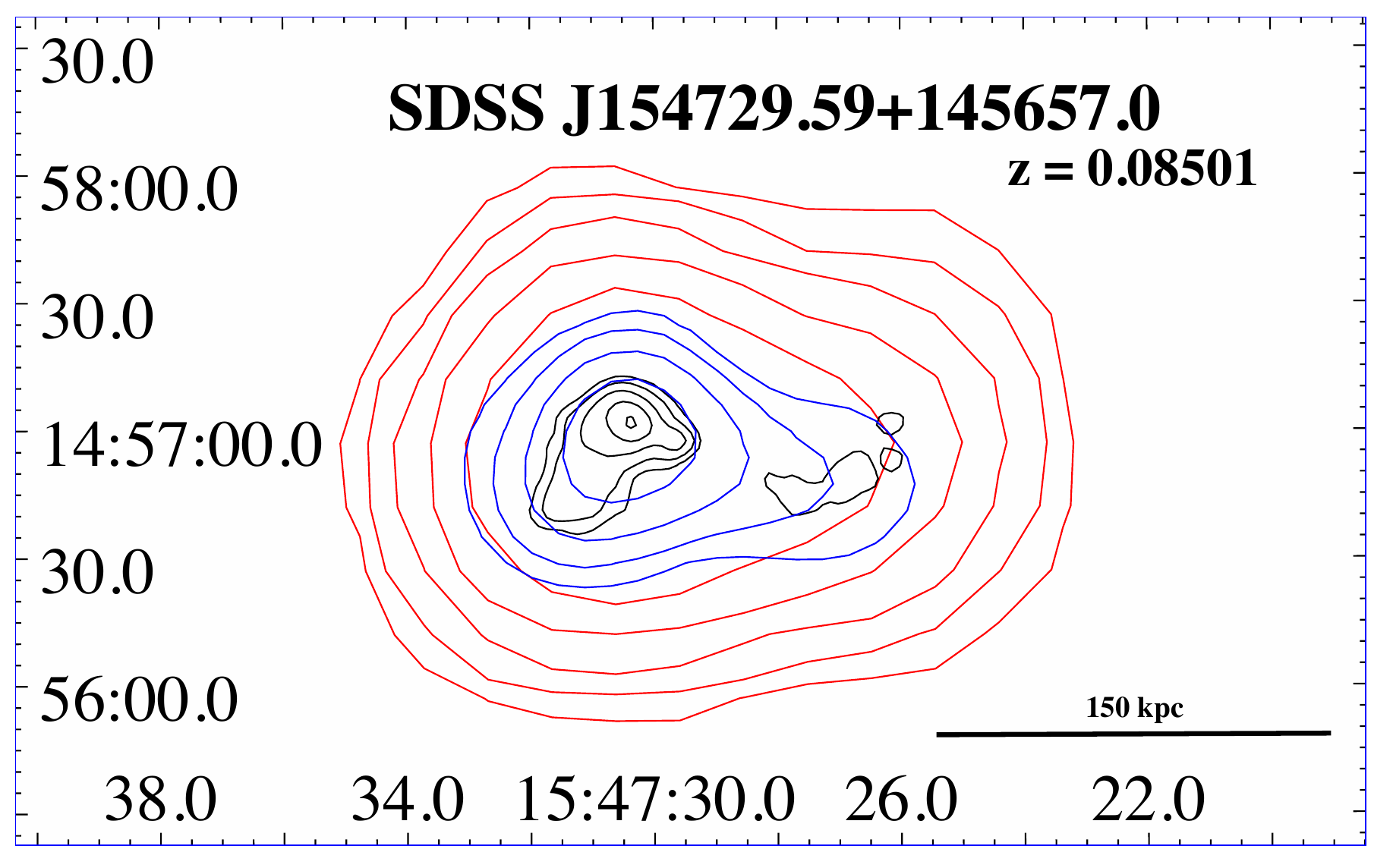} 
        \includegraphics[height=5.cm,width=6.cm,angle=0]{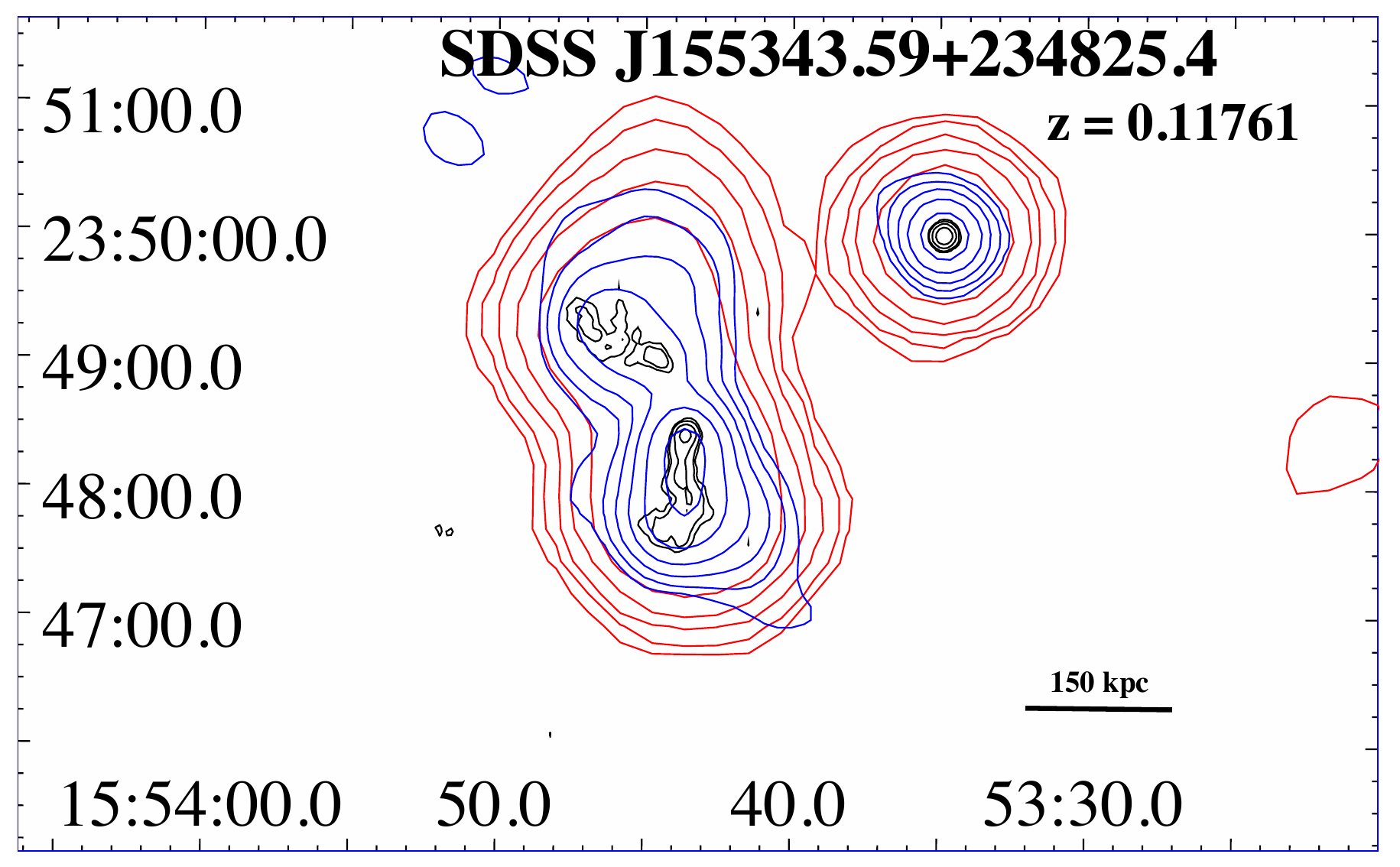} 
        \includegraphics[height=5.cm,width=6.cm,angle=0]{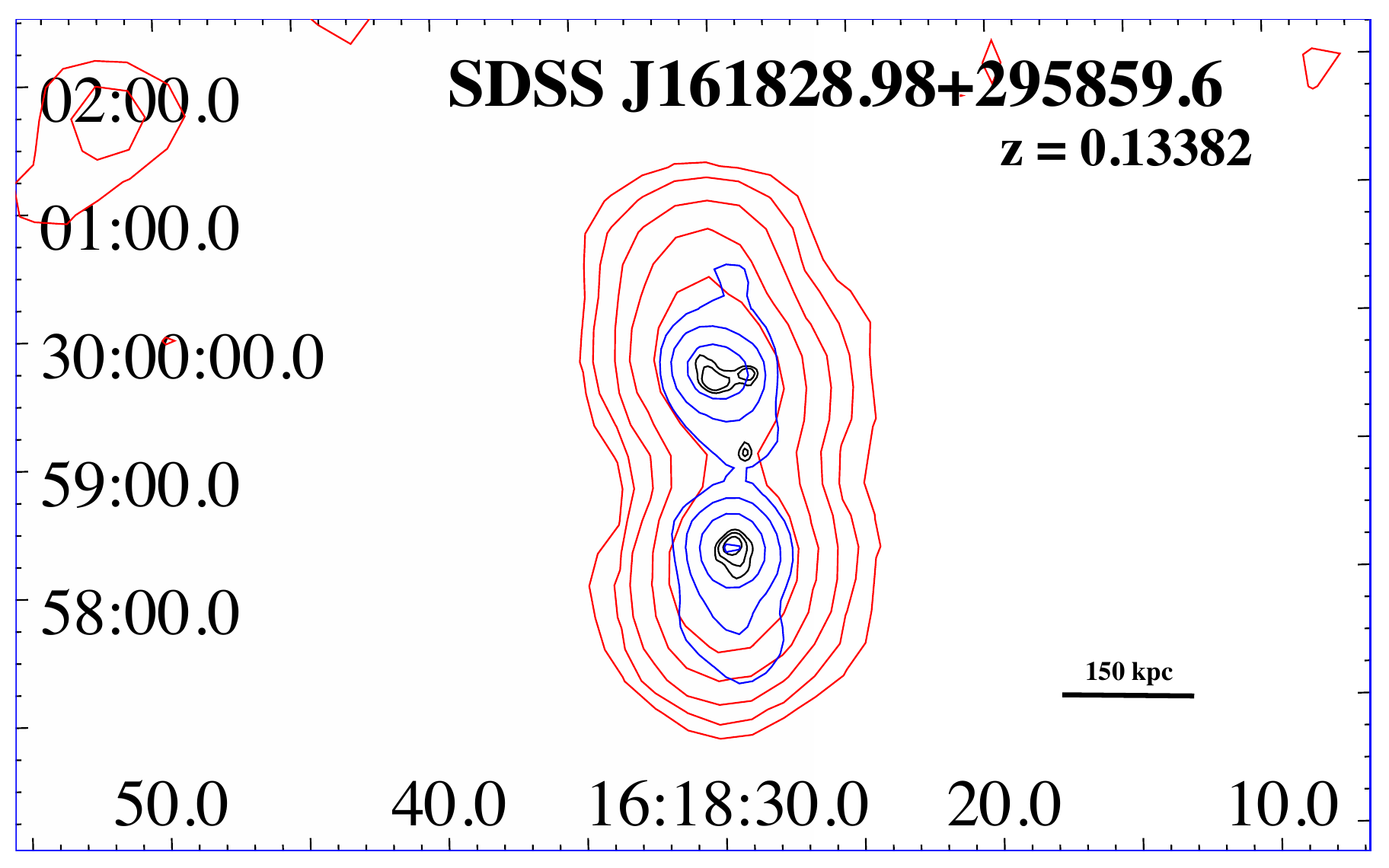} 
        \includegraphics[height=5.cm,width=6.cm,angle=0]{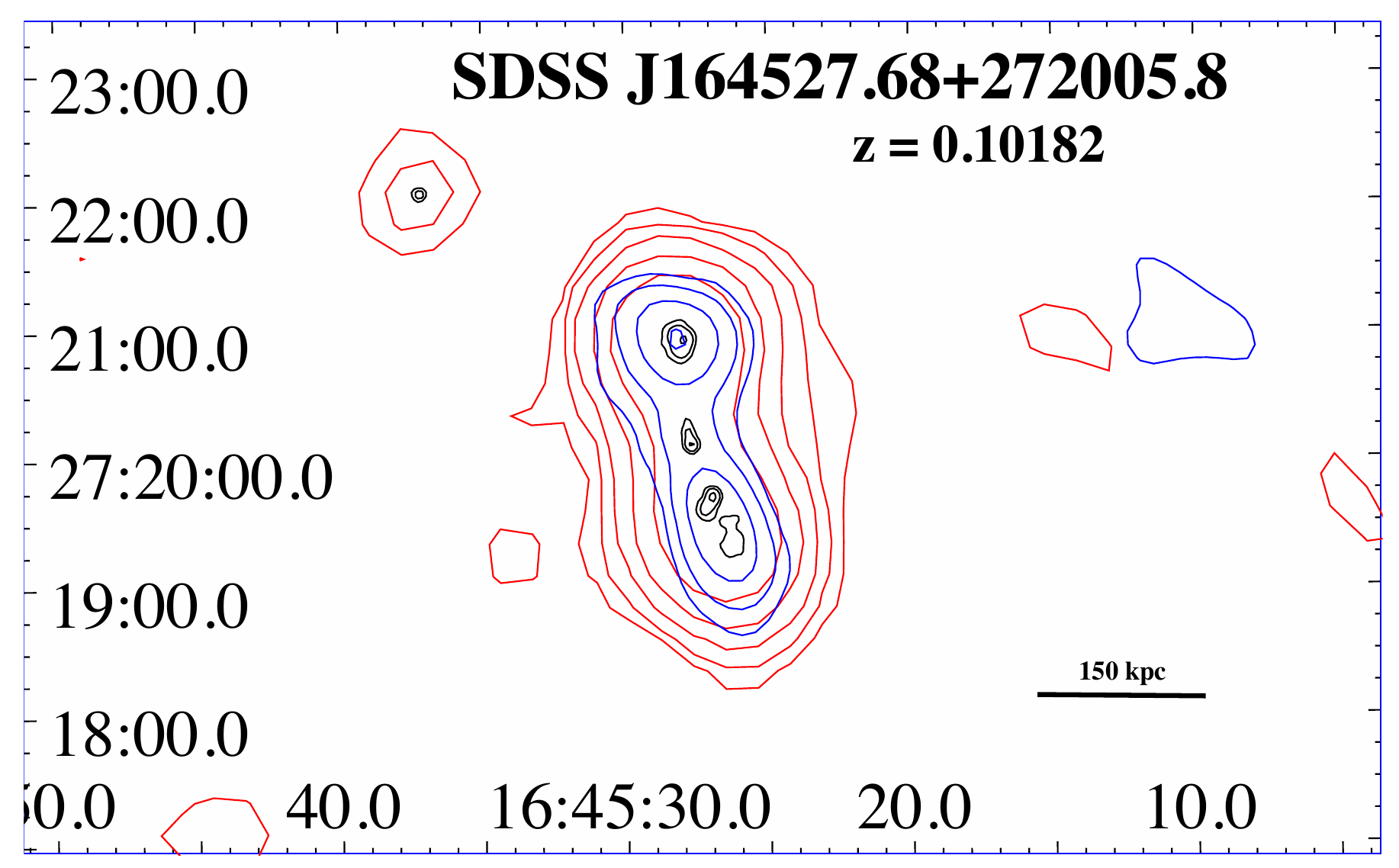} 
        \includegraphics[height=5.cm,width=6.cm,angle=0]{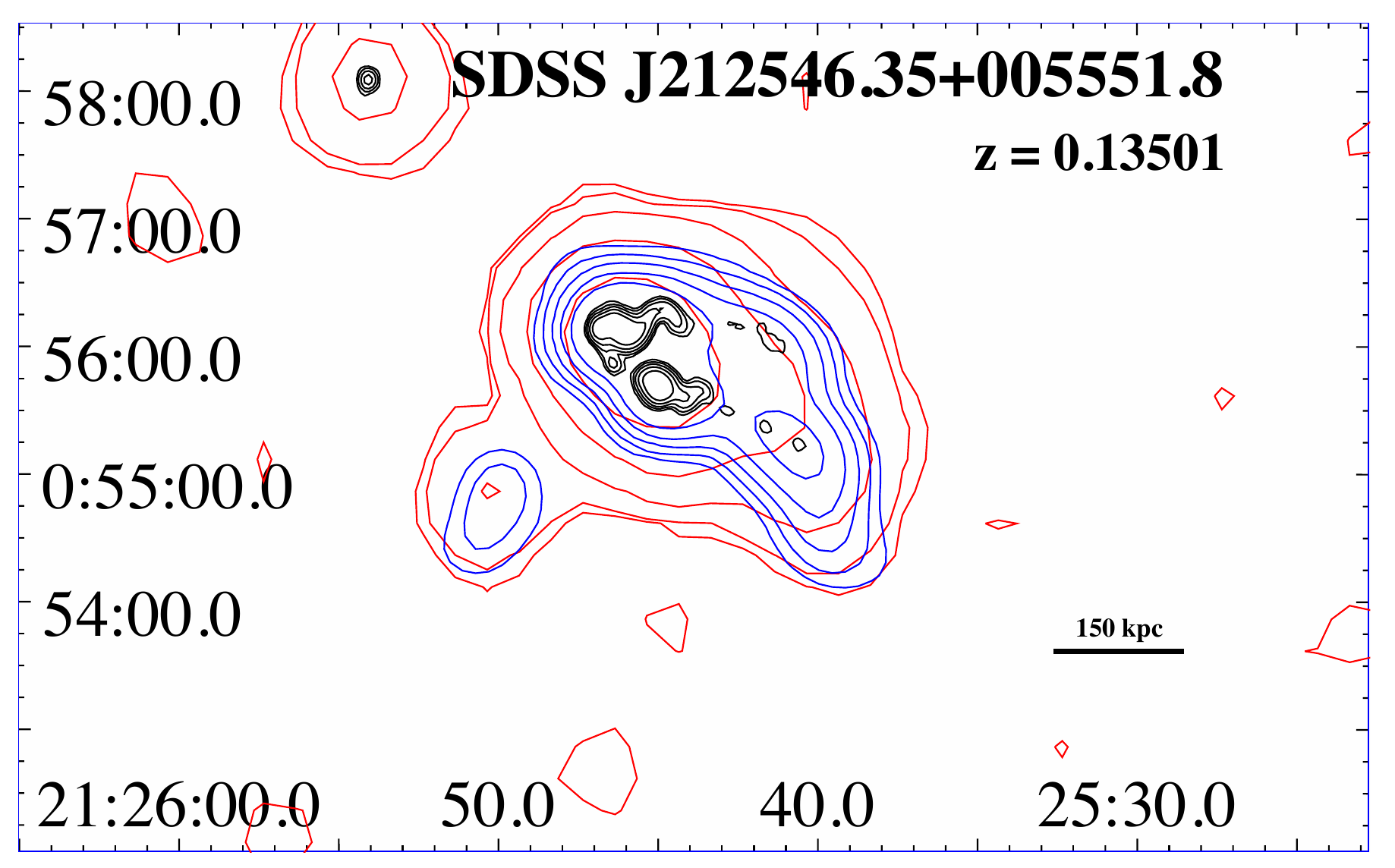} 
        \includegraphics[height=5.cm,width=6.cm,angle=0]{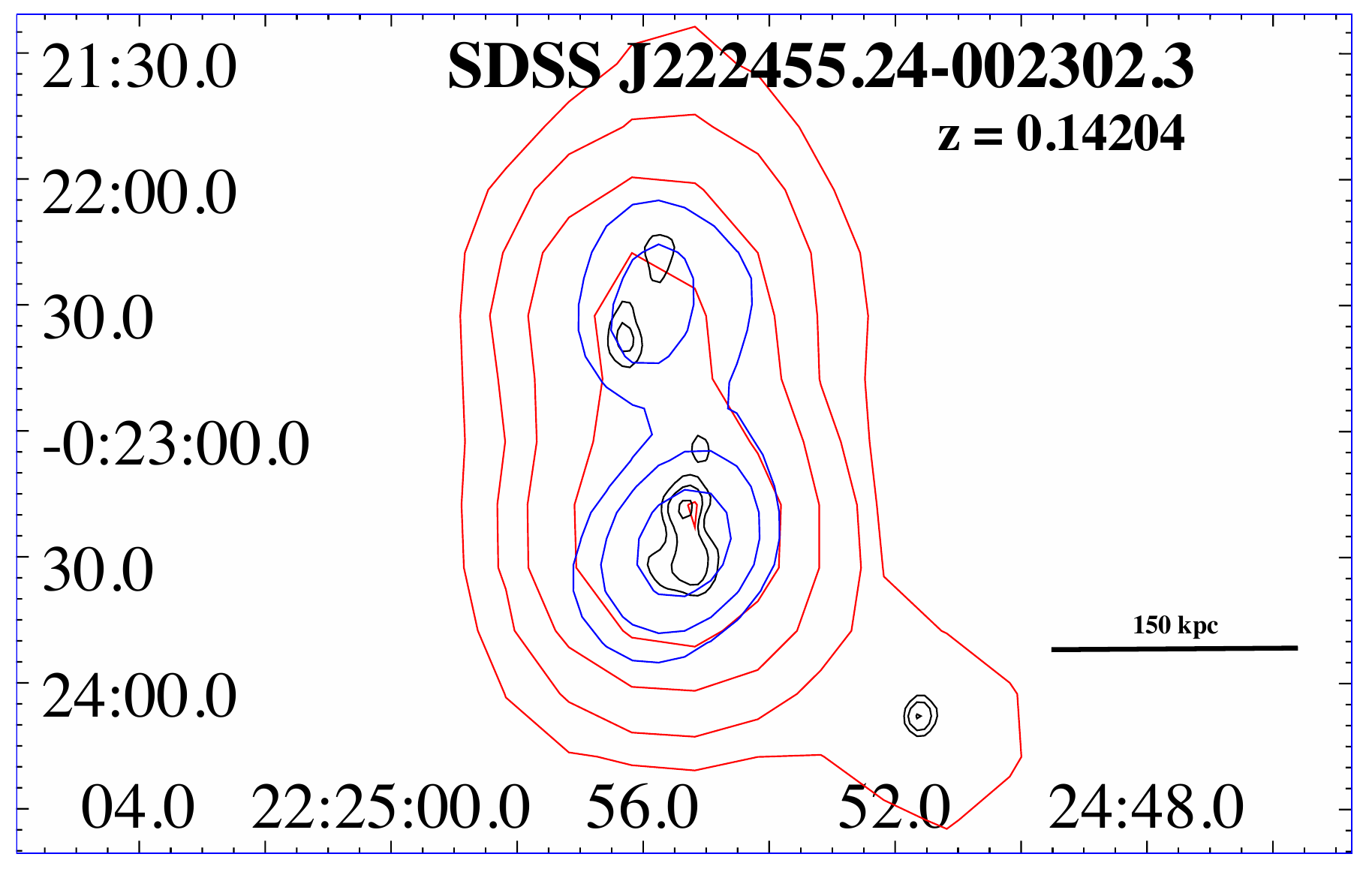} 
        \caption{(continued)}
\end{figure*}

\end{appendix}

\end{document}